\documentclass[a4paper,11pt]{article}
\pdfoutput=1 

\usepackage{jheppub} 

\usepackage[T1]{fontenc} 
\usepackage{amsmath,amssymb,graphicx}
\usepackage{subfigure}
\usepackage{multirow}
\usepackage{dcolumn}
\usepackage{color}
\usepackage{bm}
\usepackage{siunitx}
\usepackage{slashed}
\usepackage{natbib}
\usepackage{feynmp-auto}
\newcommand{\bea}{\begin{eqnarray}}
\newcommand{\eea}{\end{eqnarray}}
\newcommand{\beq}{\begin{eqnarray}}
\newcommand{\eeq}{\end{eqnarray}} 

\newcommand{\power}[2]{{#1}^{#2}}

\def\missE{\slashed E} 

\title{\boldmath New Physics in multi-Higgs boson final states}

\author[a]{Wolfgang Kilian}
\author[b,c]{Sichun Sun}
\author[d,e]{Qi-Shu Yan}
\author[f]{Xiaoran Zhao}
\author[a]{Zhijie Zhao}

\affiliation[a]{Department of Physics, University of Siegen, 57072 Siegen, Germany}
\affiliation[b]{Jockey Club Institute for Advanced Study, Hong Kong University of Science and Technology, Clear Water Bay, Hong Kong}
\affiliation[c]{Department of Physics, National Taiwan University, Taipei, Taiwan}
\affiliation[d]{School of Physics Sciences, University of Chinese Academy of Sciences, Beijing 100039, China}
\affiliation[e]{Center for future high energy physics, Chinese Academy of Sciences, Beijing 100039, China}
\affiliation[f]{Centre for Cosmology, Particle Physics and Phenomenology (CP3), Universit\'{e} catholique de Louvain, Chemin du Cyclotron, 2, B-1348 Louvain-la-Neuve, Belgium}

\emailAdd{kilian@physik.uni-siegen.de}
\emailAdd{sichunssun@gmail.com}
\emailAdd{yanqishu@ucas.ac.cn}
\emailAdd{xiaoran.zhao@uclouvain.be}
\emailAdd{zhao@physik.uni-siegen.de}

\preprint{SI-HEP-2017-05}

\abstract{We explore the potential for the discovery of the triple-Higgs signal
in the
$2b2l^{\pm}4j+\slashed E$ decay channel at a $100$ TeV hadron
collider.  We consider both the Standard Model and generic new-physics
contributions, 
described by an effective Lagrangian that includes higher-dimensional
operators.  The selected subset of operators is
motivated by composite-Higgs and Higgs-inflation models. In
the Standard Model, we perform both a parton-level and a 
detector-level analysis.  Although the parton-level results are
encouraging, the detector-level results demonstrate that this mode is
really challenging.  However, sizable contributions from new effective operators
can largely increase the cross section and/or modify the kinematics of
the Higgs bosons in the final state.  Taking into account the
projected constraints from single and double Higgs-boson production,
we propose benchmark points in the new physics models for the measurement of the triple-Higgs boson final state for future collider projects.}

\begin{document} 
\maketitle
\flushbottom

\section{Introduction}

After the discovery of the Higgs boson $h$ ($125$ GeV) at the
LHC~\cite{Aad:2012tfa,Chatrchyan:2012xdj}, measurements of the
Higgs self-couplings become crucial for our understanding of
fundamental particle physics. In the Standard Model (SM), the
Higgs boson has three types of interaction: (1)
the interactions with electroweak gauge bosons ($W^\pm$ and $Z$); (2)
the Yukawa interactions with fermions; (3) the triple and quartic
self-interactions.  A measurement of the last type of interaction
would complete the phenomenological reconstruction of the Higgs
potential~\cite{Dawson:2013bba} and thus should lift our knowledge
about electroweak 
symmetry breaking (EWSB) to a new level.  Furthermore, Higgs
self-interactions could be related to the problems of
baryogenesis~\cite{Trodden:1998ym} and vacuum stability~\cite{Ellis:2009tp,Degrassi:2012ry,Rojas:2015yzm}. 

In the SM, the Higgs potential is written as
\beq
  V(H^\dagger H) &=& -\mu^2(H^\dagger H) + \frac{\lambda}{4} (H^\dagger H)^2,
  \label{eq1}
\eeq
where $H=(G^+,\frac{1}{\sqrt{2}}(v+h+i G^0))^T$ is the Higgs
doublet, and $G^\pm,G^0$ are the unphysical Goldstone bosons
associated with spontaneous EWSB in a renormalizable gauge. This
potential has a minimum for the Higgs-field vacuum expectation value
$v=2 |\mu|/\lambda \approx 246$ GeV.  After 
EWSB and switching to unitarity gauge, the Higgs self-interactions
take the following form
\begin{eqnarray}
  V_\text{self} &=& \frac{\lambda}{4} v h^3+\frac{1}{16} \lambda  h^4,
  \label{eq2}
\end{eqnarray}
which corresponds to a triple-Higgs self-coupling $g_{hhh}=\frac{3}{2}
\lambda v$ and a quartic Higgs self-coupling $g_{hhhh}=\frac{3}{2}
\lambda$, respectively. The parameter $\lambda$ can be determined by
measuring the Higgs mass $m_h$, since $\lambda=\frac{2
  m_h^2}{v^2}$. In the SM, the Higgs potential is thus completely
fixed after the measurement of $m_h\approx 125$ GeV. However, the
story could be different if new physics can contribute to the Higgs
self-interactions.  Independently measuring the triple and quartic
couplings of the Higgs boson via double and triple-Higgs final states
is an essential project for future collider experiments.

Deviations from the SM that manifest themselves prominently in double
and triple-Higgs final-state processes are expected for various
new-physics scenarios. In order to study the Higgs
potential in a largely model-independent way, we will parameterize new
physics beyond the SM (BSM) in terms of an effective field theory
(EFT).  This systematic method captures the essence of a wide class of
BSM models.  It is well suited to collider studies that require
exclusive Monte Carlo simulations.

For concreteness, we will describe BSM Higgs physics in terms of the
strongly-interacting light Higgs (SILH) version~\cite{Giudice:2007fh}
of the EFT approach~\cite{Buchmuller:1985jz,Grzadkowski:2010es}.  The
operators in this choice of basis are designed to directly correspond
to low-energy effects of specific BSM Higgs-sector realizations,
including composite Higgs models~\cite{Kaplan:1983fs,
  Degrande:2012gr,Agashe:2004rs,Giudice:2007fh,Contino:2013kra} and
the Higgs inflation model~\cite{He:2015spf}.  We will consider
operators up to dimension 6.  Nonvanishing coefficients for some of
those, such as $\partial^\mu (H^\dagger H)\partial_\mu (H^\dagger H)
$, can substantially enhance multi-Higgs production rates and/or
modify final-state kinematics.

In the SM, the leading order (LO) for the production of one or more
Higgs bosons in gluon-gluon fusion involves one-loop diagrams.  The
calculation of higher order corrections becomes quite a
challenge. Most of these calculations~\cite{Dawson:1990zj,
  Dawson:1998py,Harlander:2002wh,Anastasiou:2002yz,Ravindran:2003um,deFlorian:2013jea,Anastasiou:2015ema,Degrassi:2016vss,deFlorian:2016uhr,deFlorian:2016sit}
are based on effective-theory methods, working in the limit of
infinite top-quark mass.  Regarding effects of finite top-quark mass,
only NLO QCD corrections to single-Higgs production are known
analytically~\cite{Spira:1995rr,Djouadi:1991tka}. One way to estimate
finite top-quark mass effects is series expansion, which can work well
for single-Higgs
production~\cite{Degrassi:2016vss,Harlander:2009mq,Pak:2009dg} but
converges poorly for double-Higgs production~\cite{Grigo:2015dia}.
Recently, NLO QCD corrections for double-Higgs production with full
top-quark mass dependence have been calculated
numerically~\cite{Borowka:2016ypz,Borowka:2016ehy}.  The results show
large differences in kinematical distributions compared to the
prediction of the infinite top-mass limit.

The feasibility of an analysis of double-Higgs production at the LHC
has become a hot
topic~\cite{Plehn:1996wb,
  Baur:2002rb,Baglio:2012np,Li:2013flc,Bhattacherjee:2014bca},
because this process probes the triple coupling $g_{hhh}$. The
dominant mode for double-Higgs production is gluon fusion via a box or
triangle loop of quarks. Various decay channels have been studied in
the literature, such as
$WWWW$~\cite{Baur:2002qd},
$b\bar{b}\gamma\gamma$~\cite{Baur:2003gp,Yao:2013ika,Kling:2016lay},
$b\bar{b}WW$~\cite{Papaefstathiou:2012qe},
$b\bar{b}\tau\tau$~\cite{Baur:2003gpa,Dolan:2012rv,Barr:2013tda},
$b\bar{b}\mu\mu$~\cite{Baur:2003gp},
$WW\gamma\gamma$~\cite{Lu:2015qqa} and
$b\bar{b}b\bar{b}$~\cite{Baur:2003gpa,deLima:2014dta,Behr:2015oqq}. It has been argued that the
triple self-coupling can be measured within $40\%$ accuracy at the
high luminosity LHC ($3$ ab$^{-1}$) with $14$ TeV
energy~\cite{Barger:2013jfa}, but recently more detailed studies have shed
doubt on this estimate~\cite{ATLAS:2016-Higgs-pair}.  At a future
$100$ TeV hadron collider~\cite{cern:fcc,ihep:sppc}, the rate for
double-Higgs production will be 
significantly higher.  The prospects for a measurement at such a
machine have been investigated
in Refs.~\cite{Barr:2014sga,Papaefstathiou:2015iba,Li:2015yia,Zhao:2016tai,Kling:2016lay,Contino:2016spe}.

The triple-Higgs self-coupling $g_{hhh}$ can also be measured at a
future lepton collider through the double Higgs-strahlung process
$e^+e^-\to Zhh$ or the vector-boson fusion process $e^+e^-\to
\nu\bar{\nu}hh$. It has been shown that $g_{hhh}$ can be measured within
$27\%$ accuracy at the luminosity-upgraded
ILC~\cite{Fujii:2015jha}. At a low-energy machine, such as the $250$ GeV
CEPC, the triple self-coupling could be determined indirectly via the loop
corrections to the $ZZh$ vertex~\cite{McCullough:2013rea,Sun:2015oea}.

By contrast, a measurement of the quartic self-coupling $g_{hhhh}$ is
a real challenge at the LHC, since at $\sqrt{s}=14$ TeV the cross
section of $gg\to hhh$ is only $\mathcal{O}(0.01)$
fb~\cite{Plehn:2005nk,Binoth:2006ym}. Alternatively, one can consider
$pp\to Zhhh$, but that cross section is also
tiny~\cite{Dicus:2016rpf}.  This problem cannot be solved at a lepton
collider either, because the cross section
for $e^+e^-\to Zhhh$ is only $\mathcal{O}(0.1)$ ab at a $\sqrt{s}=1$
TeV machine~\cite{Djouadi:1999gv}, too small for a measurement.
 
The proposals for future $pp$ colliders have motivated the study of
the process $gg\to hhh$ at high energy. The cross section of $gg\to
hhh$ at a $100$ TeV hadron collider can be estimated to be about $3$ fb
if NLO corrections are accounted for~\cite{Maltoni:2014eza}, which makes it
at least possible to observe the final states of this process. The
discovery potential of decay channels $hhh\to
b\bar{b}b\bar{b}\gamma\gamma$~\cite{Papaefstathiou:2015paa,Chen:2015gva}
and $hhh\to b\bar{b}b\bar{b}\tau\tau$~\cite{Fuks:2015hna} has
been explored.  It turns out that the discovery of three-Higgs
final state through these channels is challenging, and an
extreme high quality detector is needed.

In this paper, we investigate the sensitivity of the decay channel
$hhh\to b\bar{b}WW^*WW^*\to 2b2l^\pm 4j+\missE$, which has not been
carefully analyzed in the literature before. We also examine how new
physics can contribute to triple-Higgs production. We will consider
the effects of a set of dimension-6 effective operators to the cross
section and kinematics of Higgs bosons in the final state. Especially,
we extend the study of Ref.~\cite{He:2015spf} to the triple-Higgs
production case, where the effects of derivative operators on the
kinematics of Higgs bosons in double-Higgs production were
explored. We also study the projected bounds for all relevant
couplings in the EFT at the LHC and at a future 100 TeV $pp$ collider.

This paper is organized as follows. In
Sec.~\ref{Sec:EFT}, we briefly introduce the
EFT Lagrangian as appropriate for our study and relate our
parameterization to particular models that are of interest in the
context of new Higgs-sector BSM physics.  In
Sec.~\ref{Sec:analysis}, we present a Monte Carlo
(MC) analysis of $hhh\to b\bar{b}WW^*WW^*\to 2b2l^\pm 4j+\missE$ in
the SM, and investigate the discovery potential and identify challenges
of this channel. In Sec.~\ref{Sec:hhh-EFT}, we
describe the calculation of triple-Higgs production in the context of
the EFT with dimension-six operators in detail and
present our numerical results.  We conclude this paper with a
discussion of our findings in
Sec.~\ref{Sec:conc}.

\section{Effective Lagrangian up to dimension-6 operators}
\label{Sec:EFT}
It has been accepted for a long time that new-physics effects
associated with a characteristic scale higher than the energy of the
processes under study, can be conveniently expressed in terms of a
low-energy EFT.  This is a local Lagrangian which includes an infinite
series of operators of dimension greater than four, constructed as
monomials of fields and organized in terms of the canonical
dimension.  The operators may incorporate only the unbroken Lorentz,
electromagnetic and colour symmetries~\cite{Burgess:1992gz}.  However,
our knowledge of flavor data, electroweak precision data, and Higgs
properties strongly suggests to furthermore implement the power counting of
EWSB and thus build operators out of
classically gauge-invariant combinations under the full electroweak
symmetry.  Up to dimension four, this
reproduces the SM.  The set of operators up to dimension six
was introduced in Ref.~\cite{Buchmuller:1985jz} and has been reworked
to a minimal basis in Ref.~\cite{Grzadkowski:2010es}.  Adopting this as
a phenomenological model implies rather generic assumptions on
the flavor and gauge structure of the underlying fundamental theory.

In the present context, we are more specifically interested in the
possibility that Higgs self-couplings act as primary probes to
new-physics effects, while other SM fields are affected only by
secondary 
corrections.  This notion is realized by scenarios where the Higgs
field acts as the only SM field with sizable couplings to a new
sector.  Specific models with this property have been proposed, e.g.,
in Refs.~\cite{Kaplan:1983fs,Agashe:2004rs}. A general discussion can be
found in Ref.~\cite{Giudice:2007fh} where the resulting effective
low-energy Lagrangian, expanded up to dimension six, has been
introduced as the SILH Lagrangian.  As expected, and confirmed
in Ref.~\cite{Contino:2013kra}, this Lagrangian is equivalent to the basis
of Ref.~\cite{Grzadkowski:2010es}, but the assumptions
of Ref.~\cite{Giudice:2007fh} on the underlying dynamics suggest a
hierarchy between induced tree-level and loop-level coefficients that
allows for dropping part of the operator set and thus keeping a more
economical number of phenomenological parameters.  If we follow this
line of reasoning, we can adopt the SILH Lagrangian as the basis of
the present phenomenological study.  We supply a more detailed
discussion below in Sec.~\ref{sec:SILH}.

For the actual
applications in later sections, we can focus on the
interactions of the physical Higgs field $h$, after
EWSB and expressed in unitarity gauge.  The Lagrangian
reduces to
\beq
\label{eft}
  \mathcal{L}_{EFT}
  &=&
  \mathcal{L}_{SM}+\mathcal{L}_t+\mathcal{L}_h + \mathcal{L}_{ggh}
  , \\
  \mathcal{L}_t
  &=&
  -a_1\frac{m_t}{v}\bar{t}t \, h -a_2 \frac{m_t}{2 v^2}\bar{t}t \, h^2
  -a_3\frac{m_t}{6 v^3}\bar{t}t \, h^3
  , \\
  \mathcal{L}_h
  &=&
  -\lambda_3\frac{m_h^2}{2v}h^3-\frac{\kappa_5}{2v}h\partial^\mu{h}\partial_\mu{h}
  -\lambda_4\frac{m_h^2}{8v^2}h^4-\frac{\kappa_6}{4v^2}h^2\partial^\mu{h}\partial_\mu{h}
  , \label{eq8} \, \\
  \mathcal{L}_{ggh}
  &=&
  \frac{g_s^2}{48 \pi^2}\left(c_1 \frac{h}{v} +
  c_{2} \frac{h^2}{2v^2}\right) G^a_{\mu\nu} G^{a\, \mu\nu}
\eeq
Here we confine to the CP conserving operators and omit the CP
violating operators.  In the SM, we have $a_1=\lambda_3=\lambda_4=1$
and $a_2=a_3=\kappa_5=\kappa_6=c_1= c_2=0$.  It is understood
that the corresponding terms have been removed from
$\mathcal{L}_{SM}$, such that they are not double-counted.


Another set of models that couple the Higgs sector to new physics is
provided by certain models of inflation.  As we show below in
Sec.~\ref{sec:inflation}, this effectively results in the same Higgs
Lagrangian, Eq.(\ref{eft}).  In Sec.~\ref{sec:genEFT} we briefly
review the relation to the EFT version
of Refs.~\cite{Buchmuller:1985jz,Grzadkowski:2010es} as it has been applied
to the Higgs sector in Ref.~\cite{Corbett:2012ja}.  Finally, it can be
shown that in a framework that implements a non-linear realization of
electroweak symmetry, the result is again equivalent to SILH if
equivalent assumptions on coefficient hierarchies are
taken~\cite{Buchalla:2014eca}.

In summary, the phenomenological Lagrangian~(\ref{eft}) provides a
robust parameterization of new physics in the Higgs sector under the
condition that no new on-shell states appear in the kinematically
accessible range.

\subsection{The SILH Lagrangian in relation to composite Higgs models}
\label{sec:SILH}
The relevant part of the SILH
Lagrangian~\cite{Giudice:2007fh,Contino:2013kra}, including operators
up to dimension six, has the form
\beq
{\cal L}_\text{SILH} &&= \frac{c_H}{2f^2}\partial^\mu \left( H^\dagger H \right) \partial_\mu \left( H^\dagger H \right) 
+ \frac{c_T}{2f^2}\left (H^\dagger {\overleftrightarrow { D^\mu}} H \right)  \left(   H^\dagger{\overleftrightarrow D}_\mu H\right) \nonumber 
- \frac{c_6\lambda}{f^2}\left( H^\dagger H \right)^3 \\ 
&& + \left( \frac{c_yy_f}{f^2}H^\dagger H  {\bar f}_L Hf_R +{\rm h.c.}\right) \nonumber+\frac{c_g g_S^2}{16\pi^2f^2}\frac{y_t^2}{g_\rho^2}H^\dagger H G_{\mu\nu}^a G^{a\mu\nu} \\ 
&&+\frac{ic_Wg}{2m_\rho^2}\left( H^\dagger  \sigma^i \overleftrightarrow {D^\mu} H \right )( D^\nu  W_{\mu \nu})^i
+\frac{ic_Bg'}{2m_\rho^2}\left( H^\dagger  \overleftrightarrow {D^\mu} H \right )( \partial^\nu  B_{\mu \nu})  \nonumber \\
&& +\frac{ic_{HW} g}{16\pi^2f^2}
(D^\mu H)^\dagger \sigma^i(D^\nu H)W_{\mu \nu}^i
+\frac{ic_{HB}g^\prime}{16\pi^2f^2}
(D^\mu H)^\dagger (D^\nu H)B_{\mu \nu}
 \nonumber \\
&&+\frac{c_\gamma {g'}^2}{16\pi^2f^2}\frac{g^2}{g_\rho^2}H^\dagger H B_{\mu\nu}B^{\mu\nu}.
\label{lsilh}
\eeq
It includes all the CP-conserving gauge-invariant operators up to
dimension six with pure Higgs interactions and Higgs-gauge boson
interactions.  Some operators such as $H^\dagger H
W_{\mu\nu} W^{\mu\nu}$ are not included here since they can be
generated by integration by parts from the other operators. There are
further operators with fermions coupling to the Higgs, which are omitted here.

There is only one dimension-5 operator allowed by the SM gauge
symmetry, up to Hermitian conjugation and flavour assignments: $(H
\ell_i)^T C(H \ell_j)$.  It gives rise to the neutrino Majorana mass
and violates lepton number, so we do not include it, either.

The SM Higgs may appear as a composite pseudo Nambu-Goldstone (NG)
boson associated with some enlarged symmetry beyond the SM.  The
Lagrangian ${\cal L}_\text{SILH}$ then emerges at low energy via
spontaneous breaking of that symmetry. Since any terms in the Higgs
potential will violate the shift symmetry of this NG-boson Higgs, the
coefficients above are all suppressed by the small breaking in relation
to the compositeness scale $f$, i.e., carrying a $\xi=\frac{v^2}{f^2}$
factor. $m_\rho, g_\rho$ stand for the characteristic mass and
coupling of a strongly coupled sector, respectively, and $c_i \sim 1$.

We focus on the first five operators in
Eq.~(\ref{lsilh}), since they are the relevant operators for the
hadron-collider processes that
we want to study. The first three terms in
${\cal L}_\text{SILH}$ contribute to the Higgs potential.  They contain
only two independent terms, as can be verified by applying the
equations of motion.  After EWSB, the SILH potential reduces to the
effective potential of Eq.~(\ref{eft}). We list the 
relations between Eq.~(\ref{eft}) and Eq.~(\ref{lsilh}) in
Table~\ref{tablecomposite}. Note that we have the relation
$\kappa_{5}=\kappa_{6}$, since the associated terms come from the same
operator $\frac{c_H}{2f^2}\partial^\mu \left( H^\dagger H \right)
\partial_\mu \left( H^\dagger H \right)$. The rest of the operator
coefficients can be 
measured at future electron-positron colliders, via $W$-pair
production, $Z$-pair production, and $Z$-Higgs
production~\cite{Djouadi:2007ik,Khanpour:2017cfq}.

Regarding hadron-collider measurements, the coefficient $c_g$ is
accessible via the $pp\to h$ process at the LHC.  Run-$1$ data
have constrained $c_g/m^2_\rho\sim
10^{-6}$~\cite{Aad:2015tna}.  Bounds for the coefficients
$c_H$, $c_y$ and $c_6$ are currently much weaker~\cite{Ellis:2014jta}. It is
expected that the high-luminosity LHC will yield bounds
$c_H\xi\in [-0.044,0.035]$ and $c_y\xi\in [-0.020,0.008]$ for the top
quark~\cite{Englert:2015hrx}. The coefficient $c_H\xi$ can be further
constrained to $\mathcal{O}(10^{-3})$ at a future $e^+e^-$ collider,
and the tests for $c_y$ can be extended to $b$, $c$ quarks, and
leptons~\cite{Ge:2016zro}. The coefficient $c_6$ contributes to the
triple and 
quartic Higgs self-couplings only, so the bounds on $c_6$ will stay relatively
weak for both LHC and a future lepton collider.

We may also consider two more specific composite Higgs
models~\cite{Agashe:2004rs,Giudice:2007fh}, dubbed as MCHM4 and MCHM5, 
respectively.  Both models result in the SILH Lagrangian as their
low-energy EFT.  They contain extra fermions, which are in
representations 4 and 5 of an assumed global $SO(5)$ symmetry,
respectively. We adopt the 
notation from Ref.~\cite{Kanemura:2014kga}.   The SILH coefficient
values are
\beq
\text{MCHM4:}\quad c_H=1, \quad c_y=0, \quad c_6=1\,,\\
\text{MCHM5:}\quad c_H=1, \quad c_y=1, \quad c_6=0\,.
\eeq

The current LHC constraints and electroweak precision data imply $f
\leq 550$ GeV and $v^2/f^2 \leq 0.2$~\cite{Bellazzini:2014yua}. Later,
we will study the projected constraints from the LHC and a 100 TeV
collider.
\begin{center}
\begin{table}
  \begin{center}
  \begin{tabular}{|c|c|c|c|}
  \hline
  Parameters           &  SILH   with  Eq. (\ref{lsilh})   &MCHM4 & MCHM5      \\ 
  \hline
  $a_1$                &  $(1-\frac{3}{2}c_y\xi)(1-\frac{1}{2}c_y\xi)^{-1}(1+c_H\xi)^{-1/2}$         &$1-\frac{1}{2} \xi$ & $1-\frac{3}{2} \xi$    \\ 
  \hline
  $a_2$                &  $-3c_y\xi(1-\frac{1}{2}c_y\xi)^{-1}(1+c_H\xi)^{-1}$          &$0$ & $-3\xi$  \\ 
  \hline
  $a_3$                &  $-3c_y\xi(1-\frac{1}{2}c_y\xi)^{-1}(1+c_H\xi)^{-3/2}$        & 0 & $-3\xi$\\ 
  \hline
  \hline
  $c_1$       &  $\frac{1}{4}c_g\xi\frac{y^2_t}{g^2_\rho}$   &  $ \frac{1}{4} \xi \frac{y_t^2}{g_\rho^2}$ &$\frac{1}{4} \xi \frac{y_t^2}{g_\rho^2}$ \\
  \hline
   $c_2$      &  $c_1$  & $c_1$ & $c_1$ \\
  \hline \hline
  $\kappa_5$              &  $-2c_H\xi(1+c_H\xi)^{-3/2}$   &$-2\xi$  &$-2 \xi$  \\
  \hline
  $\kappa_6$              &  $-2c_H\xi(1+c_H\xi)^{-2}$   &$-2\xi$  &$-2 \xi$ \\
  \hline \hline
  $\lambda_3$          &  $(1+\frac{5}{2}c_6 \xi)(1+\frac{3}{2}c_6 \xi)^{-1}(1+c_H\xi)^{-1/2}$  &$1+\frac{\xi}{2}$ & $1-\frac{1 }{2} \xi$ \\
  \hline
  $\lambda_4$          &  $(1+\frac{15}{2}c_6 \xi)(1+\frac{3}{2}c_6 \xi)^{-1}(1+c_H\xi)^{-1}$   &$1+5\xi$ & $1- \xi$      \\
  \hline
  \end{tabular}
  \end{center}
    \caption{\label{tablecomposite}Parameter relationship between
      our convention and that in SILH, Eq.~\ref{lsilh}. The MCHM4 and
      MCHM5 models are from
      Refs.~\cite{Agashe:2004rs,Contino:2006qr}. Some notation is from
      Ref.~\cite{Gillioz:2012se,Giudice:2007fh}. Note that $c_1$ and $c_2$ 
      are sensitive to the detailed construction of the models. We
      consider $c_g$ as roughly of order 1.
      For the relation between our conventions and other conventions
      used in the literature, cf. Appendix \ref{appenda}.
    }
\end{table}
\end{center}

\subsection{Operators from Higgs inflation}
\label{sec:inflation}

In this section, we demonstrate how an equivalent set of dimension-six
operators arises from the standard Higgs inflation
paradigm~\cite{Bezrukov:2007ep,Bezrukov:2009db,Hamada:2015skp,Sun:2013cza, 
  Sun:2014jha}.  We incorporate a non-minimal coupling of the Higgs
field to gravity and work in unitarity gauge where $H=(0,h/\sqrt{2})$. The
gauge interactions are more complicated in this scenario; we ignore
them for now and just focus on the Higgs potential.  In the Jordan
frame, the Lagrangian has the form
\begin{align}
  \label{eq:1}
    S_{Jordan} =\int d^4x \sqrt{-g} \Bigg\{&
    - \frac{M^2+\xi h^2}{2}R
    + \frac{(\partial h)^2}{2}
    -\frac{1}{2}m_h h^2-\frac{\lambda}{4} h^4
    \Bigg\}
    \;.
\end{align}
We consider $\xi$ in the range $1\ll\sqrt{\xi}\lll10^{17}$, in
which $M \simeq M_\text{PLanck}$.

We perform a conformal transformation from the Jordan frame to the
Einstein frame,
\begin{equation}
  \label{eq:2}
  \hat{g}_{\mu\nu} = \Omega^2 g_{\mu\nu}
  \;,\quad
  \Omega^2 = 1 + \xi h^2/M_\text{Planck}^2
  \;.
\end{equation}
This transformation will give rise to derivative terms in Higgs
potentials. We furthermore redefine
\begin{equation}
  \label{eq:3}
  d\chi=\sqrt{\frac{\Omega^2+6\xi^2h^2/M_\text{Planck}^2}{\Omega^4}} dh.
\end{equation}
Then the action in the Einstein frame is given by
\begin{align}
  \label{eq:4}
    S_E =\int d^4x\sqrt{-\hat{g}} \Bigg\{
    - \frac{M_\text{Planck}^2}{2}\hat{R}
    + \frac{\partial_\mu \chi\partial^\mu \chi}{2}
    - V(\chi)
    \Bigg\},
 \end{align}
where the potential becomes
\beq
  \label{eq:5}
  V(\chi) =
  \frac{1}{\Omega(\chi)^4}\left[\frac{\lambda}{4} h(\chi)^4+\frac{1}{2}m_h h(\chi)^2\right].
\eeq

In the standard Higgs inflation paradigm, $h$ takes large values
$h\gg M_\text{Planck}/\sqrt{\xi}$ (or $\chi\gg\sqrt{6}M_\text{Planck}$)
during inflation and plays the role of the inflaton.  We have the expressions
\begin{align}\label{eq:hlarge}
  h\simeq \frac{M_\text{Planck}}{\sqrt{\xi}}\exp\left(\frac{\chi}{\sqrt{6}M_\text{Planck}}\right), \quad
  V(\chi) = \frac{\lambda M_\text{Planck}^4}{4\xi^2}
  \left(
    1+\exp\left(
      -\frac{2\chi}{\sqrt{6}M_\text{Planck}}
    \right)
  \right)^{-2}
\end{align}
This allows the potential to be exponentially flat at large $h$ to
produce a viable inflaton potential.

When the value of $h$ is near the origin as today, we can approximate
$h\simeq\chi$ and $\Omega^2\simeq1$, so the potential for the field
$\chi$ generates a potential for the SM model Higgs field plus
corrections at $O(\xi^2/M_\text{Planck}^2)$. For the purpose of this
collider study, we thus replace $\chi$ by $h$. Plugging Eq.~(\ref{eq:3})
into Eq.~(\ref{eq:4}) and omitting higher order terms, we arrive at
\begin{align}
  \label{eq:4p}
\nonumber & S_E =\int d^4x\sqrt{-\hat{g}} \bigg\{
    - \frac{M_\text{Planck}^2}{2}\hat{R}
    + \frac{\partial_\mu h \partial^\mu h}{2\Omega^2} +\frac{3 \xi}{M_\text{Planck}^2}\frac{h^2\partial_\mu h \partial^\mu h}{ \Omega^4}\\
 & - (1 - \frac{2\xi h^2}{M_\text{Planck}^2})\left[\frac{\lambda}{4} h(\chi)^4+\frac{1}{2}m_h h(\chi)^2\right]
   \bigg \} \,.
 \end{align}
Note that after EWSB, replacing $h\rightarrow h+v$ yields similar extra terms as in Eq.~(\ref{eft}).

\subsection{Alternative Parameterization of the Higgs boson self-interaction
  operators}
\label{sec:genEFT}

Another representation of the set of gauge-invariant dimension-6
operators which can modify the Higgs self-interactions, has been
studied in Ref.~\cite{Corbett:2012ja}
\begin{eqnarray}
   \mathcal{O}_{1} &=& \frac{f_1}{\Lambda^2}(D^\mu H)^\dagger HH^\dagger(D_\mu H), \\ 
   \mathcal{O}_{2} &=& \frac{f_2}{2\Lambda^2}\partial^\mu(H^\dagger H)\partial_\mu (H^\dagger H), \\
   \mathcal{O}_{3} &=& \frac{f_3}{3\Lambda^2}(H^\dagger H)^3, \\ 
   \mathcal{O}_{4} &=& \frac{f_4}{\Lambda^2}(D^\mu H)^\dagger(D_\mu H)(H^\dagger H).
   \label{eq5}
\end{eqnarray}
The operator $\mathcal{O}_1$ was considered in
Ref.~\cite{Ellis:2014jta} and can safely be neglected. In the subset
$(\mathcal{O}_2,\mathcal{O}_3,\mathcal{O}_4)$, one operator can be
eliminated by the equations of motion, so we drop $\mathcal{O}_4$. Thus
we only need to consider the operators $(\mathcal{O}_2,\mathcal{O}_3)$.

As mentioned in Ref.~\cite{He:2015spf}, the operator $\mathcal{O}_2$
induces a derivative term for the Higgs field
\begin{eqnarray}
  \mathcal{O}_2 &\to& \frac{f_2}{2\Lambda^2}(v+h)^2\partial^\mu{h}\partial_\mu{h}.
  \label{eq6}
\end{eqnarray}
Therefore the kinetic term of the Higgs field is modified to
\begin{eqnarray}
  \mathcal{L}_{kin} &=& \frac{1}{2}\left(1+\frac{f_2v^2}{\Lambda^2}\right)\partial^\mu{h}\partial_\mu{h}.
  \label{eq7}
\end{eqnarray}
This means that the Higgs field should be rescaled by $h\to\zeta h$, where $\zeta=(1+f_2v^2/\Lambda^2)^{-1/2}$.

After EWSB and choosing unitarity gauge, the Lagrangian reduces
to~(\ref{eft}) as before, where the
coefficients 
$(a_1,\lambda_3,\lambda_4,\kappa_5,\kappa_6)$ of Eq.~(\ref{eft}) can be
expressed in terms of just
two independent parameters:
\begin{eqnarray}
  \hat{x} &=& x_2\zeta^2, \\
  \hat{r} &=& -x_3\zeta^2\frac{2v^2}{3m_h^2}, 
  \label{eq9}
\end{eqnarray}
where $x_i=f_iv^2/\Lambda^2$ $(i=2,3)$. With this definition, the
rescaling factor $\zeta$ can be rewritten as
$\zeta=(1-\hat{x})^{1/2}$. The relations between our parameters and
those in Ref.~\cite{He:2015spf} are listed in Table~\ref{table5}.

\begin{center}
\begin{table}
  \begin{center}
  \begin{tabular}{|c|c|c|}
  \hline
  Our operators                             &  Operators in Ref.~\cite{He:2015spf}    &  Relations       \\ 
  \hline
  $-\frac{m_t}{v}a_1\bar{t}th$                &  $-\frac{m_t}{v}\zeta\bar{t}th$          &  $a_1=\zeta$   \\ 
  \hline
  $-\lambda_3\frac{m_h^2}{2v}h^3$             &  $-\frac{\zeta}{2v}(1+\hat{r})m_h^2h^3$    &  $\lambda_3=\zeta(1+\hat{r})$   \\ 
  \hline
  $-\lambda_4\frac{m_h^2}{8v^2}h^4$  & $-\frac{\zeta^2}{8v^2}(1+6\hat{r})m_h^2h^4$  & $\lambda_4=\zeta^2(1+6\hat{r})$ \\
  \hline
  $-\frac{1}{2v}\kappa_5h(\partial h)^2$         &  $\frac{1}{v}\hat{x}\zeta h(\partial h)^2$ &  $\kappa_5=-2\hat{x}\zeta$        \\
  \hline
  $-\frac{\kappa_6}{4v^2}h^2(\partial h)^2$  &  $\frac{\hat{x}}{2v^2}\zeta^2h^2 (\partial h)^2$  & $\kappa_6=-2\hat{x}\zeta^2$ \\
  \hline
  \end{tabular}
  \end{center}
  \caption{\label{table5}Parameter relationship between our
    convention and that in Ref.~\cite{He:2015spf}.} 
\end{table}
\end{center}

At the time when the measurements that we discuss in the present work
can be carried out, we should expect that data exist that set significant
bounds on the parameters $\hat{x}$ and~$\hat{r}$.
\begin{enumerate}
  \item $a_1$ ($\hat{x}$) is related to the direct measurement of the
    top Yukawa coupling, and its value is expected to become
    determined within $5\%$
    precision at the high-luminosity LHC~\cite{Brock:2014tja}, via
    measuring the $tth$ production rate. At a 100 TeV collider, the Yukawa
    coupling can be pinpointed down to a precision
    $1\%$~\cite{Plehn:2015cta} by measuring the ratio between the
    $tth$ and $ttZ$ 
    production rates.
  \item Another bound on $\hat{x}$ is obtained from the measurement of
    Higgs-gauge couplings~\cite{Ellis:2014jta}, since they become
    universally rescaled by $\zeta$. A future $e^+e^-$ Higgs factory
    can constrain $|\hat{x}|$ at the $1\%$
    level~\cite{Craig:2014una}. Since there are many other dimension-6
    operators which can contribute to the gauge-boson kinetic terms,
    we nevertheless take $\hat{x}$ as a free parameter in 
    our later analysis.
  \item The parameter $\hat{r}$ can only be constrained by
    double-Higgs or triple-Higgs production.  Concerning double-Higgs
    boson production, the bound will be around $40\sim 100 \%$ at the
    HL-LHC at most. At a $100$ TeV hadron collider,
    $(\hat{x},\hat{r})$ will become more strongly constrained by
    double-Higgs production. As shown in
    Ref.~\cite{He:2015spf}, the bounds on $\hat{x}$ and $\hat{r}$ will
    be of the order $2\sim 5\%$ and $4\sim 13\%$, respectively.

\end{enumerate}

\section{Detailed analysis of the $2b2l^{\pm}4j+\slashed E$ channel in
  the SM}
\label{Sec:analysis}

We study triple-Higgs production in high-energy proton-proton
collisions, $pp\to hhh$, where one Higgs boson decays into a $b\bar b$
pair while the two other Higgses decay into $WW^\ast$.  The semi-virtual
$W$ pairs can subsequently decay semileptonically, $h\to WW^*\to \ell\nu
jj$.

The dominant partonic contribution to the $pp\to hhh$ signal is gluon-gluon
fusion, $gg\to hhh$.  This process involves one-loop diagrams.  As we
did for our previous work~\cite{Chen:2015gva}, we compute the
production matrix element at LO with
MadLoop/aMC@NLO~\cite{Pittau:2012fn}.  We take the parton distribution
functions from CTEQ6l1~\cite{Pumplin:2002vw}.  For phase-space
evaluation and exclusive event generation, we interface the production
process with VBFNLO~\cite{Arnold:2008rz, Arnold:2011wj,
  Baglio:2014uba}.

Background event samples are generated by MadGraph
5~\cite{Alwall:2014hca,Hirschi:2015iia}.  Since we require a $b\bar b$
pair, the dominant background is caused by top-quark pairs in
association with electroweak bosons, namely $pp\to
h(WW^*)t\bar{t}$ and $pp\to t\bar{t}W^-W^+$.  Both classes of
processes can lead to the same final state as the signal.  To veto
further background from $Z$ bosons, we restrict the analysis to same-sign
leptons in the final state, $l^+ l^+$ or $l^-l^-$.

We list the calculated cross sections of signal and backgrounds at
$100\;\mathrm{TeV}$ in Table~\ref{table1}.  In the absence of a
complete NLO calculation for the signal, we adopt the K-factor of
$2.0$ that was obtained in Ref.~\cite{Papaefstathiou:2015paa} for Higgs
pair production.  For the $H(WW^*)t\bar{t}$ background, we use
$K=1.2$~\cite{Dawson:2003zu}.
The K-factor for $t\bar{t}W^-W^+$ at $100$ TeV is taken $1.3$ from Ref.~\cite{Zhao:2016tai}.
In the Ref. ~\cite{Maltoni:2015ena}, a K-factor around 1.2 was obtained while the total cross section $\sigma_{\textrm{NLO}}$ is given as $1.3$pb, which is around $1.4$ times larger than our LO cross section.
The derivation is mainly attributed to the choice of renormalisation and factorisation scales, i.e. our choice of K-factor equal to $1.3$ is consistent with the results given in Ref. ~\cite{Maltoni:2015ena} after taking these uncertainties into account.

We ignore all background from $h+$jets, $hh+$jets and
$W^{\pm}W^{\pm}+$jets, since the cross sections of those processes are
negligible compared to the $h(WW^*)t\bar{t}$ background.  Furthermore, we
observe that the total cross section of the background
$b\bar{b}W^-W^+W^-W^+$ is essentially exhausted by the resonant
contribution $t\bar{t}W^-W^+$.  Therefore, we approximate the former
process by the latter with subsequent top-quark decay, which
considerably simplifies the calculation.

We have three comments on the background processes $hhjj$ in the SM and new physics models.
\begin{itemize}
\item 
In SM, the $hhjj$ final state receives contribution for heavy-quark loop and vector boson fusion, while the former is dominant.
Currently, the cross section of loop-induced processes with 2 jets can be calculated by interfacing GoSam~\cite{Cullen:2014yla} or OpenLoops~\cite{Cascioli:2011va} to Madgraph5~\cite{Dolan:2015zja} or Herwig7~\cite{Bellm:2016cks}.
We use Madgraph5 to compute the cross section of top quark loop induced $pp \to h h j j$ at a 100 TeV collider. After imposing the MLM matching\cite{Mangano:2006rw} and using cuts $P_t(j)>20$ GeV and $\eta(j)<5$, we obtain an inclusive cross section $620$ fb, which is around $128$ times larger than the cross section $\sigma(hhh)$  of the signal processes $gg \to hhh$. Meanwhile, by using Madgraph5 \cite{Hirschi:2015iia}, we find that the cross section of VBF with $\sqrt{s}=100$ TeV is 34 fb.

\item It is known that when the b tagging efficiency is taken as $0.7$, the rejection rate of light jets can reach 0.1\% or so. Since we required one(two) tagged b jets in our preselection cuts, therefore the background $gg \to hh + 2$jets is suppressed by a factor $10^{-3}$ ( $10^{-6})$ or so. After imposing b taggings and the decay branching fraction of $h\to b\bar{b}$, we find that the signal cross section $\bar{b} Œb hh$ is around $0.52 (0.29) \, \sigma(hhh)$, while the cross section of  background $hh+2$jets is $ 0.13 (0.13 \times 10^{-3}) \times  \sigma(hhh)$ or so. Obviously, when $n_b\geq 2$ is imposed, it is safe to neglect this type of background in the SM.

\item In the new physics models we will consider below, the background process of $hhjj$ can have extra contributions from higher dimensional operators. When the cross section is $2 \sim 5$ magnitude orders smaller than the signal process, we can neglect it safely. In the cases when such a background is greatly enhanced or in the cases the signal process $gg\to hhh$ is greatly suppressed by the higher dimensional operators to such a degree that the cross sections of them are comparable, the background of $hhjj$ should be included in the analysis. 
\end{itemize}

Table~\ref{table1} shows a yield of $642$ signal events in
this final-state channel for $30$ ab$^{-1}$ integrated luminosity.
However, without further selection there are $\sim 10^{7}$ background
events.  Clearly, it is a challenge to observe triple-Higgs
production through this channel. In the following subsection we
discuss observables and selection methods for suppressing background
and raising the signal/background ratio to an acceptable level.

\begin{center}
\begin{table}
\begin{center}
\begin{tabular}{|c|c|c|c|}
  \hline
  Process                 & $\sigma\times BR$ (ab)   & K-factor   & Expected number of events  \\
  \hline
  Signal                  & $10.71$                  & $2.0$      & $642$                    \\
  \hline
  $h(WW^*)t\bar{t}$       & $2.55\times{10^5}$    & $1.2$      & $9.18\times{10^6}$     \\
  $t\bar{t}W^-W^+$        & $3.68\times{10^4}$    & $1.3$      & $1.55\times{10^6}$  \\
  \hline
\end{tabular}
\end{center}
\caption{Cross sections of signal and background for the
  $2b2l^{\pm}4j+\slashed E$ final state in the SM. The expected number
  of events corresponds to $30$ ab$^{-1}$ integrated
  luminosity. \label{table1}}
\end{table}
\end{center}

\subsection{Parton-level analysis}

\begin{figure}[htbp]
  \centering
  \subfigure{
  \label{Fig1.sub.1}\thesubfigure
  \includegraphics[width=0.4\textwidth]{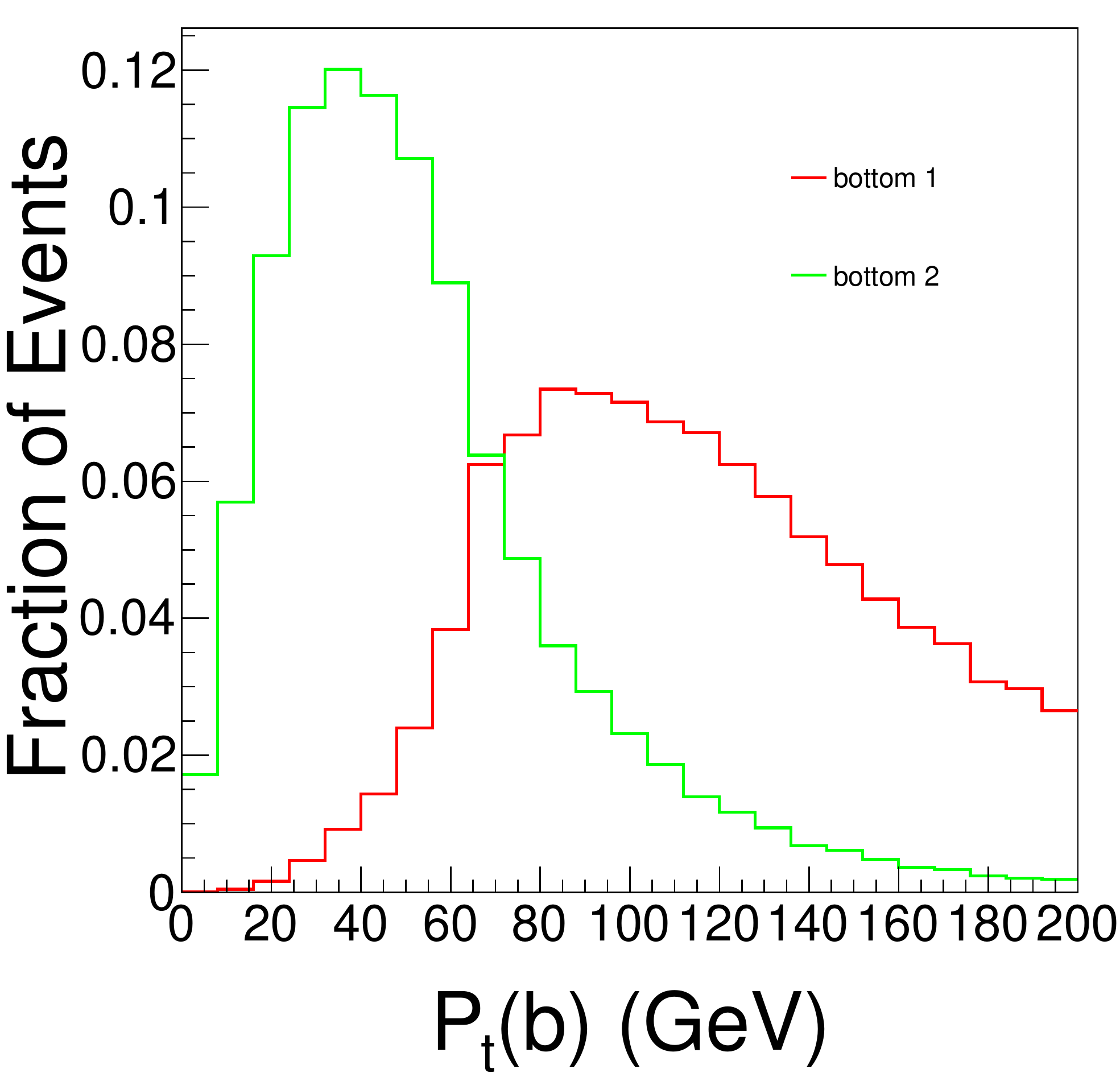}}
  \subfigure{
  \label{Fig1.sub.2}\thesubfigure
  \includegraphics[width=0.4\textwidth]{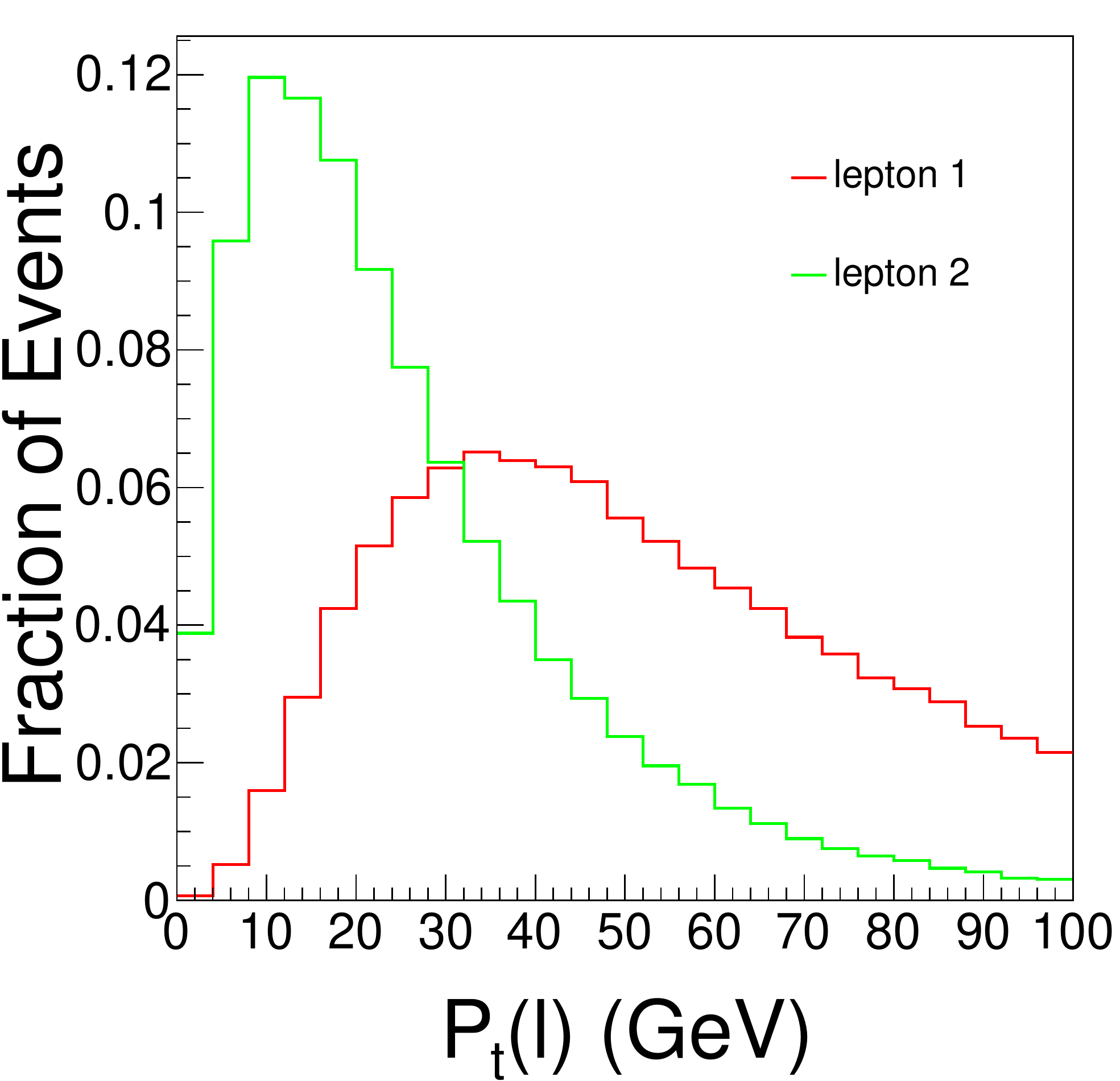}}
  \subfigure{
  \label{Fig1.sub.3}\thesubfigure
  \includegraphics[width=0.4\textwidth]{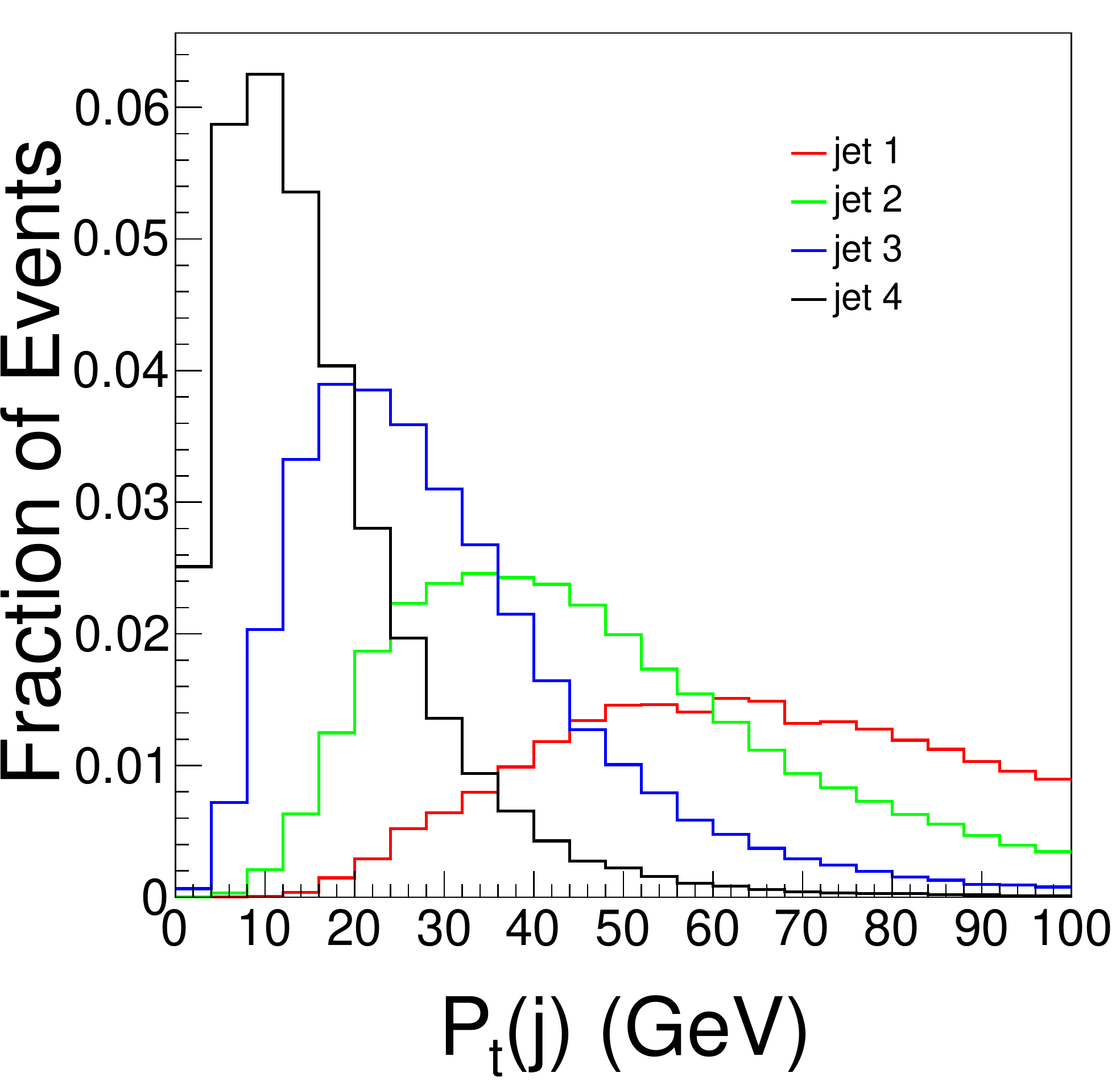}}
  \subfigure{
  \label{Fig1.sub.4}\thesubfigure
  \includegraphics[width=0.4\textwidth]{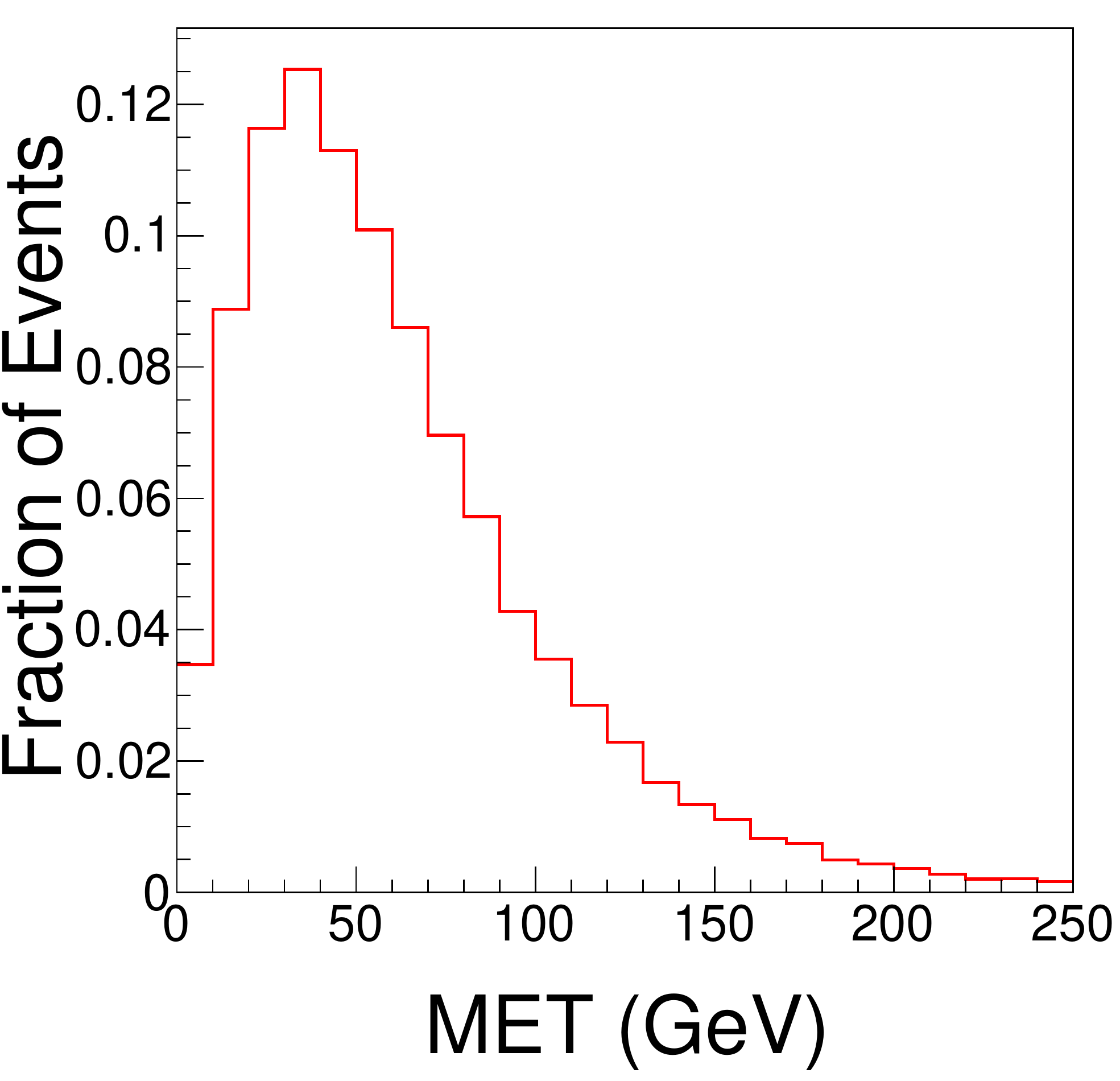}}
  \caption{Distributions of (a) the transverse momentum of $b$
    quarks, (b) the transverse momentum of leptons, (c) the transverse
    momentum of light quarks (labeled by $j$), and (d) missing transverse
    energy in the signal events.}\label{fig1}
\end{figure}

We simulate the Higgs boson decays that lead to the final state
$2b2l^{\pm}4j+\slashed E$ by using the DECAY package provided by MadGraph
5. Here we do not consider any parton shower effects, which will be discussed in section~\ref{detector}. 
The transverse momentum ($P_t$) distributions of the visible
particles and missing transverse energy (MET) are shown in
Fig.~\ref{fig1}. In this figure, the objects are sorted by $P_t$. On the
one hand, one can expect that the $b$ quarks are harder than the light
quarks, since they originate from a Higgs boson decay directly. On the
other hand, the decay chain $h\to WW^*\to jjl\nu$ leads to soft
leptons and light jets, especially when they are coming from the
off-shell $W$ bosons.

In Fig.~\ref{Fig1.sub.2} and Fig.~\ref{Fig1.sub.3}, we observe that
the $P_t$ distributions of the softest leptons and jets peak around
$10$ GeV, which might make it a challenge to successfully reconstruct
these objects with the currently planned detectors. Since the signal
contains only two neutrinos, MET should not be too large.  As illustrated by
Fig.~\ref{Fig1.sub.4}, MET peaks around $50$ GeV,
somewhat below half the Higgs boson mass.

Because there are two unobserved neutrinos in the final state, their
mothers being either on-shell or off-shell W bosons, it is not
convenient to fully reconstruct the Higgs bosons.  A partial
reconstruction should nevertheless be possible.  In order to extract
this information, it is crucial to correctly associate the mother
Higgs bosons with their decay products.  Here we encounter a problem
of combinatorics, which leads to a 12-fold ambiguity.  To simplify the
problem, we assume that both $b$ quarks can be tagged correctly, so
only the light quarks can be reassigned and the ambiguity reduces to
6-fold.

To find the correct combination of the visible particles from Higgs
boson decays, we examine the following four alternative reconstruction methods at
parton level:

\begin{enumerate}
 \item The decay chain $h\to WW^*\to jjl\nu$ suggests that the lepton
   and the hadronically decayed $W$ boson should have a small angular
   separation $\Delta R(l,W_{jj})$. Since there are two Higgs bosons
   with this decay chain, the sum of $\Delta R_1(l,W_{jj})+\Delta
   R_2(l,W_{jj})$ should be minimal. We choose a combination with
   minimal value of this observable.
 \item The semileptonic Higgs invariant masses can be computed from
   the visible particles; we denote them as $m^\text{vis}_{h1}(l,jj)$
   and $m^\text{vis}_{h2}(l,jj)$. We choose a combination which
   minimizes their sum.
 \item We compute the $mT2$ observable as it has been defined
   in
   Refs.~\cite{Lester:1999tx,Barr:2003rg,Barr:2010zj,Barr:2011xt,Cho:2014naa}, 
   from the visible particles that originate from semileptonic Higgs decay.
   The observable can set an upper bound on the Higgs mass, so we choose a
   combination which minimizes $mT2$.
 \item Since $mT2$ should have a value close to the Higgs mass
   $m_h=126$ GeV, we choose a combination which minimizes
   $|mT2-m_h|$.
\end{enumerate}

\begin{center}
\begin{table}
  \begin{center}
  \begin{tabular}{|c|c|}
  \hline
  Methods  &  The percentage of correctness  \\ 
  \hline
  $\min[\Delta R_1(l,W_{jj})+\Delta R_2(l,W_{jj})]$ & $47.0\%$ \\ 
  \hline 
  $\min(m^\text{vis}_{h1}+m^\text{vis}_{h2})$    &  $61.2\%$\\
  \hline
  $\min(mT2)$    &  $66.8\%$\\
  \hline
  $\min|mT2-m_h|$    &  $99.98\%$\\
  \hline
  \end{tabular}
  \end{center}
  \caption{\label{table2}Methods for determining the correct
    combinations of $(l,j,j)$ and their percentages of correctness.}
\end{table}
\end{center}

These methods and their associated percentages of correct assignment in a
simulated event sample are listed in Table~\ref{table2}. The effect of
realistic $b$-tagging efficiency will be discussed in the next
subsection.  We observe that in a parton-level analysis, the method
that relies on the quantity $|mT2-m_h|$ has the best performance,
approaching $100\%$ probability for correct particle assignment in the
reconstruction.

\subsection{Detector-level analysis}\label{detector}

To obtain a hadronic event sample, we use the parton-shower and
hadronization modules of Pythia 6.4~\cite{Sjostrand:2006za}.  For jet
clustering, we use the package FASTJET~\cite{Cacciari:2011ma} with the
anti-$k_t$ algorithm~\cite{Cacciari:2008gp} and cone parameter
$R=0.5$.  To veto the large number of soft jets from initial-state
radiation, only jets with $P_t>20$ GeV are accepted.

\begin{figure}[htbp]
  \centering
  \subfigure{
  \label{Fig3.sub.1}\thesubfigure
  \includegraphics[width=0.4\textwidth]{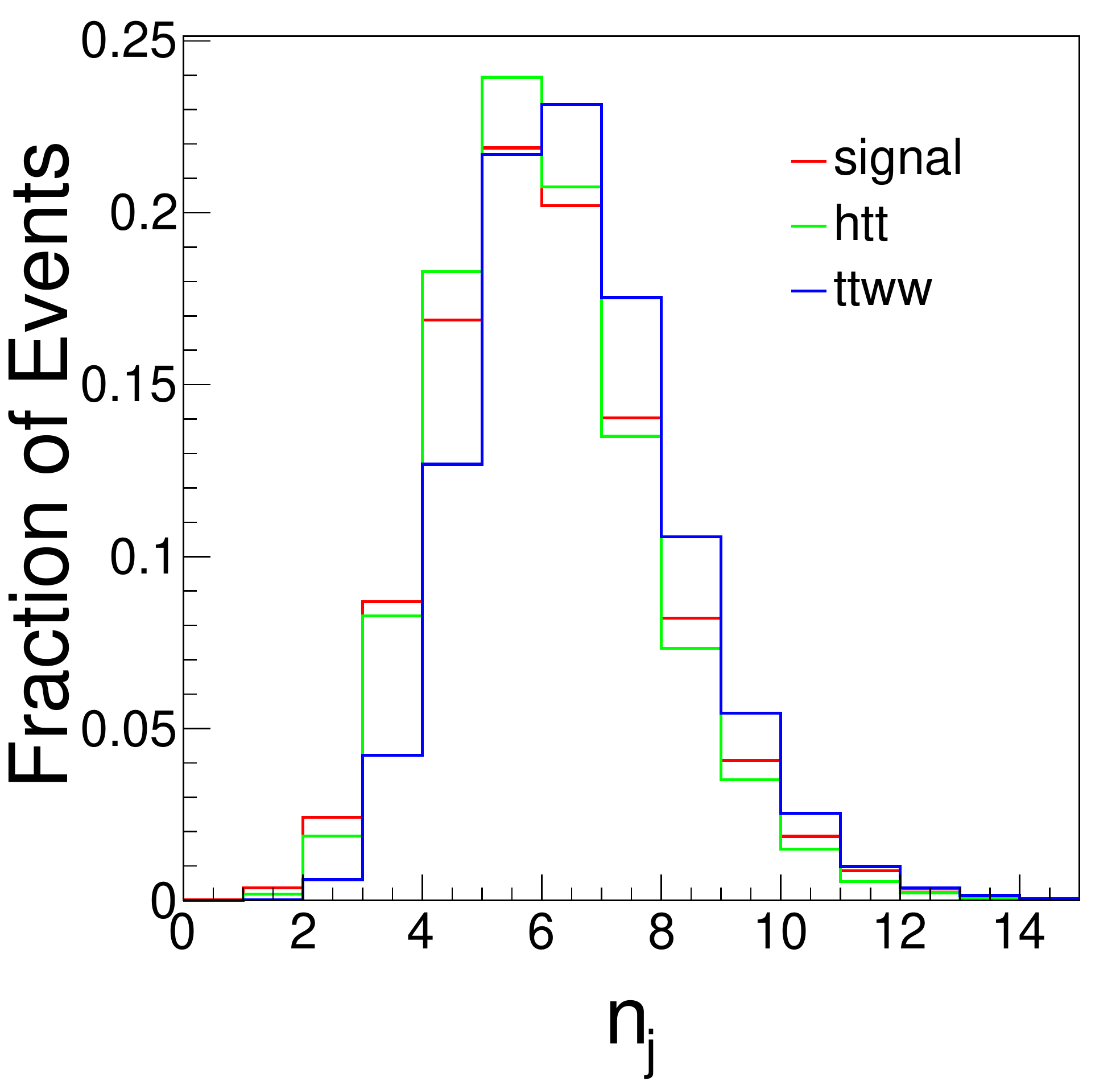}}
  \subfigure{
  \label{Fig3.sub.2}\thesubfigure
  \includegraphics[width=0.4\textwidth]{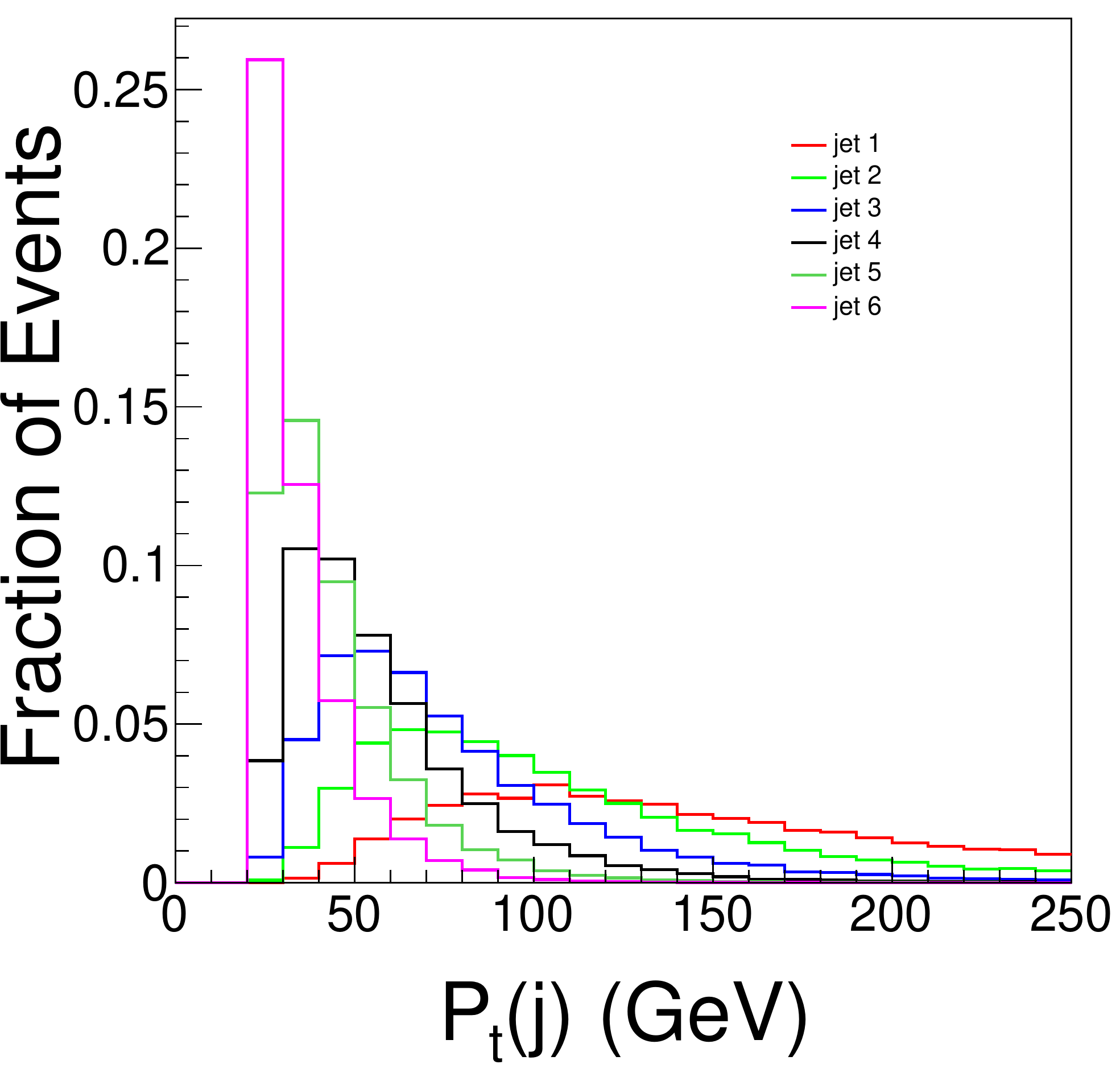}}
  \caption{Distributions of (a) the number of jets and (b) $P_t$ of the six
    leading jets of the signal.}\label{fig3} 
\end{figure}

The multiplicity distribution of jets is plotted in
Fig.~\ref{Fig3.sub.1}. Both signal and background in the MC sample provide six
jets at parton level, which explains the peak of $n_j$ around $6$ in Fig.~\ref{Fig3.sub.1}. 
In Fig.~\ref{Fig3.sub.2}, we show the $P_t$ distributions of the six
leading jets in the signal event sample. The 1st to 4th jet exhibit
similar distributions as at parton level, but the 5th and 6th
jet $P_t$ distributions have different shapes with respect to their parton-level
counterparts.

There are two simple reasons for this result: (1) the softest quark in
Fig.~\ref{Fig1.sub.3} typically has $P_t$ only around 10 GeV while most
of the low-$P_t$ jets are vetoed by our $P_t>20$ GeV cut; (2) jets
from initial-state radiation can easily be as hard as $20$ GeV at a $100$ TeV
collider. So the 5th and 6th jet are more likely produced by initial-state
radiation than by Higgs boson decays.  Fig.~\ref{fig3} illustrates
the challenge of reconstructing the soft jets generated by the
multi-Higgs signal.

\begin{figure}[htbp]
  \centering
  \subfigure{
  \label{Fig4.sub.1}\thesubfigure
  \includegraphics[width=0.4\textwidth]{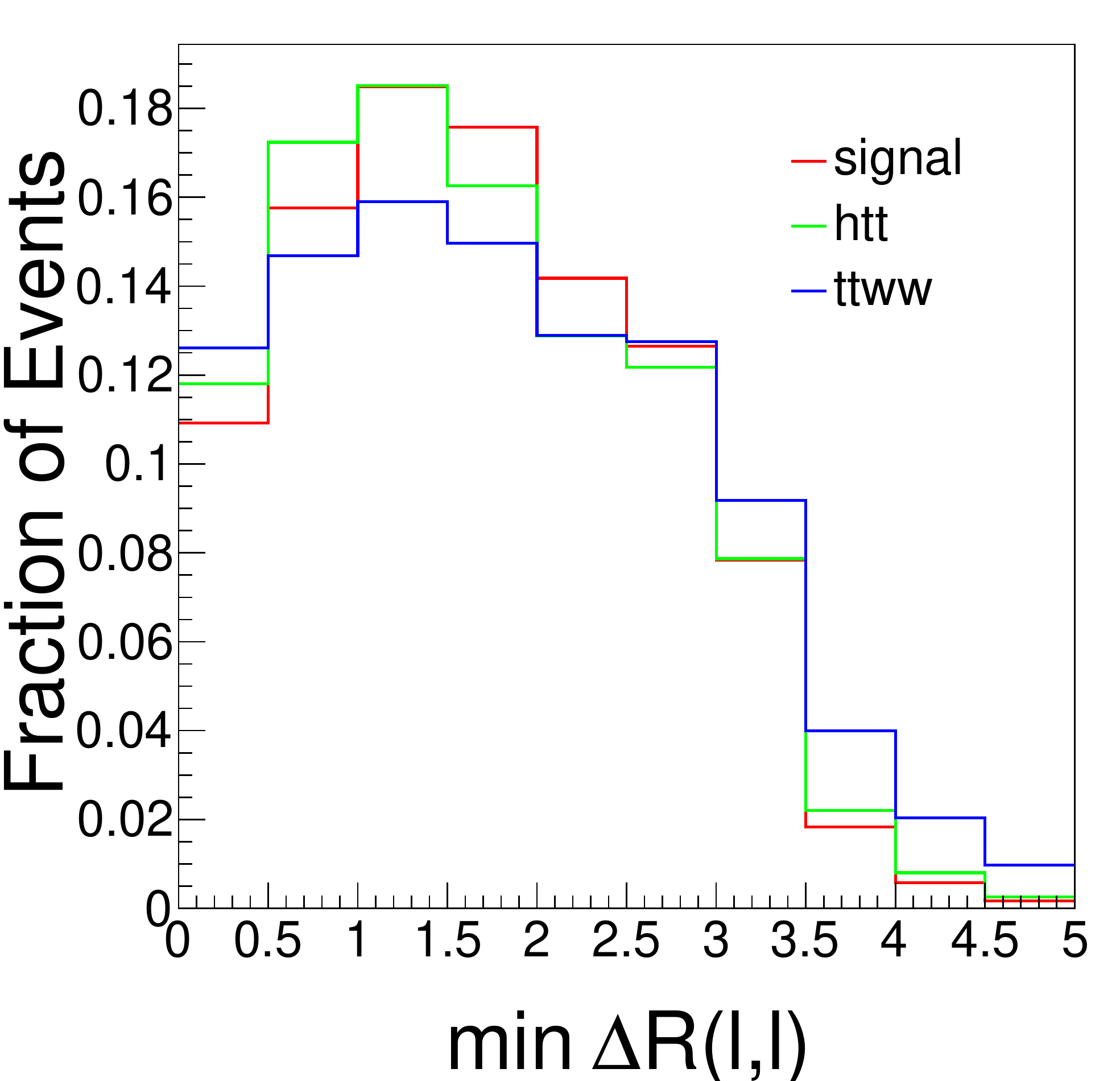}}
  \subfigure{
  \label{Fig4.sub.2}\thesubfigure
  \includegraphics[width=0.4\textwidth]{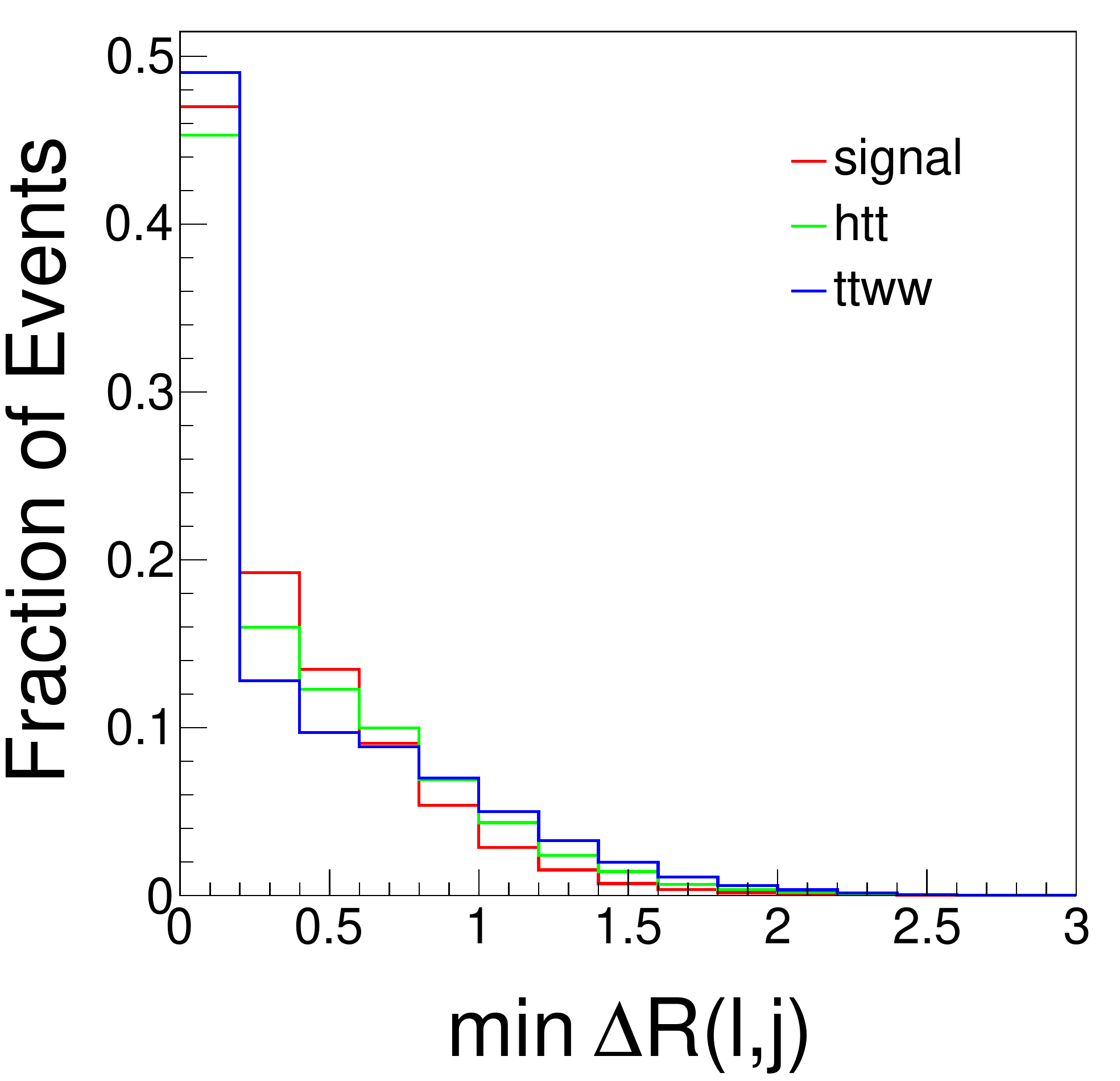}}
  \caption{Distributions of (a) the minimum angular separation between
    two leptons, and (b) the minimum angular separation between lepton
    and jet.}\label{fig4} 
\end{figure}

Another important problem is the reconstruction of leptons. We assume
that the future detector can reach a better efficiency in reconstructing
leptons than possible today ($95\%$ for $P_t>5$ GeV), so it becomes
feasible to find the soft lepton as shown in
Fig.~\ref{Fig1.sub.2}.  But in order to reject huge QCD background, we
need isolated leptons. To find a suitable isolation condition, we
investigate the angular separations between two leptons ($\Delta
R(l,l)$) and between leptons and jets ($\Delta R(l,j)$),
respectively. The minimum-value distributions of
these two observables at hadron level are displayed in
Fig.~\ref{fig4}. On the one hand, $\min \Delta R(l,l)$ tends to have a
large value, and only $10\%$ of the events have $\min \Delta R(l,l)<0.5$. On
the other hand, almost $50\%$  of the events have $\min \Delta R(l,j)<0.2$.
This makes it difficult to isolate the leptons from the jets.

To study the detector effects, we use
DELPHES~\cite{Ovyn:2009tx,deFavereau:2013fsa} to perform a detector
simulation for the generated event samples. The setup of DELPHES is
similar as in Ref.~\cite{Chen:2015gva}, with the following modifications:

\begin{enumerate}
  \item The $b$-tagging efficiency is assumed to be a constant
    $\epsilon_b=0.7$, and mistagging rates are $0.1$ and $0.001$ for
    charm and light jets, respectively. The pseudorapidity for $b$ ($c$,
    jet) is required to be $\eta<5.0$, respectively.
  \item As described above, the jets are clustered by FASTJET with a
    cut $P_{t}(j)>20$ GeV.
  \item The efficiency of lepton indentification is assumed to be $95\%$ when
    $P_{t}(l)>5$ GeV and $\eta(l)<5.0$.
  \item Isolated leptons are defined by Ref.~\cite{deFavereau:2013fsa}
\begin{equation}
  I(l) = \frac{\sum^{\Delta R<R, P_t(i)>P_t^{min}}_{i\neq l}P_t(i)}{P_t(l)}, 
\label{eq3}
\end{equation} 
where $l$ is a lepton. The sum in the numerator runs over particles with
transverse momenta above $P_t^{min}=0.1$ GeV within a cone with radius
$R=0.5$, except for $l$.  A lepton is classified as isolated if $I(l)<0.1$.
\end{enumerate}

\begin{figure}[htbp]
  \centering
  \subfigure{
  \label{Fig5.sub.1}
  \includegraphics[width=0.4\textwidth]{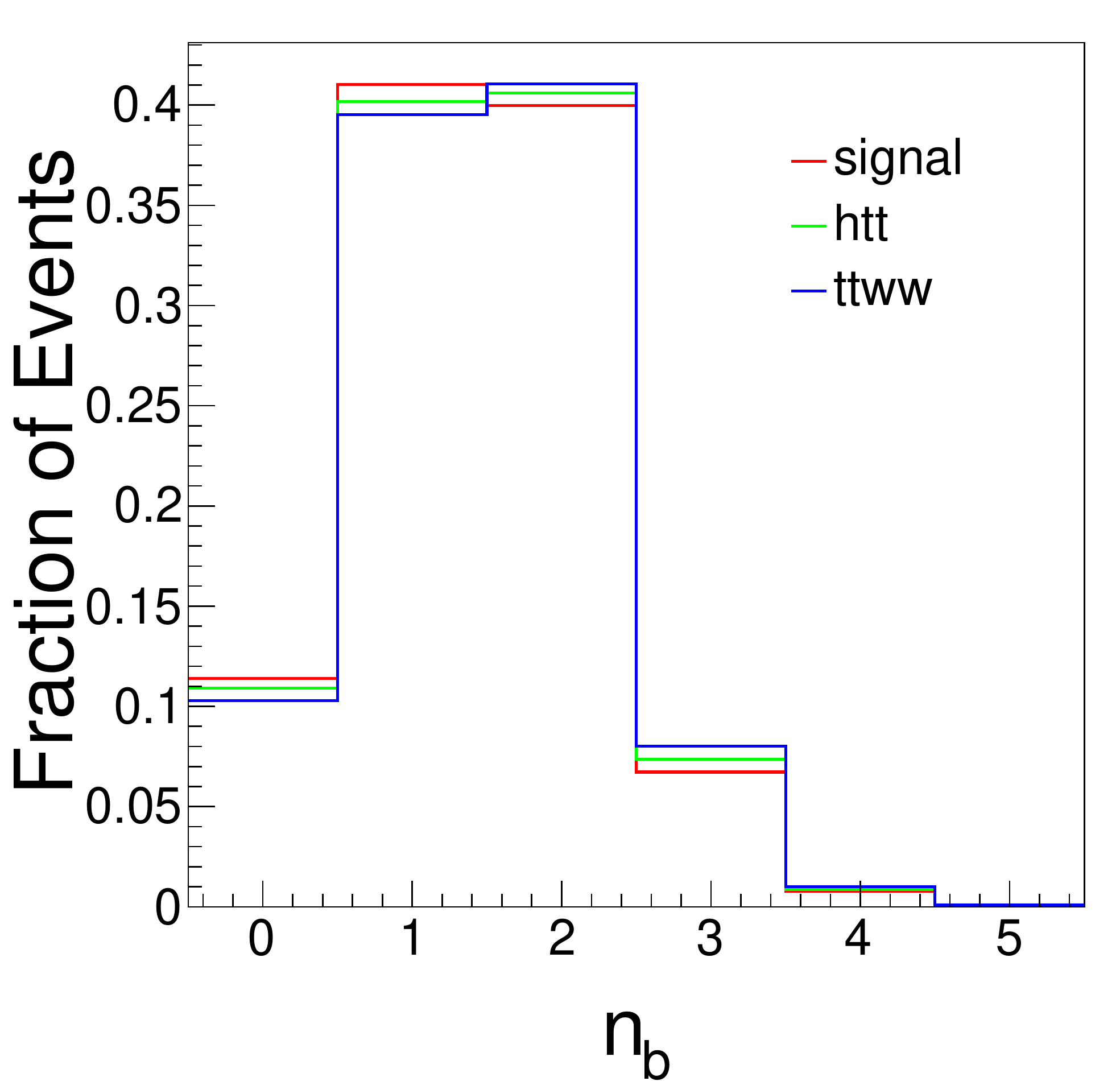}}
  \subfigure{
  \label{Fig5.sub.2}
  \includegraphics[width=0.4\textwidth]{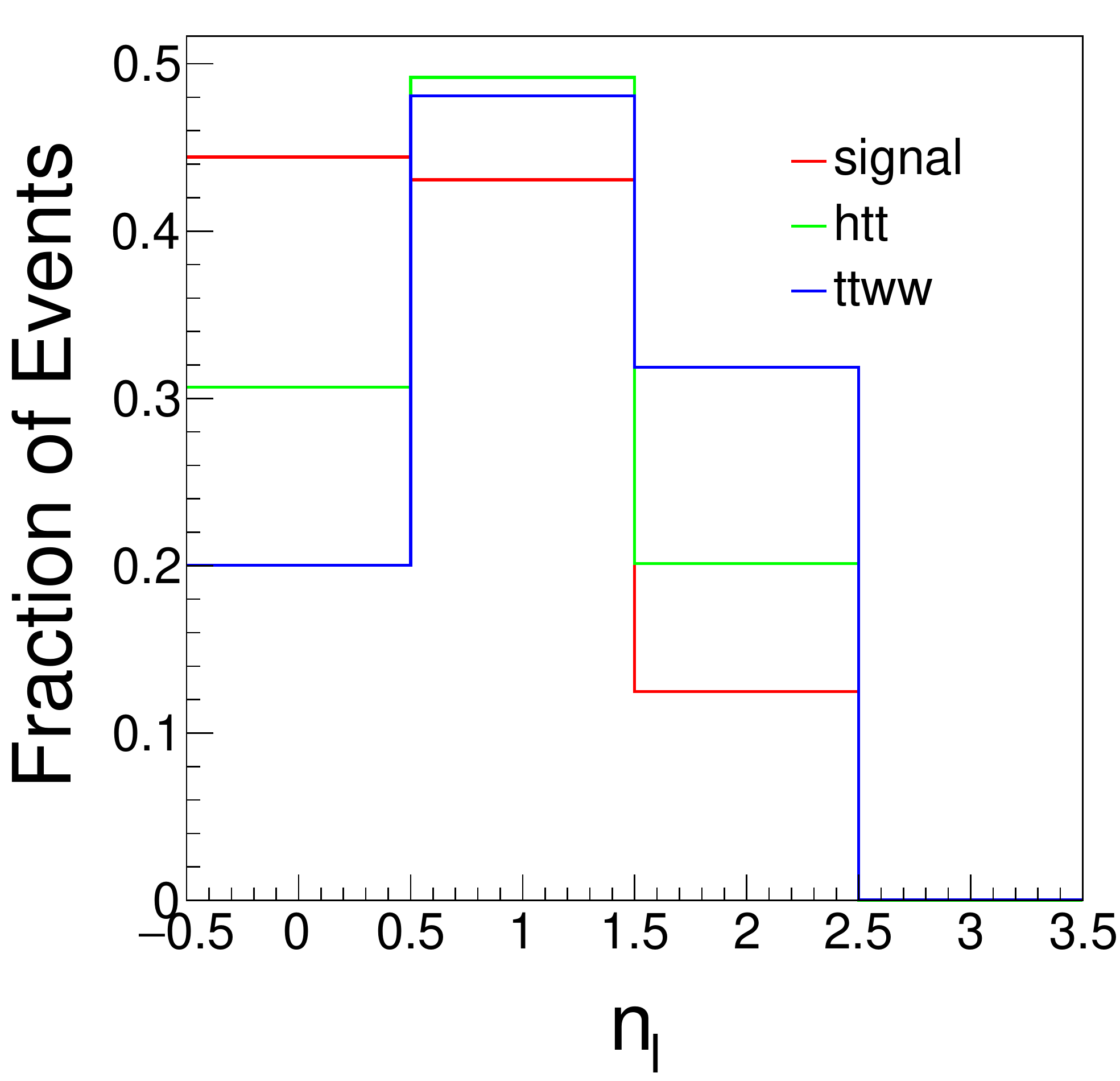}}
  \caption{Distributions of (a) the number of $b$-tagging jets and (b)
    the number of leptons.}\label{fig5} 
\end{figure}

Fig.~\ref{fig5} shows the number of $b$~jets and isolated leptons
after detector simulation. Since both signal and backgrounds
include two $b$~jets, it is easy to understand the similarity of the shapes in
Fig.~\ref{Fig5.sub.1}. However, only $10\%$ of the signal events are
found to include two leptons (Fig.~\ref{Fig5.sub.2}), which makes
it difficult to separate signal from background. The small value of
$\min \Delta R(l,j)$ in typical events explains this result.

\begin{figure}[htbp]
  \centering
  \subfigure{
  \label{Fig6.sub.1}\thesubfigure
  \includegraphics[width=0.4\textwidth]{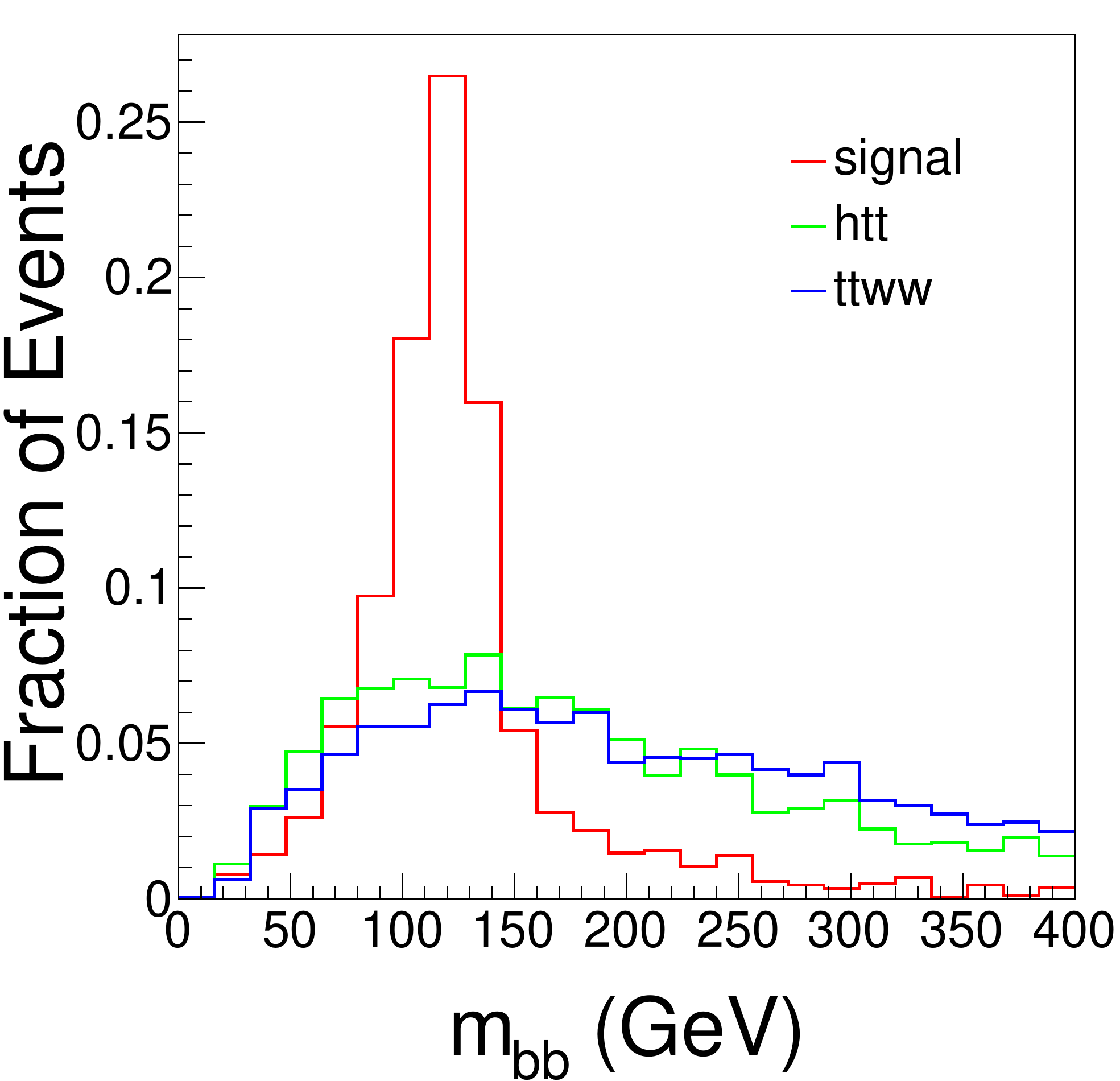}}
  \subfigure{
  \label{Fig6.sub.2}\thesubfigure
  \includegraphics[width=0.4\textwidth]{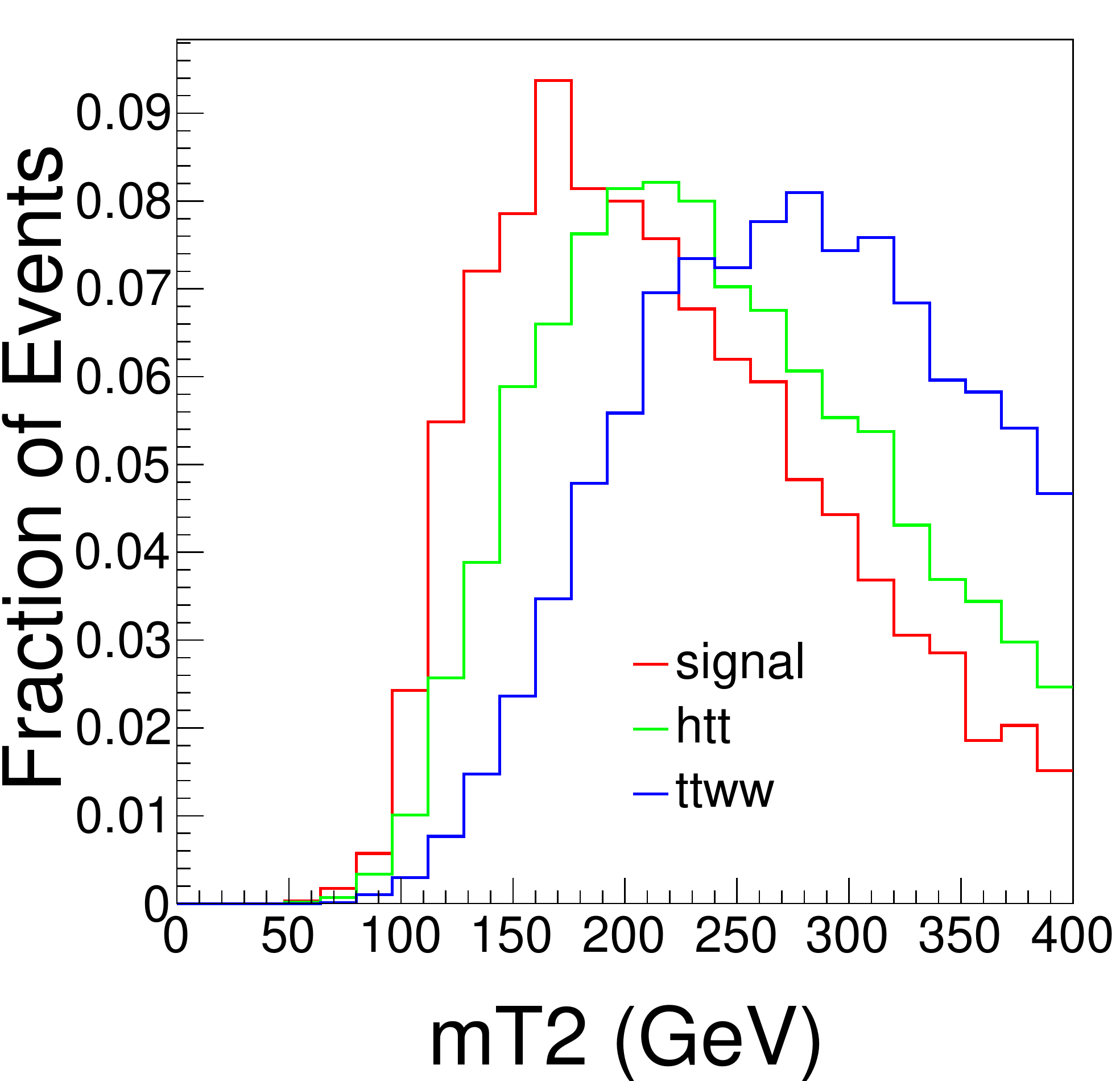}}
  \subfigure{
  \label{Fig6.sub.3}\thesubfigure
  \includegraphics[width=0.4\textwidth]{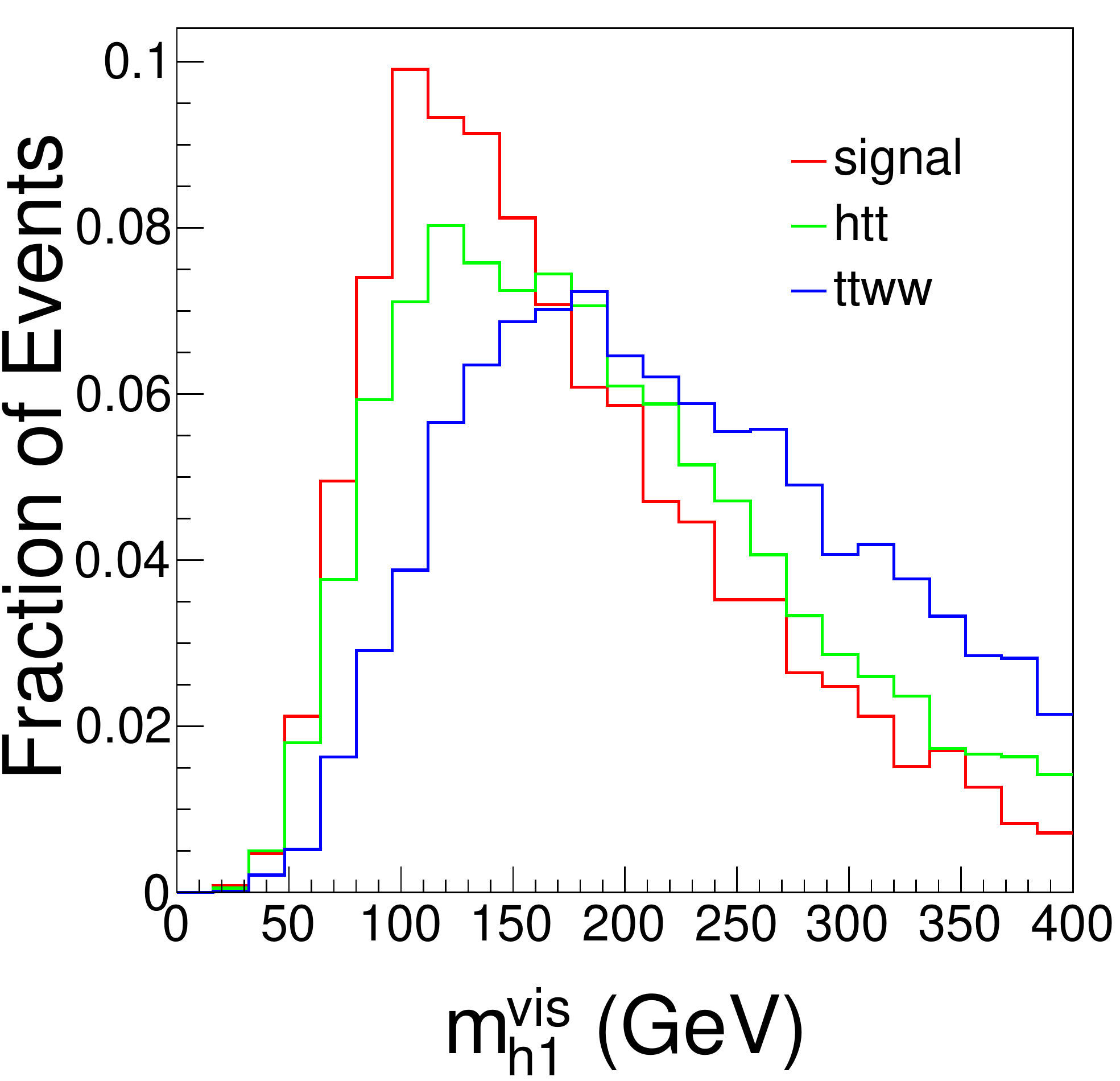}}
  \subfigure{
  \label{Fig6.sub.4}\thesubfigure
  \includegraphics[width=0.4\textwidth]{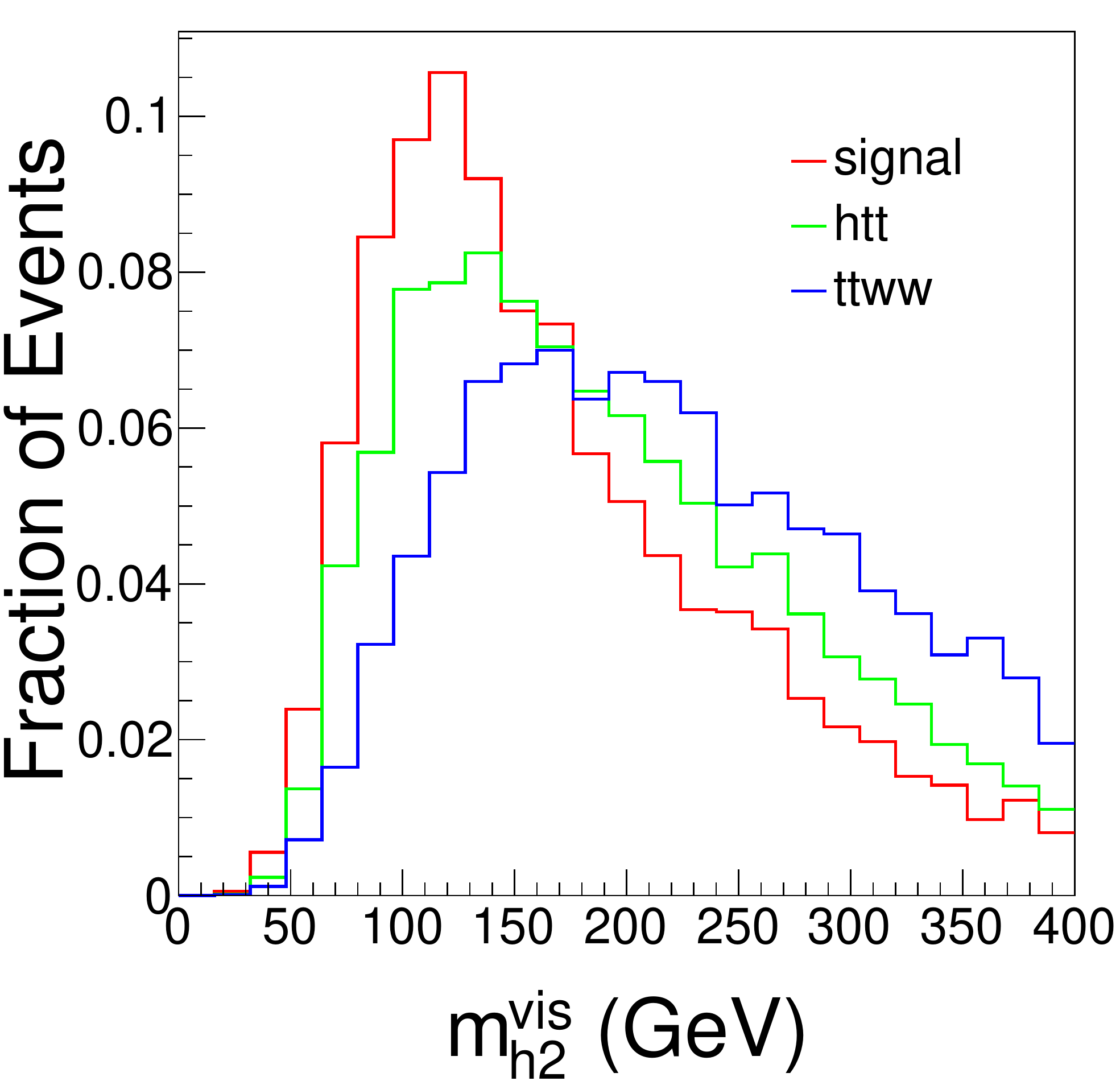}}
  \caption{Distributions of important observables at detector level:
    (a) the invariant mass of $b\bar{b}$, (b) the $mT2$
    observable, and (c) $\&$ (d) the Higgs masses as reconstructed from
    visible particles.}\label{fig6}
\end{figure}

To further enhance the signal over background ratio, we apply three preselection cuts:

\begin{enumerate}
  \item The number of $b$~jets is required to be $n_b\geq 1$. One might worry about the background $hh$+2 jets. It is found that it can only contribute around 5 events, which can further be reduced to 3 by the cut $|m_{bb} - m_h|<58$ GeV and is much smaller than the other two types of background given in Table (\ref{table4}), it is safe to omit it here. 
  \item To veto background from a $Z$ boson, we require two same-sign
    leptons, as discussed above.  Note that this also removes
    triple-Higgs signal events which decay to opposite-sign leptons.
  \item The number of light jets is required to be $n_j\geq 4$.
\end{enumerate}

We are interested in three observables: (1) the invariant mass of a
$b$-jet pair ($m_{bb}$), (2) the $mT2$ variable, and (3) the Higgs
masses ($m^\text{vis}_{h1}$ and $m^\text{vis}_{h2}$) reconstructed from the
visible objects. The distributions of these observables are displayed
in Fig.~\ref{fig6}. In Fig.~\ref{Fig6.sub.1}, the signal exhibits the
expected $m_{bb}$ peak around the Higgs mass, while the background is
non-resonant.  Regarding $mT2$, $m^\text{vis}_{h1}$, and $m^\text{vis}_{h2}$, in
the signal sample these observables should have a upper bound at the
Higgs mass.  However, many events in
Figs.~\ref{Fig6.sub.2}--\ref{Fig6.sub.4} show larger values.  As
discussed above, limitations in the reconstruction of the softest jet
together with missing lepton isolation are responsible for this
effect.

Nevertheless, we can try to suppress background by applying cuts on the above
observables. The efficiencies of each cut are listed in
Table~\ref{table4}. The significance of the signal in the cut-based
method finally amounts to just~$0.02$, which is clearly much worse
than could be expected from the parton-level calculation.  We conclude
that in the SM, a discovery of triple-Higgs production through this
channel will be extremely challenging.

\begin{center}
\begin{table}
  \begin{center}
  \begin{tabular}{|c|c|c|c|}
  \hline
                        &  Signal          &  $h(WW^*)t\bar{t}$   &  $t\bar{t}W^-W^+$       \\ 
  \hline
  Preselection          &  $24$                    &  $9.73\times{10^5}$               &  $2.59\times{10^6}$ \\ 
  \hline
  $mT2<484$ GeV         &  $23$                    &  $9.40\times 10^{5}$              &  $2.35\times 10^{5}$ \\ 
  \hline
  $|m_{bb}-m_h|<58$ GeV &  $21$              &  $6.73\times 10^{5}$              &  $1.42\times 10^{5}$ \\  
  \hline
  $m_{h}^\text{vis}<482$ GeV   &  $21$              &  $6.72\times 10^{5}$                  &  $1.42\times 10^{5}$  \\ 
  \hline
  $S/B$                           & \multicolumn{3}{c|} {$2.56\times 10^{-5}$}   \\
  \hline
  $S/\sqrt{S+B}$                  & \multicolumn{3}{c|} {$0.0231$}            \\
  \hline
  \end{tabular}
  \end{center}
  \caption{\label{table4}Efficiencies of cuts as described in the text,
    for a total integrated luminosity of $30$ ab$^{-1}$..}
\end{table}
\end{center}

\section{Triple-Higgs production with dimension-6 operators}
\label{Sec:hhh-EFT}

Given the dim prospects for observing triple Higgs production in the
pure SM, we may ask the question about SM extensions that enhance the
production rate such that the process becomes observable at a $100$
TeV collider.  In that case, such an observation would not just indicate a
significant deviation from the SM, but at the same time provide a
measurement of new BSM parameters.

We work in the context of the genuine Higgs-sector BSM models that we
have introduced above, conveniently parameterized by the SILH
Lagrangian with dimension-six operators, or, alternatively, by the
effective Higgs Lagrangian in unitarity gauge, Eq.~(\ref{eft}).  In
the SM, the production process $gg\to hhh$ proceeds via top-quark
loop diagrams coupled to Higgs bosons and involves
triple and quartic Higgs couplings, where we are obviously most
interested in the quartic-coupling contribution.  The various
anomalous couplings generated by Eq.~(\ref{eft}) modify all contributing
loop Feynman graphs, and furthermore direct Higgs-gluon couplings can
appear which are induced either from the underlying theory directly or
emerge from loop-diagram renormalization via operator mixing.
Therefore, we redo the calculation of the production process at
one-loop order with all new parameters included.

While the observation of the triple-Higgs process ultimately would
determine a particular combination of the new parameters, sizable
values for those will definitely also affect other, more easily
accessible processes such as Higgs production in association with a
top quark and double-Higgs production.  Any actual measurement of the
EFT parameters will involve a fitting procedure that takes all
available information into account.  Nevertheless, the fact that the
quartic Higgs coupling appears only in triple-Higgs production
indicates that the current process will contribute independent, and
potentially essential information.

If the triple-Higgs final state is to become observable, we have to
allow for EFT parameter values that distort the amplitudes rather
drastically, at least in the high-energy or high-$P_t$ regions of
phase space.  We should worry about unitarity of the amplitudes and
consistency of the EFT.  Physically, we expect a dampening effect from
strong rescattering of intermediate real top quarks into multiple
Higgs bosons.  While the effects of strong rescattering have
extensively been studied in the linear EFT context for vector-boson
scattering~\cite{Kilian:2014zja,Kilian:2015opv}, no results are
available for processes involving top quarks and Higgs bosons.  Power
counting suggests that the dimension-six operators that we consider in
this work are affected to a lesser extent than the dimension-eight
operators considered in Ref.~\cite{Kilian:2014zja}.  We also note that in
the SM as a weakly interacting theory, the Higgs mechanism tends to
suppress electroweak production cross sections by orders of magnitude
in relation to the bounds enforced by unitarity, so there is a
significant margin for enhancing event yields in BSM models.  For the
current study, we take the EFT unmodified over the complete parameter
space and defer a study of constraints and relations imposed by
unitarity to future work.

\subsection{Calculation}

In this section, we describe our calculation of the one-loop induced
production amplitude in the presence of the new parameters of
the unitarity-gauge effective Lagrangian, Eq.~(\ref{eft}).
We use the package
Madgraph5/aMC@NLO~\cite{Alwall:2014hca,Hirschi:2015iia} for calculating
the loop diagrams and, subsequently, evaluating phase space and
generating event samples.  To this end, we have implemented the model
described by Eq.~(\ref{eft}) as a UFO model file.  Note that our
choice of unitarity gauge for the electroweak symmetry does not affect
the QCD loop calculation, since the color symmetry is implemented in a
renormalizable gauge as usual.  The program reduces the one-loop
Feynman integrals to scalar integrals in four dimensions, employing
techniques such as the OPP method~\cite{Ossola:2006us}. The difference
between the $D$-dimensional and $4$-dimensional expressions that
arises in the calculation yields additional rational
terms~\cite{Ossola:2008xq}.  These are identified as R1 terms
associated with $D$-dimensional denominators, and R2 terms
associated with $D$-dimensional numerators.  All R1 terms 
are automatically generated as a byproduct of the reduction method,
while the R2 terms must be
calculated manually~\cite{Draggiotis:2009yb}.  We have performed this calculation and supplied
the results as effective tree-level vertices in
the UFO model file~\cite{Degrande:2014vpa}.

In particular, we obtain the R2 terms that amount to contact
interactions of a pair of gluons with one to three Higgs bosons:
\begin{align}
    \includegraphics[width=0.10\textwidth]{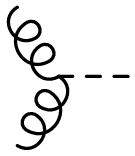}
    =& -i \frac{g_s^2m_t^2\delta^{a b}g_{\mu_1\mu_2}}{8\pi^2v} a_1\\
    \includegraphics[width=0.10\textwidth]{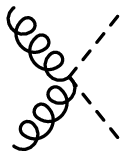}
    =& -i \frac{g_s^2m_t^2\delta^{a b}g_{\mu_1\mu_2}}{8\pi^2v^2}(a_1^2+a_2)\\
    \includegraphics[width=0.10\textwidth]{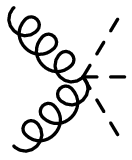}
    =& -i \frac{g_s^2m_t^2\delta^{a b}g_{\mu_1\mu_2}}{8\pi^2v^3}(a_3+3a_1a_2)
\end{align}
The coefficients depend on the EFT parameters $a_1,a_2$, and $a_3$.
Since these terms are required to restore the exact QCD symmetries in
the calculated amplitude, by themselves they manifestly violate gauge
invariance.  We have verified that the complete renormalized one-loop
result does respect gauge invariance, a convenient cross-check of the
calculation.

Besides these loop-induced contributions, gluon fusion into Higgs
bosons also receives contribution from contact interactions between
gluons and Higgs bosons that do not exist in the SM.  As mentioned
above, the inclusion of such contact interactions is required by
loop-induced operator mixing in the EFT, but could also originate from
independent BSM contributions.  Technically, we implement them as
independent R2 terms, so that Madgraph5/aMC@NLO will sum them together
with loop-induced contribution.


\subsection{Cross sections of $gg \to hhh$ and Kinematics}

The amplitudes of the process $pp\to hhh$ are constructed from the
Feynman diagrams in Fig.~\ref{hhhme}.  For illustrating the method, we
take the terms that depend on $a_1$, $\kappa_5$, $\kappa_6$,
$\lambda_3$ and $\lambda_4$.  Complete results are given in the
Appendix~\ref{appendix}.

Each of the following top-quark loop diagrams has a different dependency
on the top-Yukawa couplings and Higgs-boson self-couplings,
\begin{align}
    \includegraphics[width=0.13\textwidth]{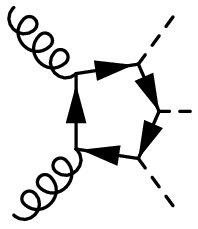}&\propto a_1^3
&\includegraphics[width=0.13\textwidth]{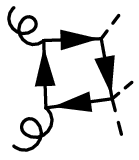}\propto a_1^2\lambda_3/a_1^2\kappa_5 \nonumber \\
    \includegraphics[width=0.13\textwidth]{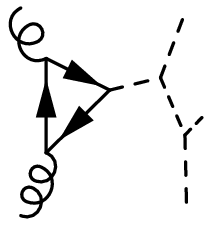}&\propto a_1 \lambda_3^2/a_1\lambda_3\kappa_5/a_1 \kappa_5^2
&\includegraphics[width=0.13\textwidth]{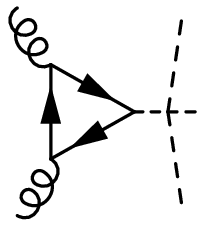}\propto a_1 \lambda_4/a_1 \kappa_6
    \label{hhhme}
\end{align}
and the corresponding matrix element is proportional to the following terms
\bea
M(gg \to hhh) & \propto & f_1 a_1^3 + f_2 a_1^2 \lambda_3 + f_3 a_1^2 \kappa_5 + f_4 a_1 \lambda_3^2 + f_5 a_1 \lambda_3 \kappa_5 \nonumber \\  & + &f_6 a_1 \kappa_5^2 + f_7 a_1 \lambda_4 + f_8 a_1 \kappa_6\,,
\eea
where $f_i$ are form factors, which after loop integration depend on
the external momenta, partly in form of Higgs-boson propagators. After
squaring the matrix element and integrating over phase space, we arrive at the
total cross section which can be parameterised as below
\begin{align}
    \begin{split}
        \sigma(pp\to hhh)&=t_1 a_1^6+t_2 a_1^5\lambda_3+t_3 a_1^5\kappa_5
        +t_4 a_1^4\lambda_3^2+t_5 a_1^4\lambda_3\kappa_5 \\& +t_6 a_1^4\kappa_5^2
        +t_7 a_1^4\lambda_4+t_8 a_1^4\kappa_6
        +t_9 a_1^3\lambda_3^3+t_{10} a_1^3\lambda_3^2\kappa_5 \\& +t_{11}a_1^3\lambda_3\kappa_5^2+t_{12}a_1^3\kappa_5^3 
        +t_{13}a_1^3\lambda_3\lambda_4+t_{14}a_1^3\lambda_3\kappa_6+t_{15}a_1^3\kappa_5\lambda_4 \\& +t_{16}a_1^3\kappa_5\kappa_6
        +t_{17}a_1^2\lambda_3^4+t_{18}a_1^2\lambda_3^3\kappa_5+t_{19}a_1^2\lambda_3^2\kappa_5^2
        +t_{20} a_1^2\lambda_3\kappa_5^3 \\ & +t_{21}a_1^2\kappa_5^4
        +t_{22}a_1^2\lambda_3^2\lambda_4+t_{23}a_1^2\lambda_3^2\kappa_6+t_{24}a_1^2\lambda_3\kappa_5\lambda_4
        +t_{25}a_1^2\lambda_3\kappa_5\kappa_6 \\& +t_{26}a_1^2\kappa_5^2\lambda_4+t_{27}a_1^2\kappa_5^2\kappa_6
        +t_{28}a_1^2\lambda_4^2+t_{29}a_1^2\lambda_4\kappa_6+t_{30}a_1^2\kappa_6^2 \,.
        \label{xsechhh1}
\end{split}
\end{align}
To determine the integrated form factors $t_1\ldots t_{30}$, we calculate
the total cross section at 480 selected points in the space of
parameters $(a_1,\lambda_3,\lambda_4,\kappa_5,\kappa_6)$, then obtain
the numerical values of these coefficients $t_{1}\ldots t_{30}$ via linear
regression. The resulting values are shown in
Table~\ref{numxsechhh}.  The complete set of results that accounts for
all effective operators is provided in the Appendix ~\ref{appendix}.

\begin{center}
\begin{table}{}
\begin{center}
\begin{tabular}{|c|c|c|c|c|c|c|c|c|c|}
    \hline
   $t_1$ & $t_2$ & $t_3$ & $t_4$ & $t_5$ &  $t_6$ & $t_7$ & $t_8$ & $t_{9}$ & $t_{10}$ \\
    7.57 & -7.79 & -13.9 & 4.33 & 14.7 & 12.3 &  0.13 & -0.79 & -0.95 & -7.63 \\
    \hline \hline
 $t_{11}$ & $t_{12}$ & $t_{13}$ & $t_{14}$ & $t_{15}$ & $t_{16}$ & $t_{17}$ & $t_{18}$ & $t_{19}$ & $t_{20}$\\
     -18.8 & -16.4 & -0.63 & -3.16 & -1.07 & -6.47 & 0.09 & 1.12 & 5.61 & 13.6 \\
    \hline \hline
     $t_{21}$ & $t_{22}$ & $t_{23}$  &  $t_{24}$ & $t_{25}$ & $t_{26}$ & $t_{27}$ & $t_{28}$ & $t_{29}$ & $t_{30}$ \\
   17.2 & 0.12 & 0.55 & 0.85 & 5.38 & 1.34 & 14.7 & 0.04 & 0.54 & 3.22 \\
    \hline
\end{tabular}
\end{center}
    \caption{ Numerical values of $t_{1}\ldots t_{30}$ for a 100 TeV
      hadron collider, for use in Eq.~(\ref{xsechhh1}).\label{numxsechhh}} 
\end{table}
\end{center}

To study the parameter dependence of the cross section of $gg \to hhh$
in one of the more specific models introduced in Section~\ref{sec:genEFT}, we
simply replace 
$(a_1,\lambda_3,\lambda_4,\kappa_5,\kappa_6)$ by $(\hat{r},\hat{x})$
according to Table~\ref{table5}, so we obtain Eq.~(\ref{xsechhhrx}) below.
The numerical results for $\hat{t}_{1}\ldots \hat{t}_{14}$ are listed
in Table~\ref{numxsechhhrx}, and the total cross section has the value
$\sigma_{SM}^{hhh}=5.84$~fb.  (This includes a K factor of 2.0,
following Ref.~\cite{Maltoni:2014eza}).
\begin{align}
    \begin{split}
        \sigma(pp\to hhh)&=\sigma_{SM}^{hhh}(1-\hat{x})^3(1+\hat{t}_1\hat{x}+\hat{t}_2\hat{r}
       +\hat{t}_3\hat{x}^2+\hat{t}_4\hat{x}\hat{r}+\hat{t}_5\hat{r}^2 \\
        &+\hat{t}_6\hat{x}^3+\hat{t}_7\hat{x}^2\hat{r}+\hat{t}_8\hat{x}\hat{r}^2+\hat{t}_{9}\hat{r}^3 \\&  +\hat{t}_{10}\hat{x}^4+\hat{t}_{11}\hat{x}^3\hat{r}+\hat{t}_{12}\hat{x}^2\hat{r}^2
    +\hat{t}_{13}\hat{x}\hat{r}^3+\hat{t}_{14}\hat{r}^4)
        \label{xsechhhrx}
    \end{split}
\end{align}

\begin{center}
\begin{table}{}
\begin{center}
\begin{tabular}{|*{14}{@{\hskip2pt}c@{\hskip2pt}|}}
    \hline
    $\hat{t}_1$ & $\hat{t}_2$ & $\hat{t}_3$ & $\hat{t}_4$ & $\hat{t}_5$ & $\hat{t}_6$ & $\hat{t}_7$ & $\hat{t}_8$ & $\hat{t}_9$ & $\hat{t}_{10}$ & $\hat{t}_{11}$ & $\hat{t}_{12}$ & $\hat{t}_{13}$ & $\hat{t}_{14}$ \\ 
    6.02 &-1.29 &3.51 &-2.40 &0.48 &-32.5 &8.07 & 
    -0.96 &0.05 &94.20 &-37.20 &7.69 &-0.77 &0.03\\
    \hline
\end{tabular}
\end{center}
\caption{\label{numxsechhhrx} Numerical values of integrated form factors }
\end{table}
\end{center}

\begin{figure}[htbp]
  \centering
  \subfigure{
  \label{Fig7.sub.1}\thesubfigure
  \includegraphics[width=0.45\textwidth]{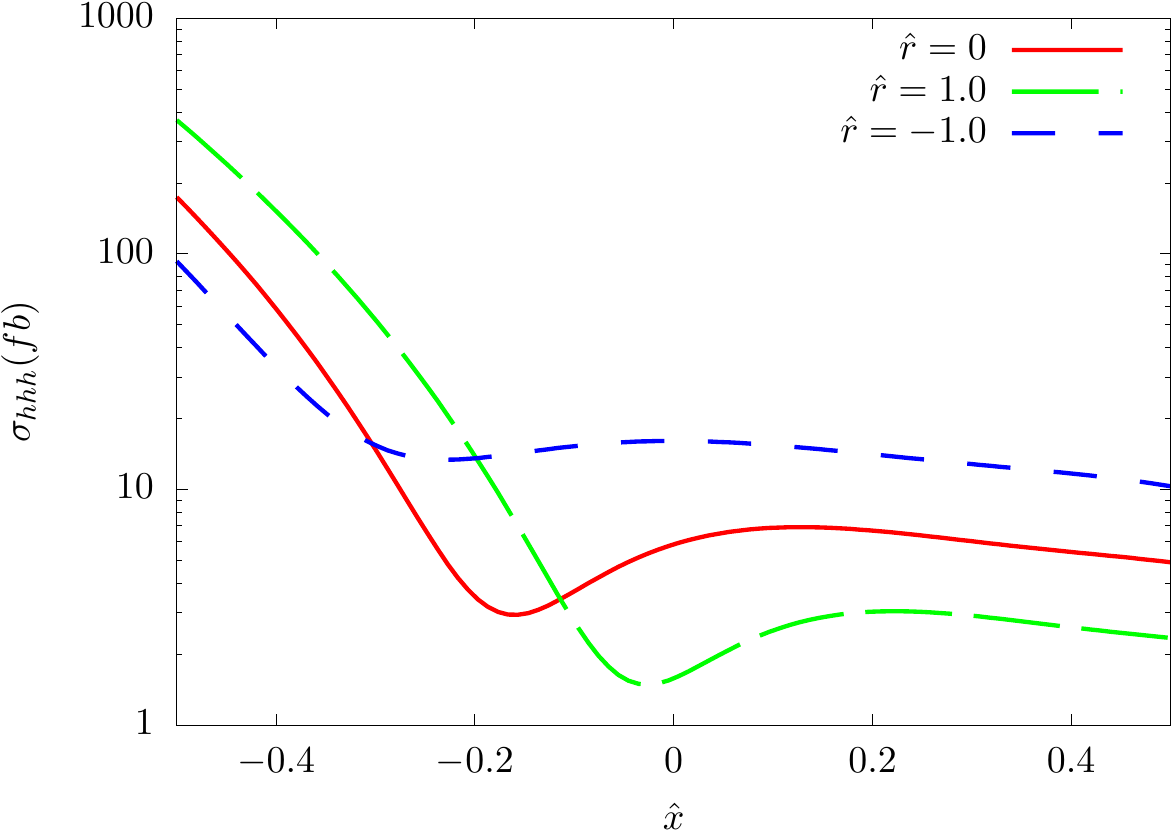}}
  \subfigure{
  \label{Fig7.sub.2}\thesubfigure
  \includegraphics[width=0.45\textwidth]{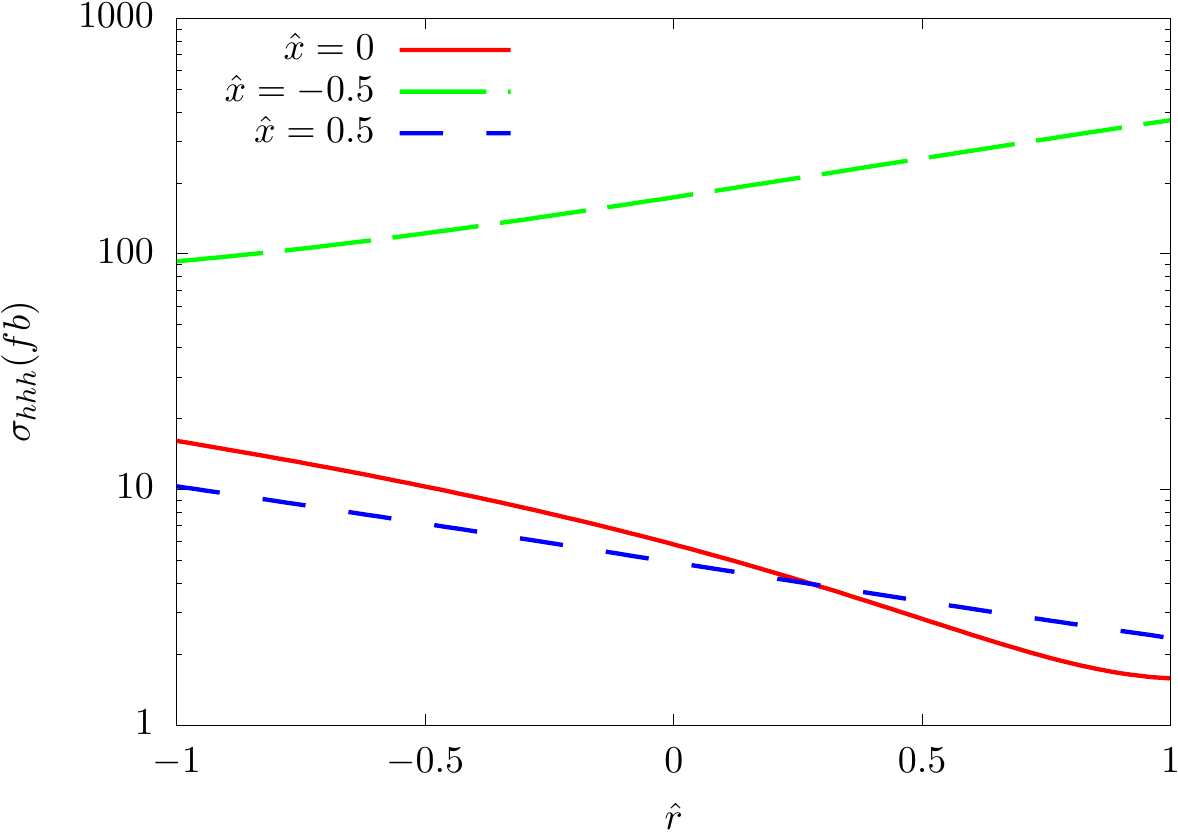}}
  \caption{Dependence of the cross section on (a) $\hat{x}$ and (b)
    $\hat{r}$. The other observable is kept fixed, as indicated by the
    curve labels.}\label{fig7}
\end{figure}
\begin{figure}[htbp]
  \centering
  \subfigure{
  \label{Fig8.sub.1}\thesubfigure
  \includegraphics[width=0.4\textwidth]{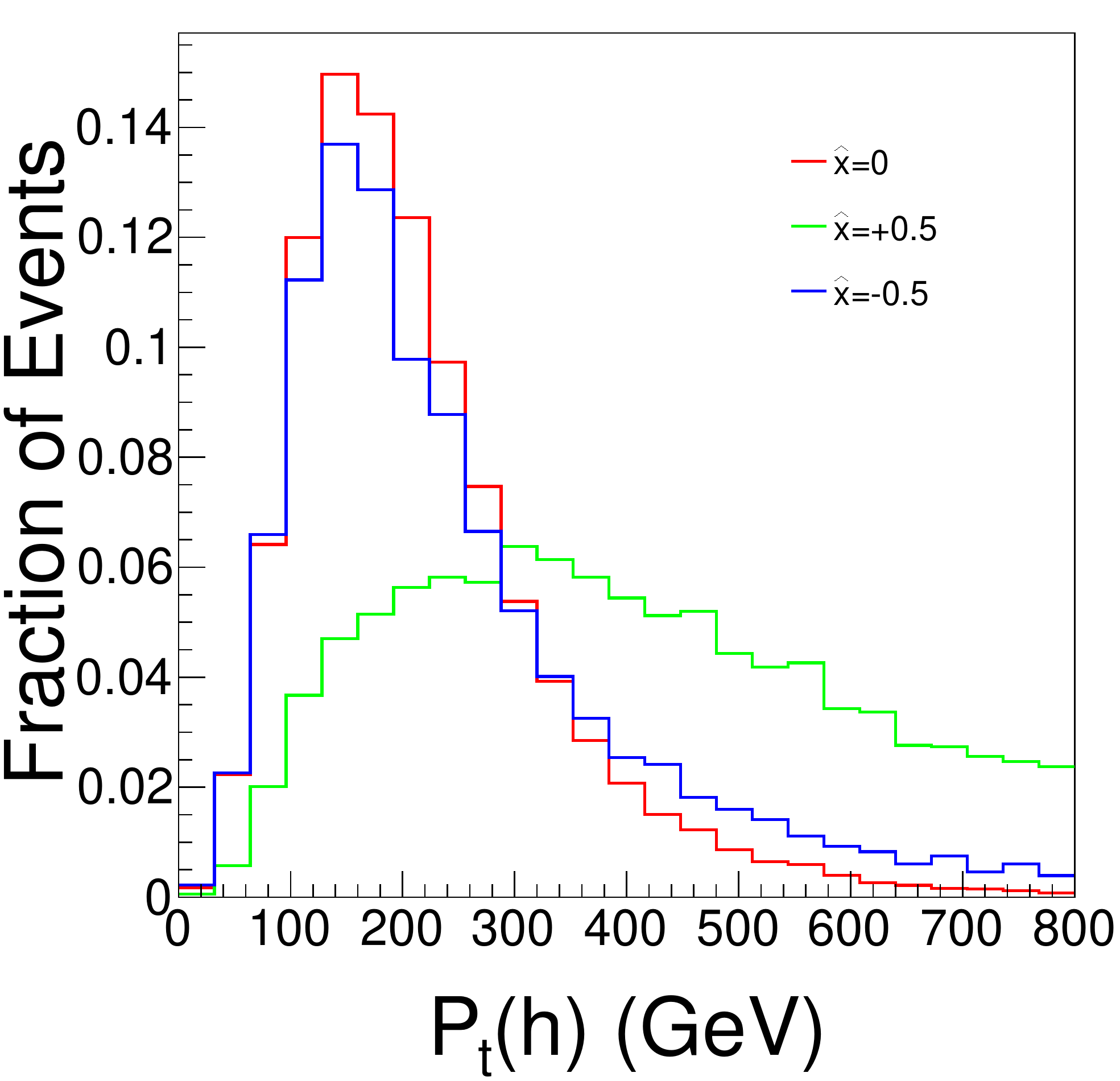}}
  \subfigure{
  \label{Fig8.sub.2}\thesubfigure
  \includegraphics[width=0.4\textwidth]{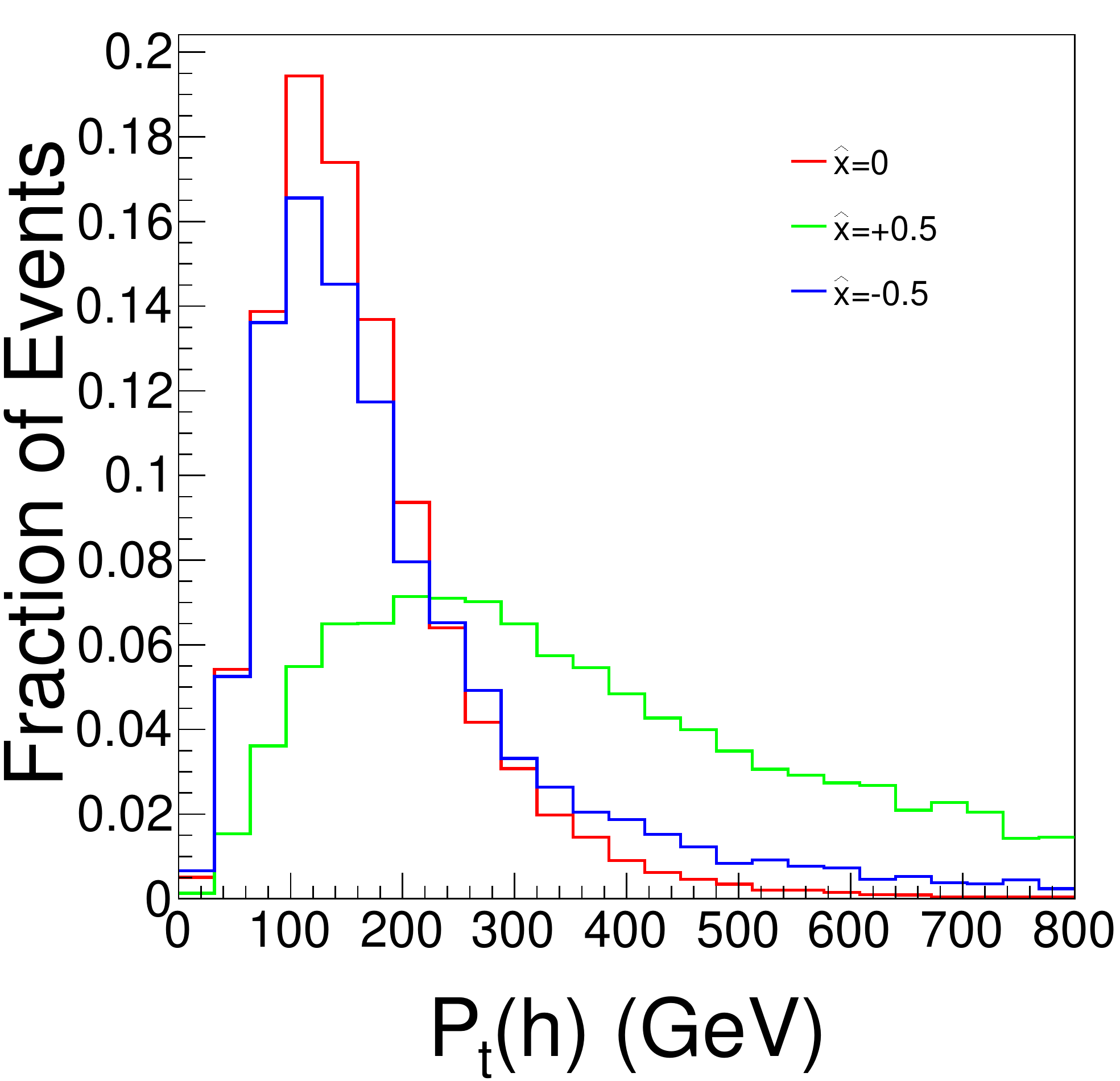}}
  \subfigure{
  \label{Fig8.sub.3}\thesubfigure
  \includegraphics[width=0.4\textwidth]{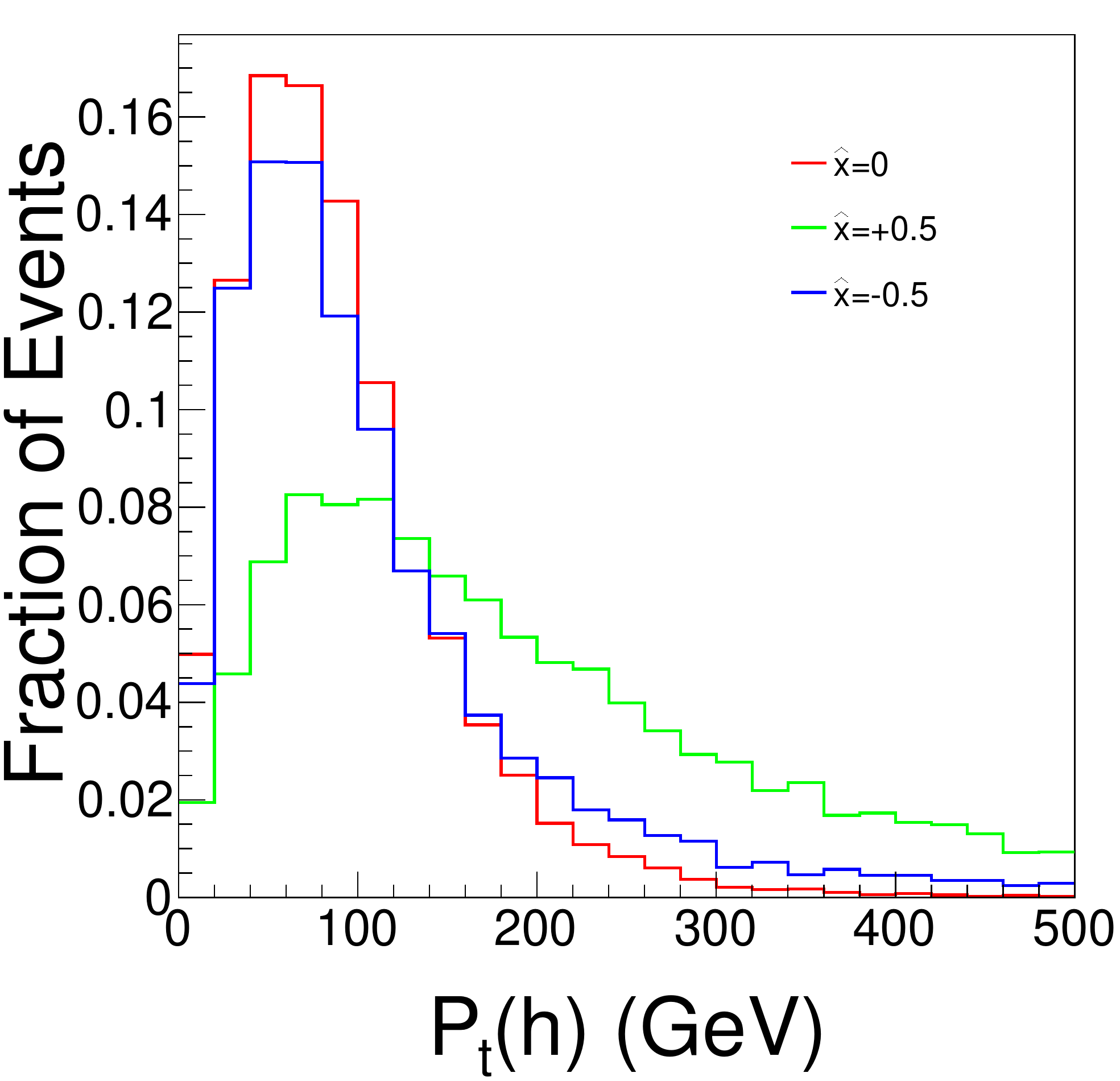}}
  \subfigure{
  \label{Fig8.sub.4}\thesubfigure
  \includegraphics[width=0.4\textwidth]{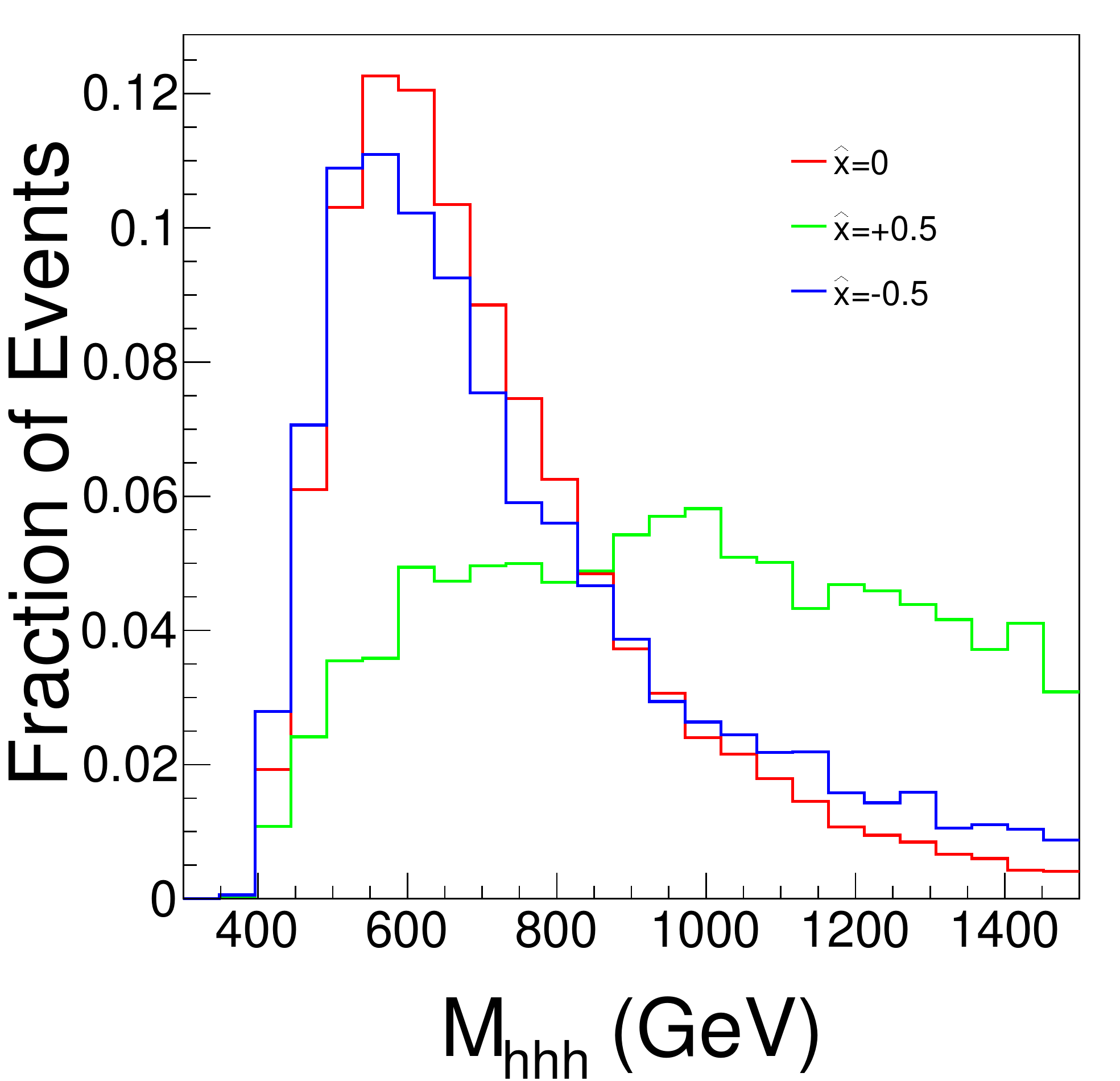}}
  \caption{$P_{t}$ distributions of (a) the leading Higgs, (b) the
    sub-leading Higgs, and (c) the softest Higgs.  In (d), we show the
    distribution of the invariant mass of the triple-Higgs
    system.  We plot results for three values of $\hat{x}$: $-0.5$, $0$ and
    $+0.5$, where $\hat{r}$ is fixed to zero.}\label{fig8}
\end{figure}
A visual representation of the cross-section dependence on the
parameters $(\hat{x},\hat{r})$ is shown in Fig.~\ref{fig7}.  We read
off that the cross section can exceed the SM value by two orders of
magnitude for reasonable variations of $(\hat{x},\hat{r})$.  In
particular, if $\hat r$ is fixed (Fig.~\ref{Fig7.sub.1}), the cross section
increases in the $\hat{x}<0$ region.  In this region, all of the
dependent parameters $\lambda_3$, $\lambda_4$, $\kappa_5$, and
$\kappa_6$ have the same sign, and the derivative couplings can
greatly enhance the cross section.  By contrast, in the $\hat{x}>0$
region, the contributions of $\lambda_3$ and $\lambda_4$ cancel
against the terms with $\kappa_5$ and $\kappa_6$.

The complementary plot Fig.~\ref{Fig7.sub.2} shows the dependence on
$\hat r$ for fixed $\hat x$.  The cross section changes only mildly
with $\hat r$ as long as $\hat x$ is small or positive, and for $\hat
r>0$ it actually undershoots the SM value.  We recall that the
dominant contribution to triple-Higgs production originates from the
diagram with a pentagon top-quark loop~\cite{Chen:2015gva}.  This
part does not depend on the Higgs self-couplings which enter the
parameter $\hat{r}$.  Only if the latter contributions become sizable
and the interference is constructive, we expect a
large enhancement of the cross section.

Beyond the effect on the total cross section, we may look at
distortions of kinematical distributions.  The $P_t$ distributions of
the three Higgs bosons are shown in Fig.~\ref{Fig8.sub.1},
Fig.~\ref{Fig8.sub.2}, and Fig.~\ref{Fig8.sub.3}, for three different
values of $\hat{x}$: $-0.5$, $0$ and $+0.5$, respectively.  We
observe that the distributions change significantly with respect to
the SM reference value if $\hat{x}=+0.5$, especially in the large
$P_t$ region.  The distortion happens in the parameter region where
the total cross section is not enhanced by a large factor, and it is
helpful for the $2b2l^\pm4j+\missE$ channel since it should improve
the reconstruction of the softest jet.  For $\hat{x}=-0.5$ the
distributions do not change that much, but the analysis would benefit
from the remarkable cross-section enhancement in that region.  We also
show the invariant mass distribution of the three Higgs bosons
(Fig.~\ref{Fig8.sub.4}); this is also modified by the derivative
operator.

\begin{figure}[htbp]
  \centering
  \subfigure{
  \label{Fig9.sub.1}\thesubfigure
  \includegraphics[width=0.4\textwidth]{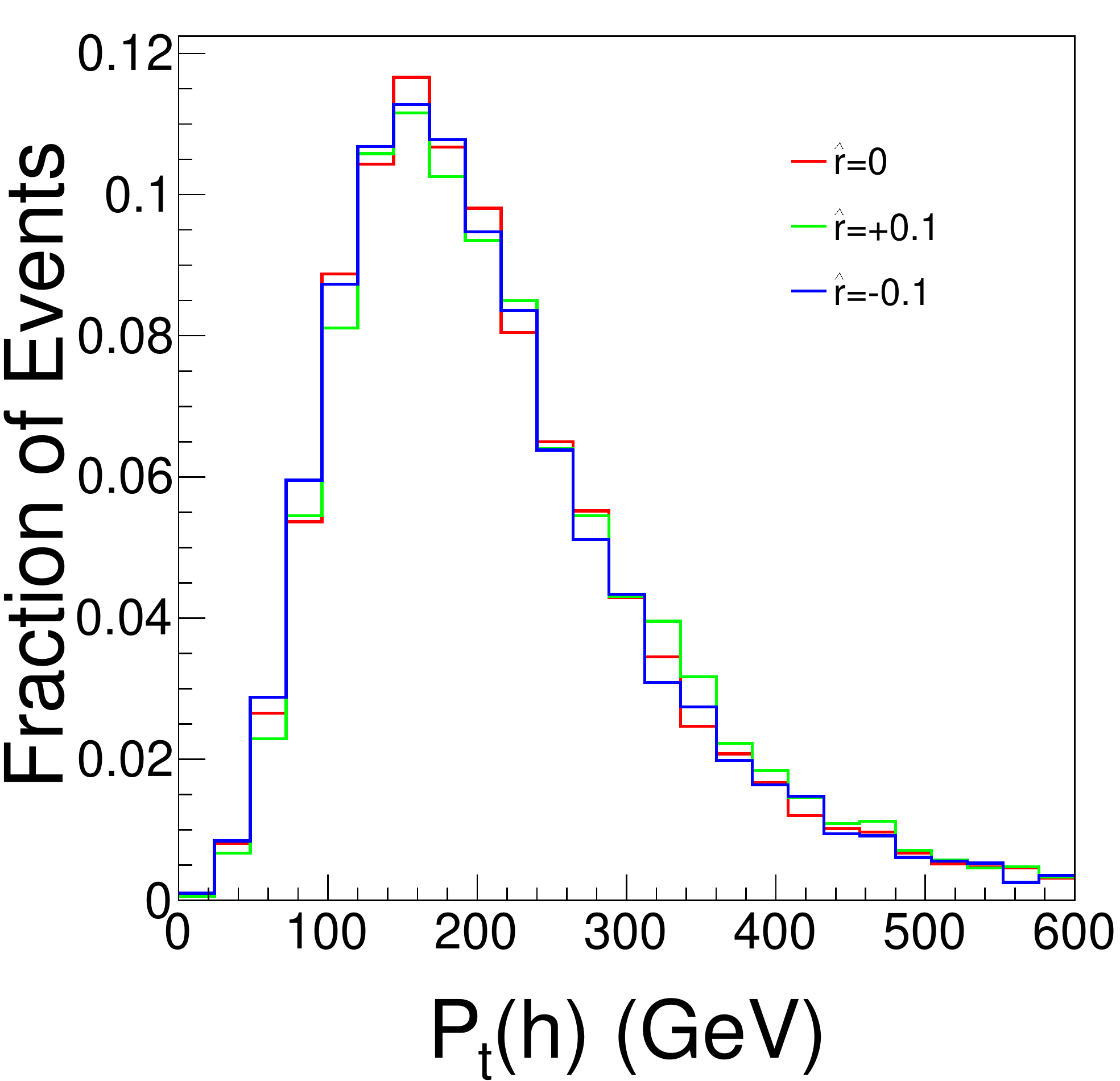}}
  \subfigure{
  \label{Fig9.sub.2}\thesubfigure
  \includegraphics[width=0.4\textwidth]{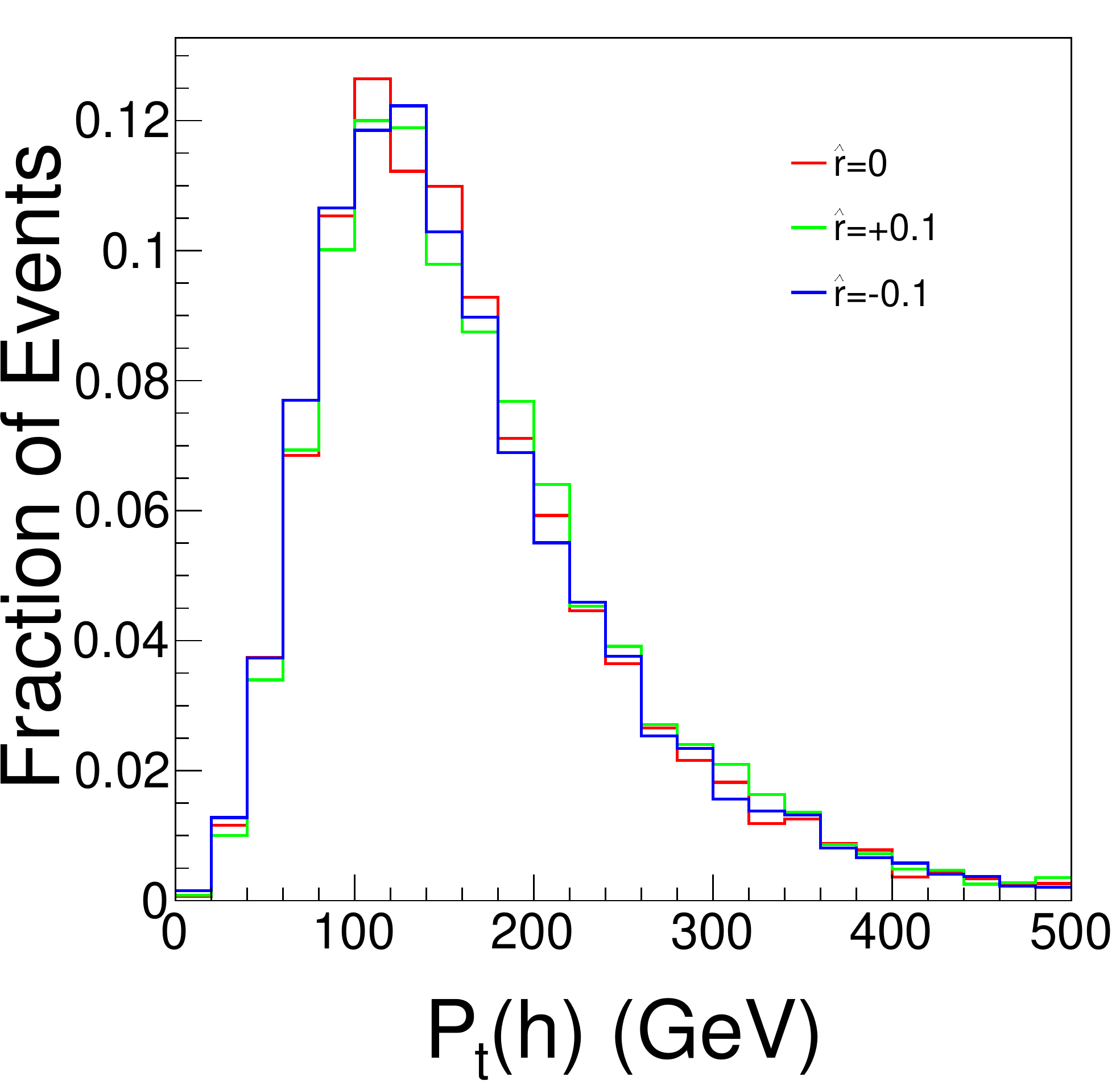}}
  \subfigure{
  \label{Fig9.sub.3}\thesubfigure
  \includegraphics[width=0.4\textwidth]{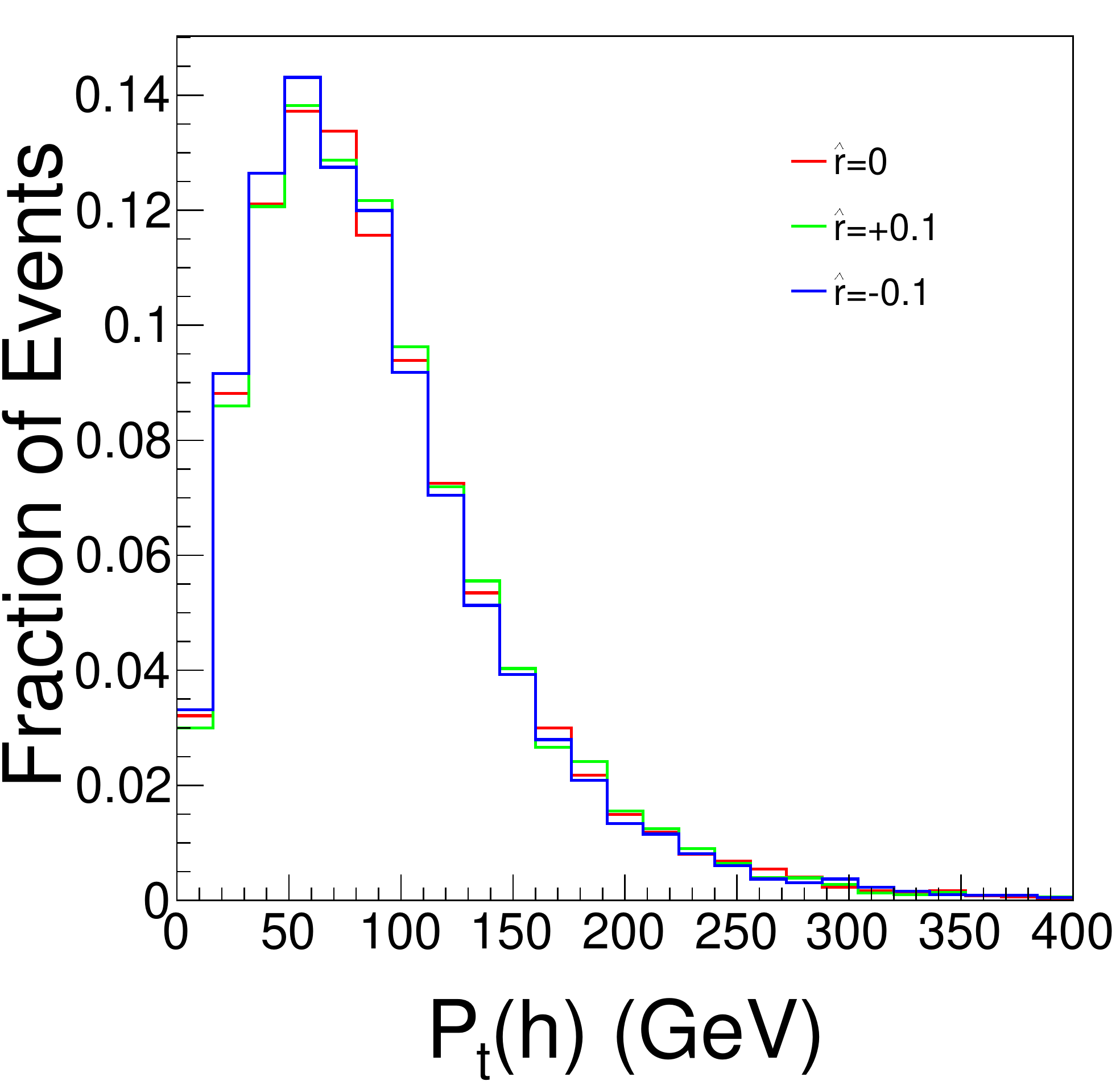}}
  \subfigure{
  \label{Fig9.sub.4}\thesubfigure
  \includegraphics[width=0.4\textwidth]{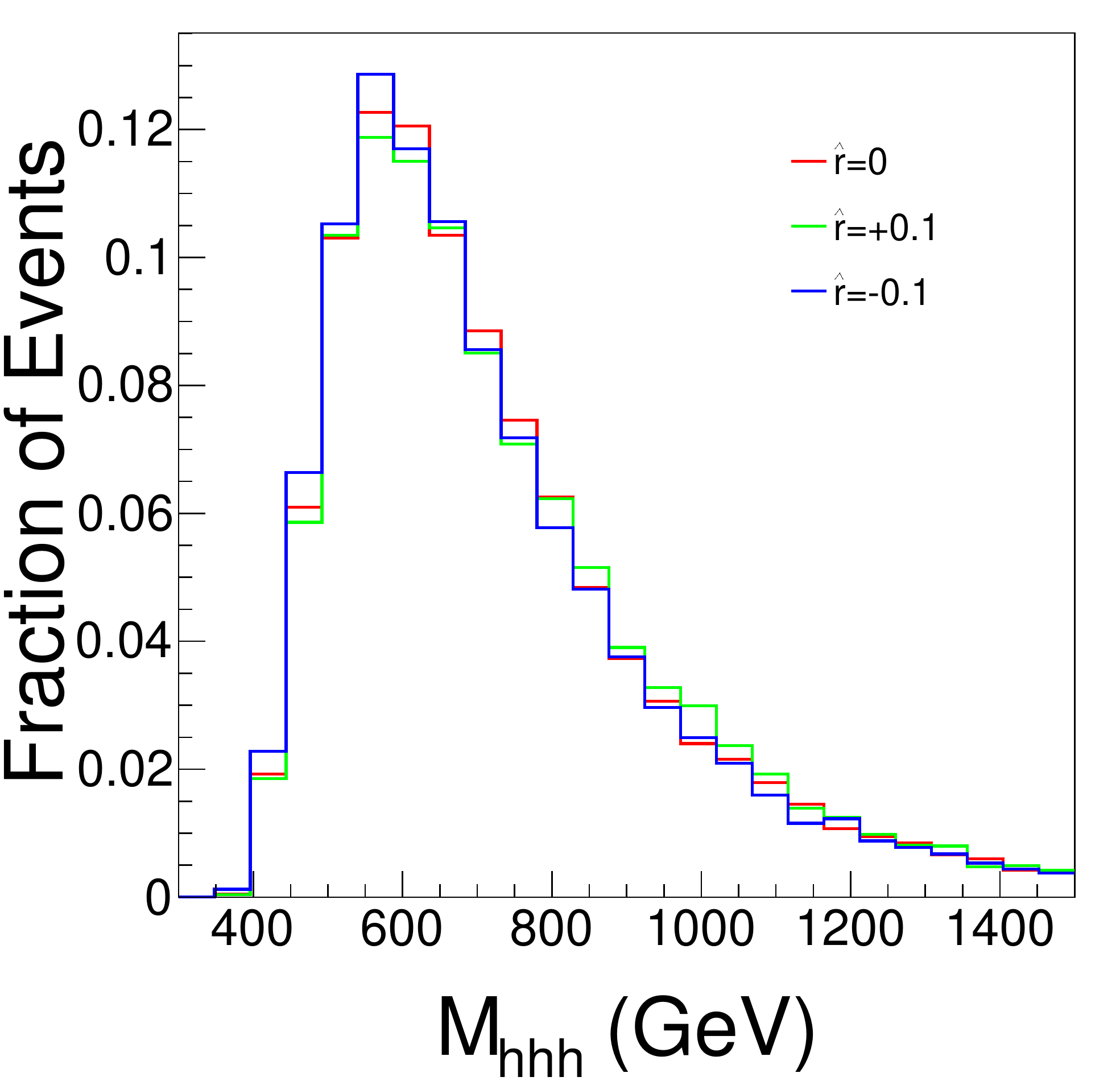}}
  \caption{$P_{t}$ distributions of (a) the leading Higgs, (b) the
    sub-leading Higgs, and (c) the softest Higgs.  In (d), we show the
    distribution of the invariant mass of the triple-Higgs
    system.  We plot results for three values of $\hat{r}$: $-0.1$, $0$ and
    $+0.1$, where $\hat{x}$ is fixed to zero.}\label{fig9}
\end{figure}

In Fig.~\ref{fig9}, we show the same observables as in
Fig.~\ref{fig8}; this time $\hat{r}$ is varied and $\hat{x}$ is fixed
to zero.  The distributions do not actually depend on $\hat{r}$, since
the parameter affects only $\lambda_3$ and $\lambda_4$ which are not
associated with derivative couplings.

\subsection{Correlations between $gg\to hhh$ and single and
  double-Higgs production}
A measurement of triple-Higgs production would not occur in a vacuum.
Apart from the genuine quartic Higgs couplings, all parameters of the
EFT also enter other Higgs processes, and the discussion in
Sec.~\ref{Sec:EFT} suggests that in typical strongly-interacting
models, all parameters would receive BSM contributions.  We can expect
that a measurement or exclusion limit on the triple-Higgs process
would add information on top of the amount of Higgs-physics data
gained up to that point, and all results should be combined
in the interpretation 
towards BSM physics.  Therefore, in this section we study correlations
between $gg \to h h h$ and $gg \to h$, $gg \to h h$ in particular.
We phrase the problem in terms of the following questions: 
\begin{itemize}
\item  To what extent can $a_1$ and $c_1$ be determined from $gg \to
  h$ at the LHC (14 TeV) and at a 100 TeV collider?
\item To what extent can $a_2$, $c_2$, $\lambda_3$, $\kappa_5$ be
  determined from $gg \to hh $ at the LHC and at a 100 TeV Collider?
\item To what extent can $a_3$, $\lambda_4$, $\kappa_6$ be determined
  from $gg \to hhh$ at a 100 TeV collider, including channels not
  considered in this paper?
\end{itemize}

In Table~\ref{tablexs} we quote estimates for the theoretical and
projected experimental uncertainties for the processes $gg\to h$, $gg
\to hh$ and $gg\to hhh$.  For convenience, we list the parameters that
enter these processes in Table~\ref{thhh}.  The theoretical
uncertainties are obtained by summing the squared uncertainties in
PDF, renormalisation scales, and $\alpha_s$, based on current
knowledge.  For the process $gg \to h $, the experimental
uncertainties are mainly statistical ones which pertain to the Higgs
decay $h \to \gamma \gamma$.  The projected experimental bound for $gg
\to h h$ at the LHC is taken from the studies of the $b\bar{b} \gamma
\gamma$ final state~\cite{He:2015spf} and $3\ell 2
j+$MET~\cite{Li:2015yia}.  We also quote the expected experimental
bound for 
$gg \to h h h $ which is expected from the analysis of $4b2\gamma$
final states~\cite{Chen:2015gva} at $100$~TeV.  We emphasize that
these estimates are derived from phenomenological studies; full
simulation and experience gained in the analysis of actual data may
change the conclusions significantly, such as in the expectations for
the observation of Higgs-pair production at the
LHC~\cite{Aad:2015tna}.
\begin{center}
\begin{table}
  \begin{center}
  \begin{tabular}{|c||c|c|c||c|c|c||}
  \hline
  Process                &  $\sigma(14$ $TeV)$ (fb) & err.[th] &  err.[exp] &  $\sigma(100$ $TeV)$ (fb) & err.[th]  & err. [exp]    \\ 
  \hline
  $gg\to h$              &  $4.968\times 10^{4}$ & $^{+7.5\%}_{-9.0\%}$ & $\pm 1\%$          &  $8.02\times 10^{5}$ & $^{+7.5\%}_{-9.0\%}$ & $\pm 0.1\%$  \\ 
  \hline
  $gg\to hh$             &  $45.05$ & $^{+7.3\%}_{-8.4\%}$ & $ < 120 fb$   &  $1749$ & $^{+5.7\%}_{-6.6\%}$ & $\pm 5 \%$   \\ 
  \hline
  $gg\to hhh$            &  $0.0892$ & $^{+8.0\%}_{-6.8\%} $ & $-$  & $4.82$ & $^{+4.1\%}_{-3.7\%}$ & $ < 30 fb $\\
  \hline
  \end{tabular}
  \end{center}
  \caption{\label{tablexs} Cross sections of the processes $gg\to h$,
    $gg\to hh$ and $gg\to hhh$ at $14$ TeV and $100$ TeV hadron
    colliders, respectively. The $14$ TeV cross section of $gg\to h$
    is taken from Ref.~\cite{xsworkinggroup}; the other values are
    taken from Ref.~\cite{Contino:2016spe}. The cross sections for $gg\to
    h$ and $gg\to hh$ are the NNLO results, while the cross sections
    for $gg\to hhh$ are the NLO results.}
\end{table}
\end{center}

\begin{center}
\begin{table}
    \begin{center}
    \begin{tabular}{|c|c|c|c|c|}
        \hline
        &  $gg \to h$ & $gg \to h h$ & $gg \to h h h$ \\
         \hline
        Parameters                     & $a_1$, $c_1$ &  $a_1$, $c_1$  &  $a_1$, $c_1$  \\
            involved                & - &  $a_2$, $c_2$, $\lambda_3$, $\kappa_5$  &   $a_2$, $c_2$, $\lambda_3$, $\kappa_5$ \\
                             & - &  -  &   $a_3$, $\lambda_4$, $\kappa_6$ \\
         \hline
    \end{tabular}
    \end{center}
    \caption{ \label{thhh} Parameters that contribute to the
      particular Higgs-production processes.}
\end{table}
\end{center}

We first consider Higgs-pair production in gluon fusion, the dominant
contribution to the process $pp\to hh$.  The dependence of the total
cross section on the EFT parameters can be written as in
Eq.~(\ref{xsechh1}).   Explicitly, we have
\begin{align}
    \sigma(pp\to hh)=&f_1 a_1^4+f_2 a_1^3\lambda_3+f_3 a_1^3\kappa_5+ f_4 a_1^2\lambda_3^2+f_5 a_1^2\lambda_3\kappa_5 \nonumber \\ + & f_6 a_1^2\kappa_5^2 
    + f_7 a_1^2a_2+ f_8 a_1\lambda_3a_2+ f_9 a_1\kappa_5a_2+ f_{10} a_2^2 \,.
    \label{xsechh1}
\end{align}
The numerical values for the coefficients $f_1-f_{10}$ at a $100$~TeV
hadron collider are
provided in Table~\ref{numxsechh}.
\begin{center}
\begin{table}
    \begin{center}
    \begin{tabular}{|c|c|c|c|c|c|c|c|c|c|}
        \hline
        $f_1$ & $f_2$ & $f_3$ & $f_4$ & $f_5$ & $f_6$ & $f_7$ & $f_8$ & $f_9$ & $f_{10}$ \\
        1.56 & -0.94 & -2.14 & 0.18 & 0.69 & 0.87 & -3.34 & 1.01 & 2.77 & 2.26 \\
        \hline
    \end{tabular}
    \end{center}
    \caption{ Numerical results for $f_1-f_{10}$(in pb) at a 100 TeV
      hadron collider.\label{numxsechh}}
\end{table}
\end{center}

By substituting $(a_1,\lambda_3,\lambda_4,\kappa_5)$ into
$(\hat{r},\hat{x})$ according to Table~\ref{table5}, we obtain
\begin{align}
    \sigma(pp\to hh)=& \sigma_{SM}^{hh} (1 - \hat{x})^2  (1 + \hat{f_1} \hat{x} + \hat{f_2} \hat{r} + \hat{f_3} \hat{x}^2 + \hat{f_4} \hat{x} \hat{r}  + \hat{f_5} \hat{r}^2 ),
    \label{xsechhrx}
\end{align}
where $\sigma^{hh}_{SM}=1.75$ pb.  (We insert a NNLO K factor of
2.17~\cite{deFlorian:2013jea}.)  Table~\ref{xrsechh} lists the
numerical values for $\hat f_i$ in this expansion.

\begin{center}
\begin{table}
    \begin{center}
    \begin{tabular}{|c|c|c|c|c|}
        \hline
        $\hat{f_1}$ & $\hat{f_2}$ & $\hat{f_3}$ & $\hat{f_4}$ & $\hat{f_5}$  \\
       -3.63 & -0.72 & 4.32 & 1.72 & 0.23 \\
        \hline
    \end{tabular}
    \end{center}
    \caption{ Numerical results for $\hat{f_1}-\hat{f_{5}}$ at a 100 TeV hadron collider.\label{xrsechh}}
\end{table}
\end{center}

\begin{figure}[htbp]
  \centering
  \subfigure{
  \label{1ha1c114}\thesubfigure
  \includegraphics[width=0.4\textwidth]{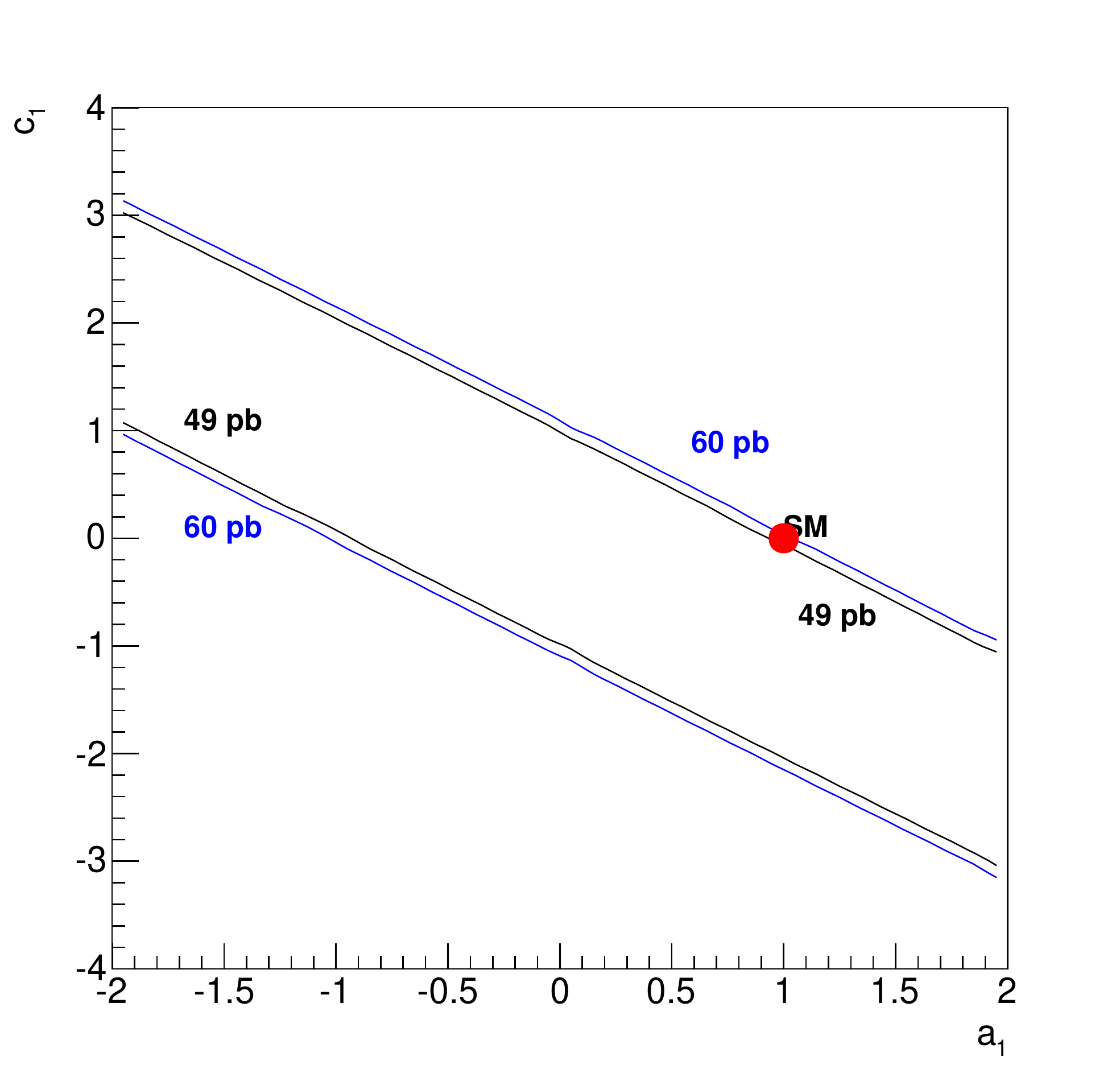}}
  \subfigure{
  \label{1ha1c1100}\thesubfigure
  \includegraphics[width=0.4\textwidth]{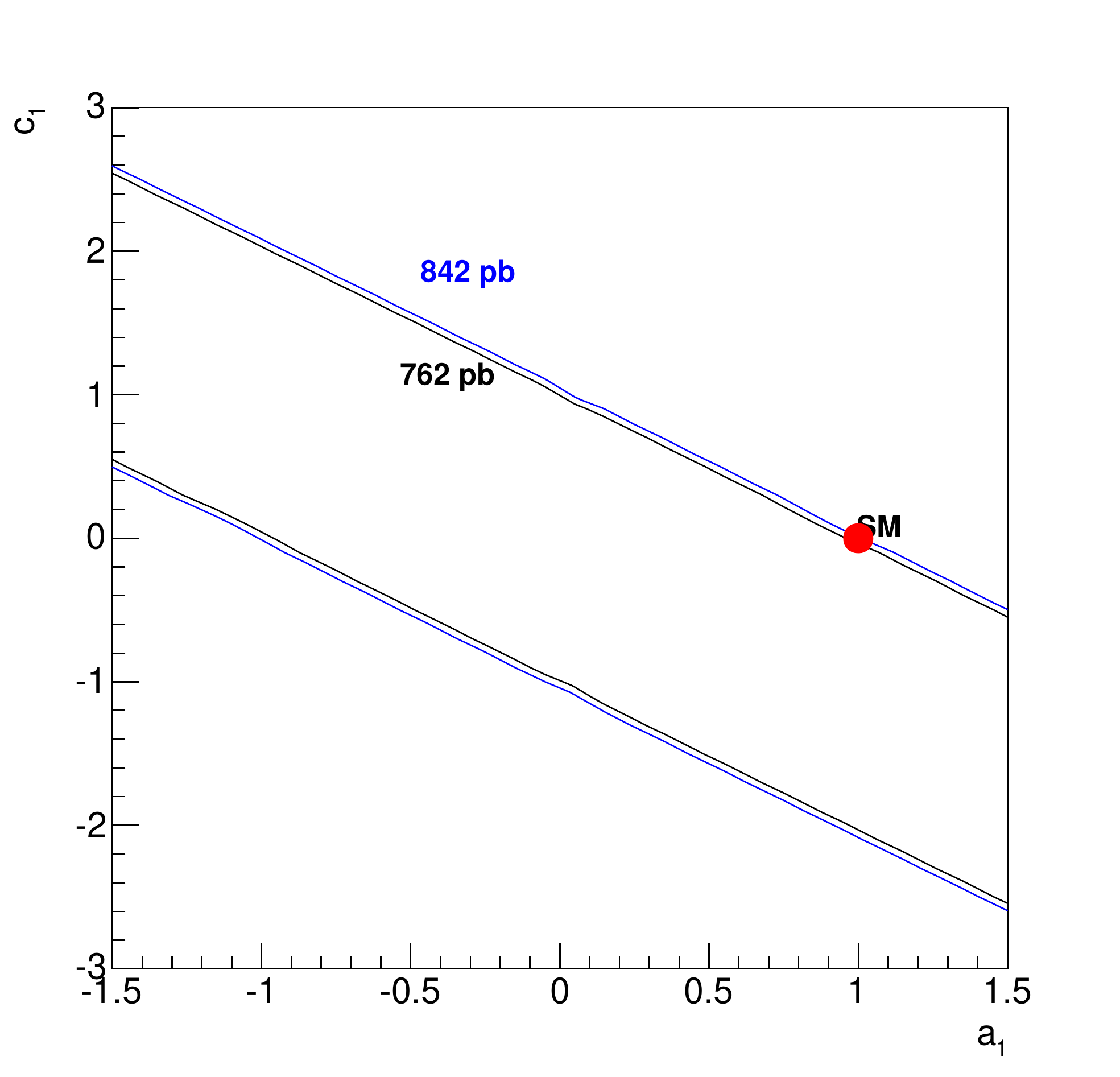}}
  \subfigure{
  \label{a1bd14}\thesubfigure
  \includegraphics[width=0.4\textwidth]{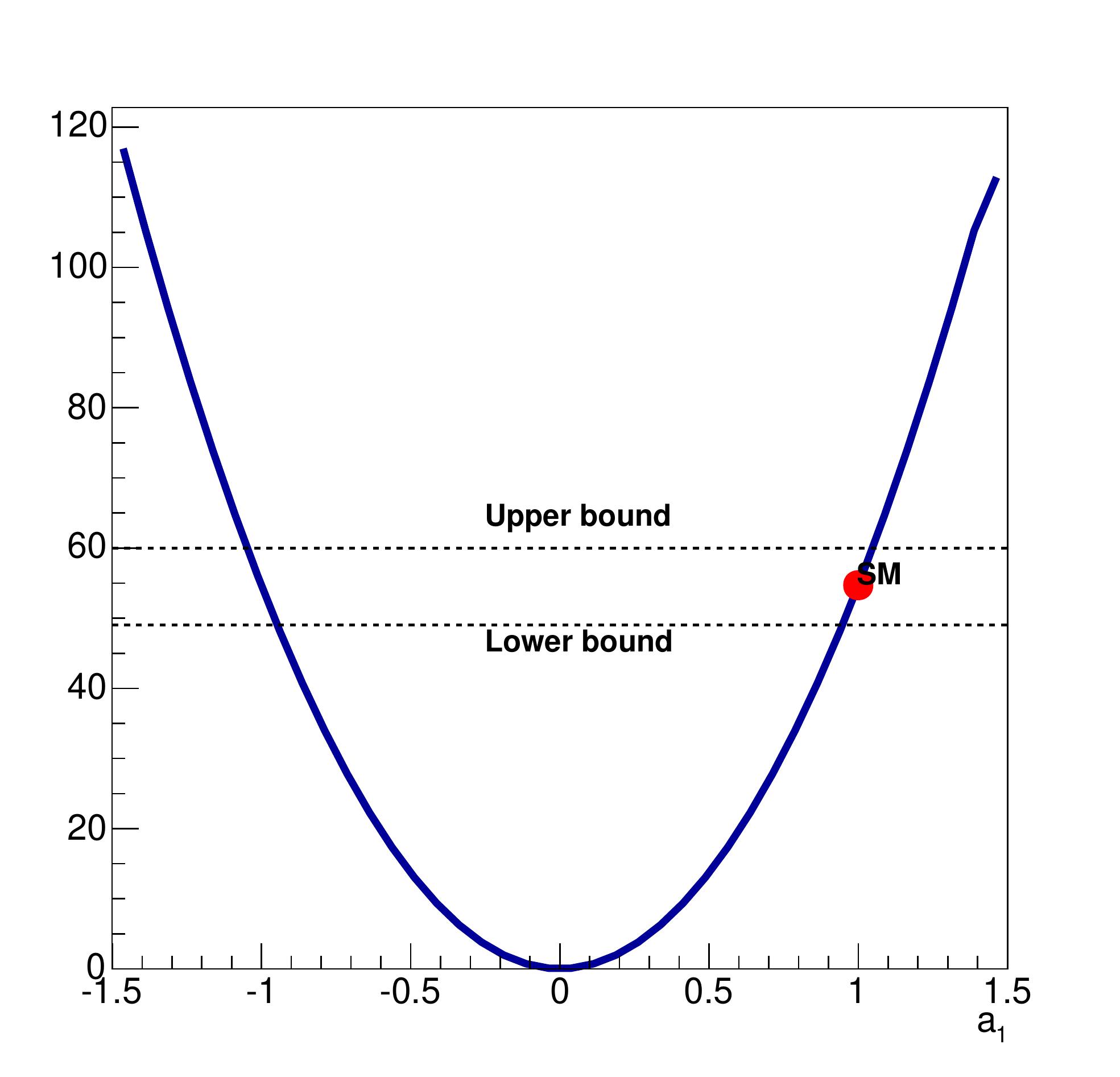}}
  \subfigure{
  \label{a1bd100}\thesubfigure
  \includegraphics[width=0.4\textwidth]{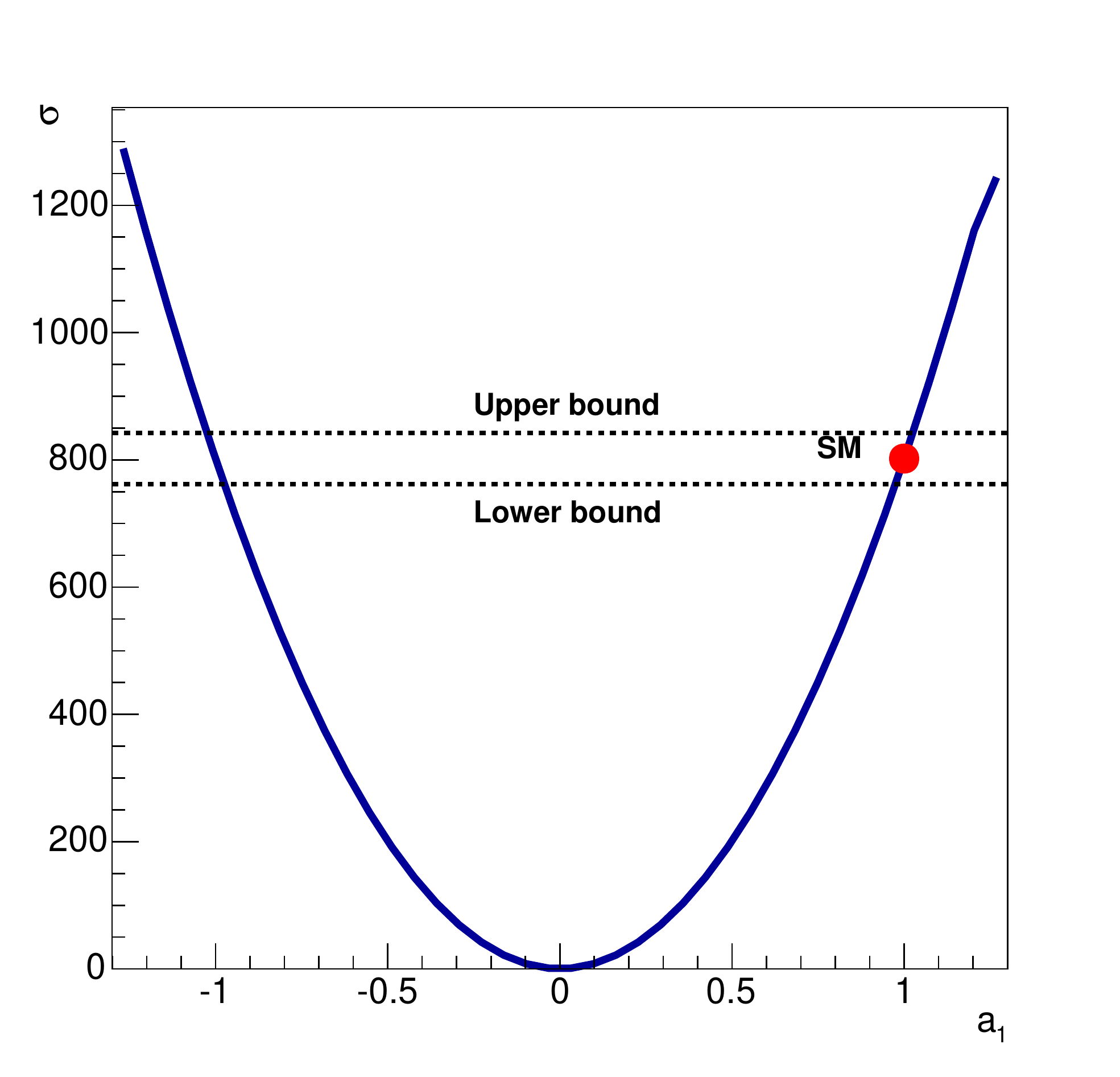}}
  \subfigure{
  \label{c1bd14}\thesubfigure
  \includegraphics[width=0.4\textwidth]{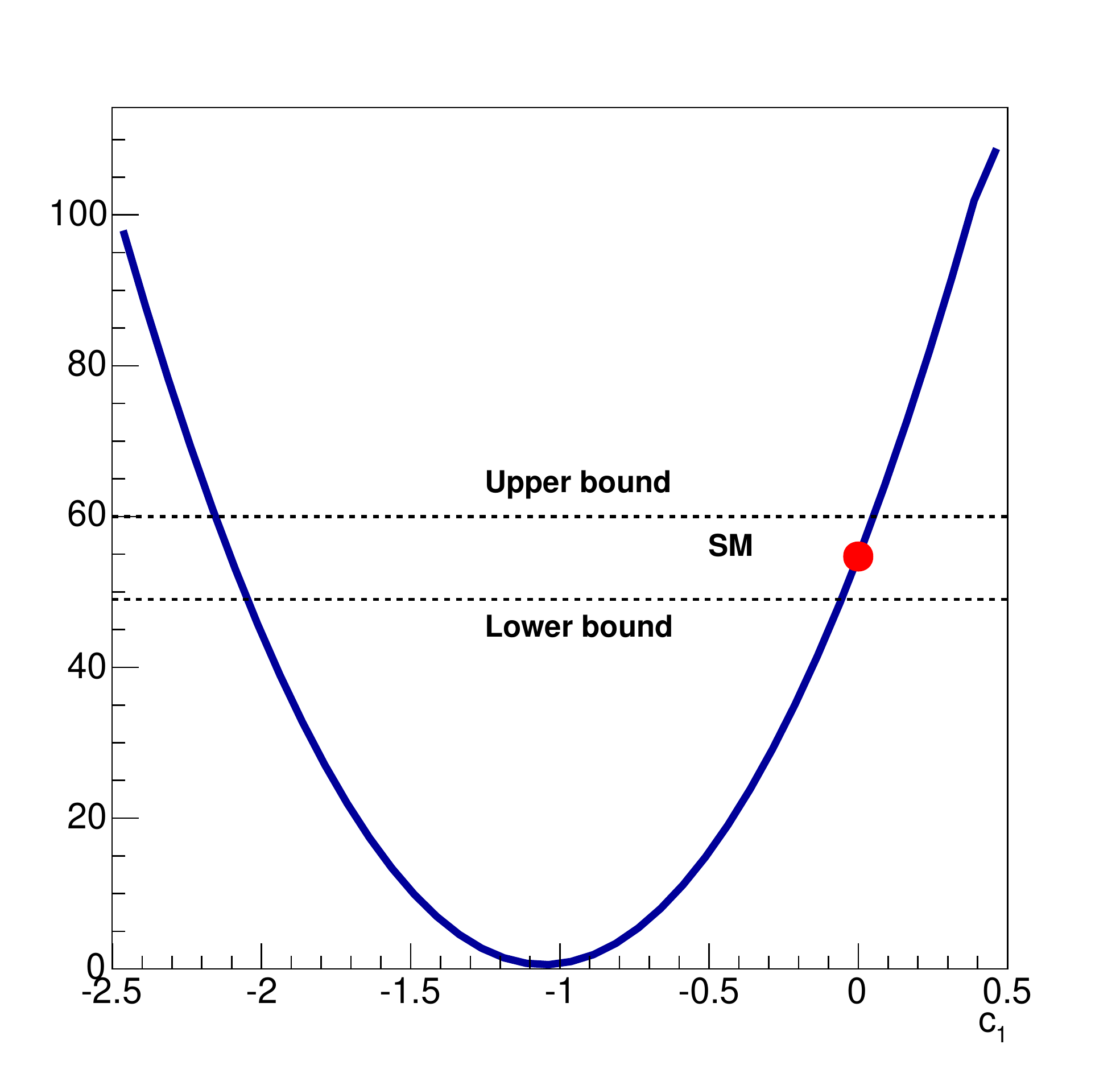}}
  \subfigure{
  \label{c1bd100}\thesubfigure
  \includegraphics[width=0.4\textwidth]{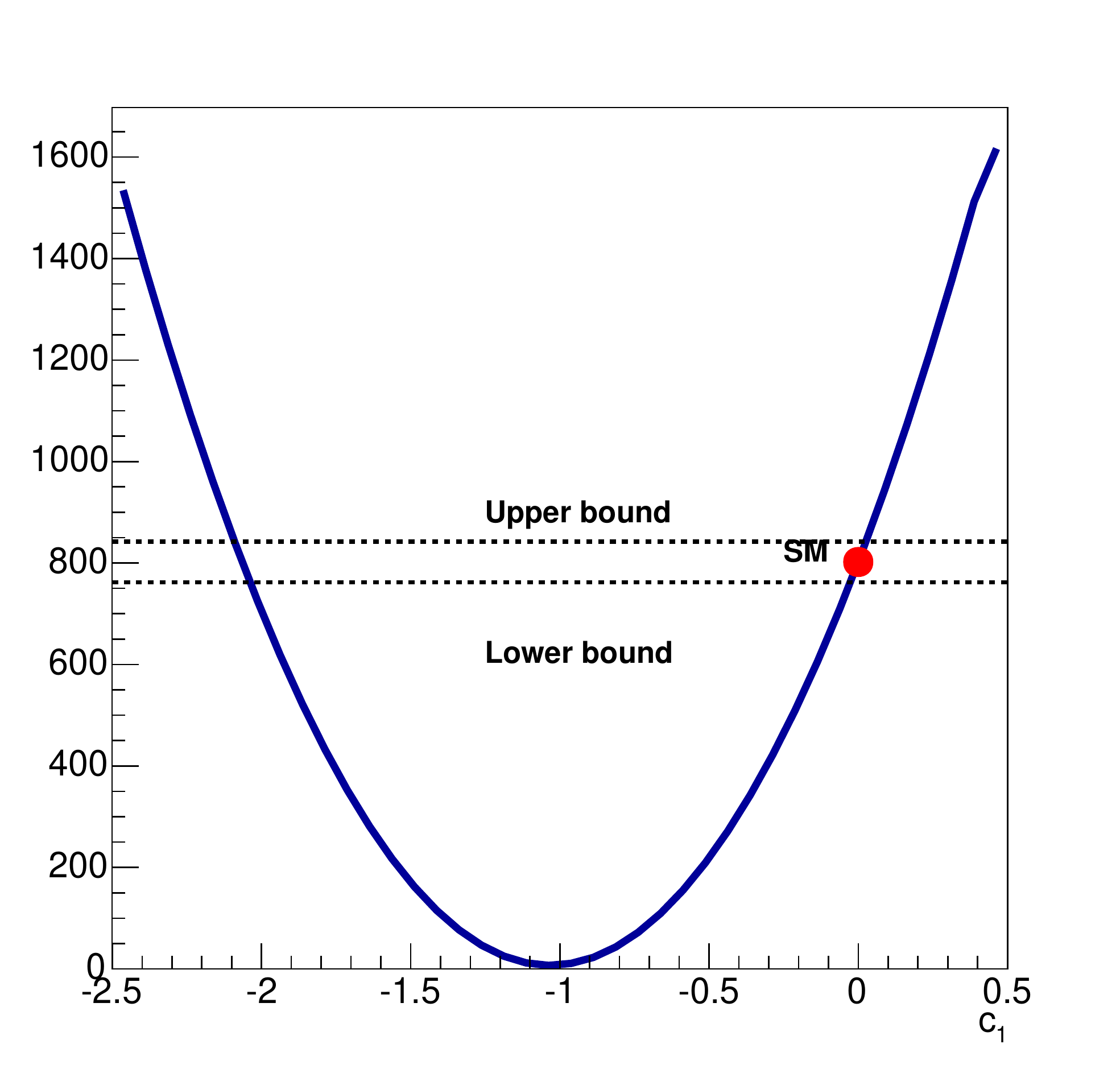}}
  \caption{Upper row: correlations between $a_1$ and $c_1$ extracted from the
    process $gg \to h$ for the LHC 14 TeV (a) and for a 100 TeV $pp$
    collider (b), respectively.  Middle row: individual bounds on $a_1$ (c) and
    $c_1$ (e) for the LHC 14 TeV (the total uncertainties are assumed
    to be $10\%$), respectively.   Lower row: the analogous results
    for a 100 TeV $pp$ collider (the total uncertainties are
    assumed to be $5\%$). }\label{gg1hbds}
\end{figure}

Turning to single-Higgs production, in Fig.~\ref{gg1hbds} we show
projections on the bounds of $a_1$ and $c_1$ from the process $g g \to
h$ at the LHC ($14$~TeV) and at a $100$ TeV collider, respectively.
Assuming a measurement result equal to the SM prediction, the allowed ranges for
$a_1$ and $c_1$ are highly correlated and are confined to be two
narrow bands, one of which containing the SM reference point.  The
other band is centered on a mirror solution $a_1=-1$, $c_1=0$.

At the 14 TeV LHC, both theoretical and experimental (statistical)
uncertainties are relevant and have to be taken into account.  By
contrast, at a 100 TeV collider the main uncertainties will come from
theory, while statistical uncertainties will become less than
$0.1\%$. Compared with the ultimate LHC bounds, the projected accuracy
on the determination of $a_1$ and $c_1$ from a 100 TeV collider will
improve by a factor~$2$.  If theoretical and parametric
uncertainties can also be improved in the future, we can expect the
bounds on $a_1$ and $c_1$ to tighten even further.

The correlation of the parameters in Fig.~\ref{gg1hbds}a and
Fig.~\ref{gg1hbds}b indicates an obvious shortcoming of a simple
cross-section analysis.  The degeneracy of solutions can be lifted by
examining the kinematics of the Higgs boson in the final state, and by
adding the measurement of $gg\to hh$, as we will explain later.

A future $e^+e^-$ collider can improve the measurement of $a_1$ and
$c_1$ beyond the reach of the LHC.  This is mainly due to QCD entering
the processes only beyond leading order, so statistical uncertainties
will likely dominate.  For instance, at the CEPC a combination of the
parameters $c_1$ and $a_1$ can be determined within $1\%$ for a
collision energy $\sqrt{s}=240-250$ GeV and an integrated luminosity
5 ab$^{-1}$~\cite{Ge:2016zro}.  At the ILC, the parameter $a_1$ can be
determined within $10\%$ for $\sqrt{s}=500$ GeV and an integrated
luminosity of 1 ab$^{-1}$, via the process $e^+ e^- \to t \bar{t}
h$~\cite{Tian:2016qlk}.  We add the result that at a 100 TeV hadron
collider 
with integrated luminosity 30 ab$^{-1}$, $a_1$ can be further constrained by
the measurement of $tt h$ to within $1\%$~\cite{Plehn:2015cta}.

\begin{figure}[htbp]
  \centering
  \subfigure{
  \label{2ha1c114}\thesubfigure
  \includegraphics[width=0.4\textwidth]{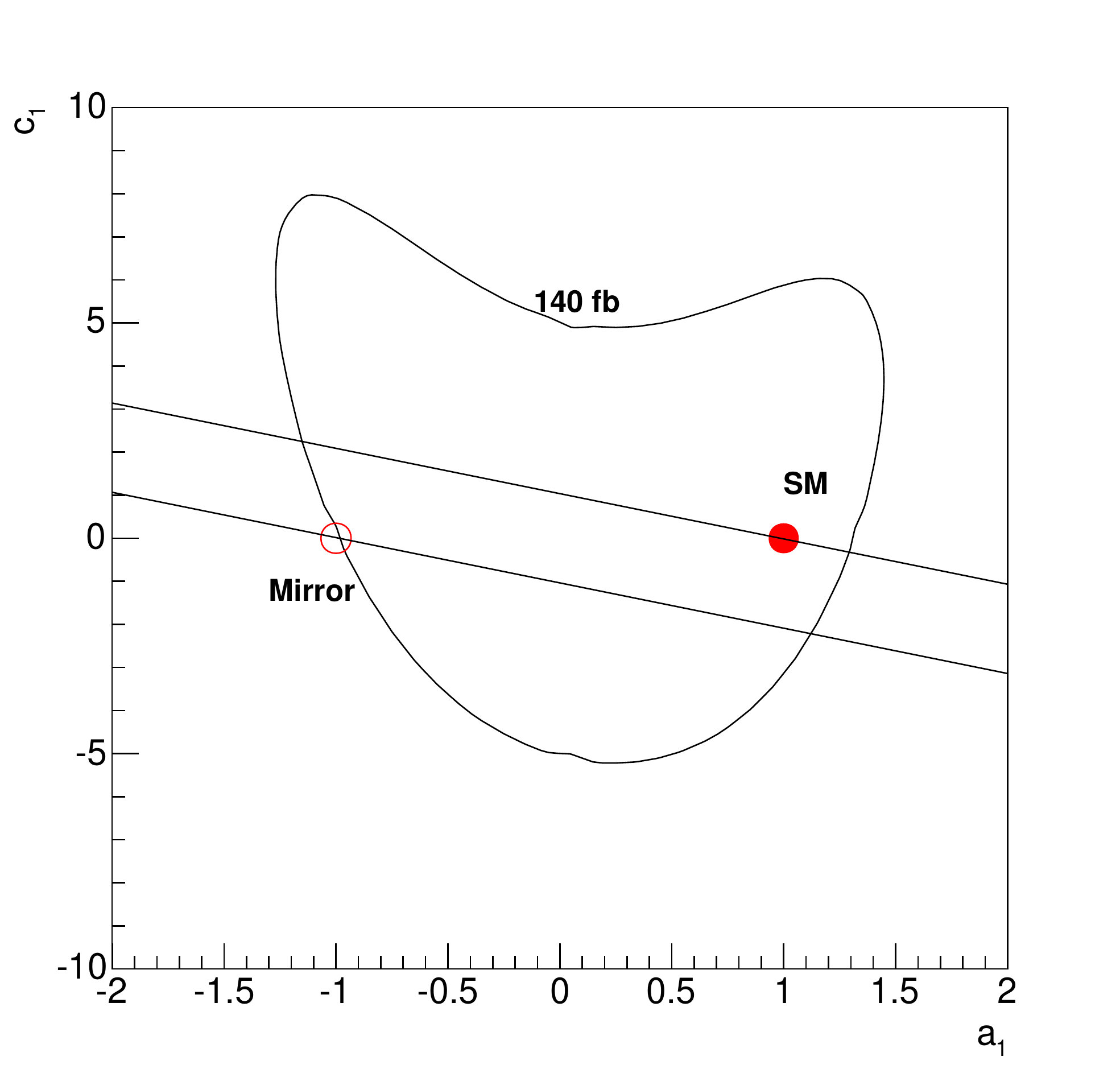}}
  \subfigure{
  \label{2ha1c1100}\thesubfigure
  \includegraphics[width=0.4\textwidth]{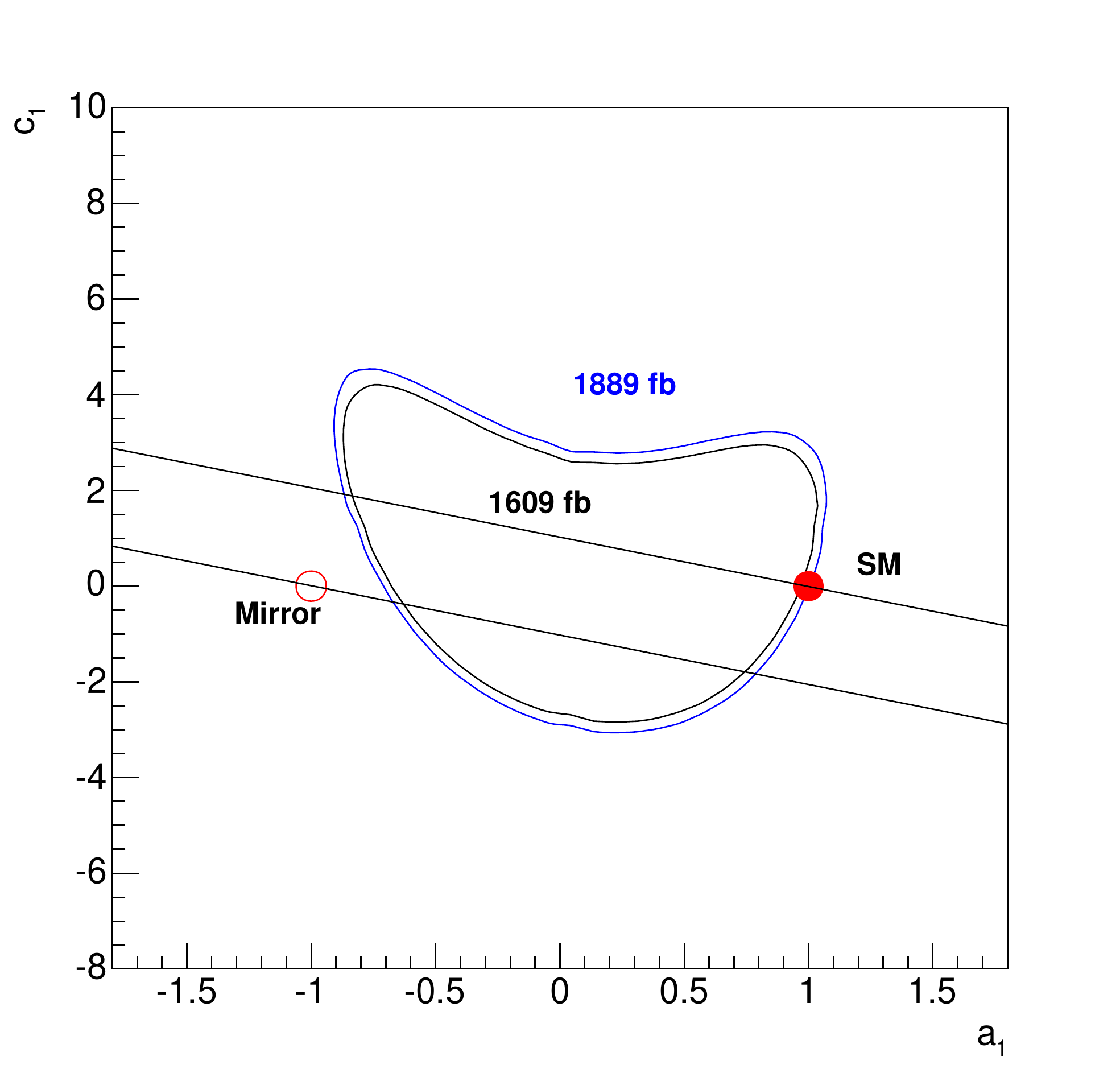}}
  \caption{Projected exclusion bounds in $a_1$-$c_1$ plane extracted
    from the process $gg \to hh$ for both (a) the LHC 14 TeV and (b) a
    100 TeV collider.  In each plot, the straight lines indicate the
    solutions for the cross
    section of $gg \to h$, assuming a measurement consistemt with the
    SM. The mirror region of the SM point, the solution with $a_1 =
    -1$ and $c_1=0$, is
    denoted by a circle, respectively. }\label{gg2h14a1c1bds}
\end{figure}

In Fig.~\ref{gg2h14a1c1bds}, we display the expected bounds in the
$a_1$-$c_1$ plane for both the LHC 14 TeV and a 100 TeV collider.
Adding in $gg \to hh $, the linear degeneracy in
the $a_1$-$c_1$ plane that follows from the measurement of $gg \to h$
is cut down to a limited region, even for the LHC 14 TeV.  With a 100
TeV collider, the degenerate solutions for both $g g \to h$ and $gg
\to h h$ shrink to four small regions, cf. TABLE.~\ref{benchmarks}, and
it becomes possible to exclude the mirror solution.

In this situation, we may include the cross section of the process $gg
\to hhh$ and distinguish the second point from the rest.  The second
point has a production rate for the process $gg \to hhh$ that is large
enough to be observed.  (Here we would like to remind the reader that
when we examine the correlations of $a_1$ and $c_1$, we set the other
free parameters to their SM values.)
\begin{center}
\begin{table}
    \begin{center}
    \begin{tabular}{|c||c|c||c|c|c|}
        \hline
 No.       & $a_1$ & $c_1$ & $\sigma(gg\to h)$ [pb] &  $\sigma(gg \to h h)$ [fb]& $\sigma(gg \to h h h)$ [fb] \\
        \hline
        1 & 0.99 & -0.01& 771 & 1710 & 5.90 \\
        \hline
        2 & -0.86 & 1.94 & 839.6 & 1685 & 29.7 \\
        \hline
        3 & 0.78 & -1.82 & 763 & 1747 & 6.23 \\
        \hline
        4 & -0.66 & -0.37 & 817.8 & 1690 & 5.74\\
        \hline
    \end{tabular}
    \end{center}
    \caption{Four representative points in the four parameter regions
      and the corresponding cross sections for Higgs production at a
      100 TeV collider.\label{benchmarks}} 
\end{table}
\end{center}

\begin{figure}[htbp]
  \centering
  \subfigure{
  \label{2ha2l314}\thesubfigure
  \includegraphics[width=0.4\textwidth]{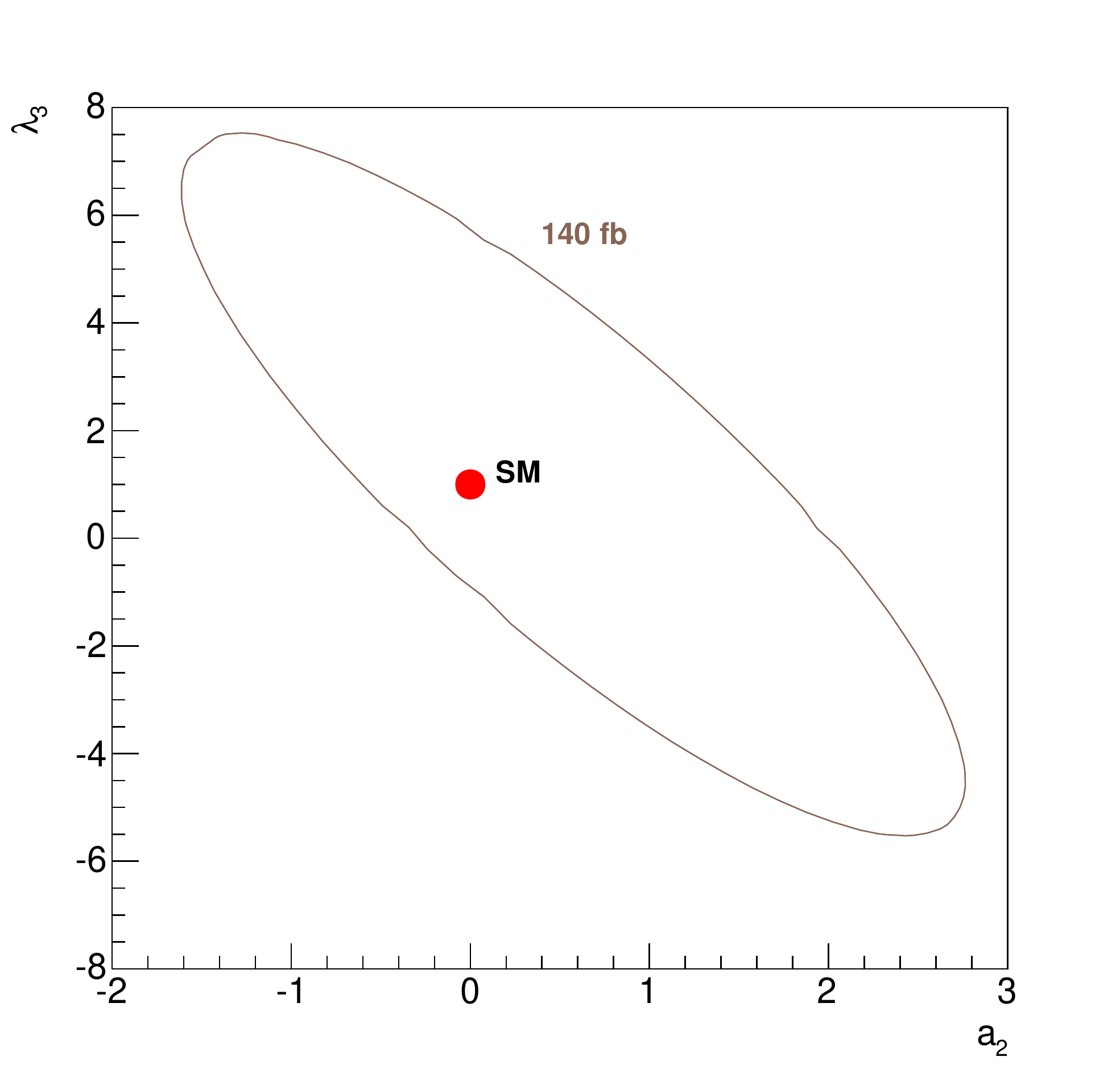}}
  \subfigure{
  \label{2hc1l314}\thesubfigure
  \includegraphics[width=0.4\textwidth]{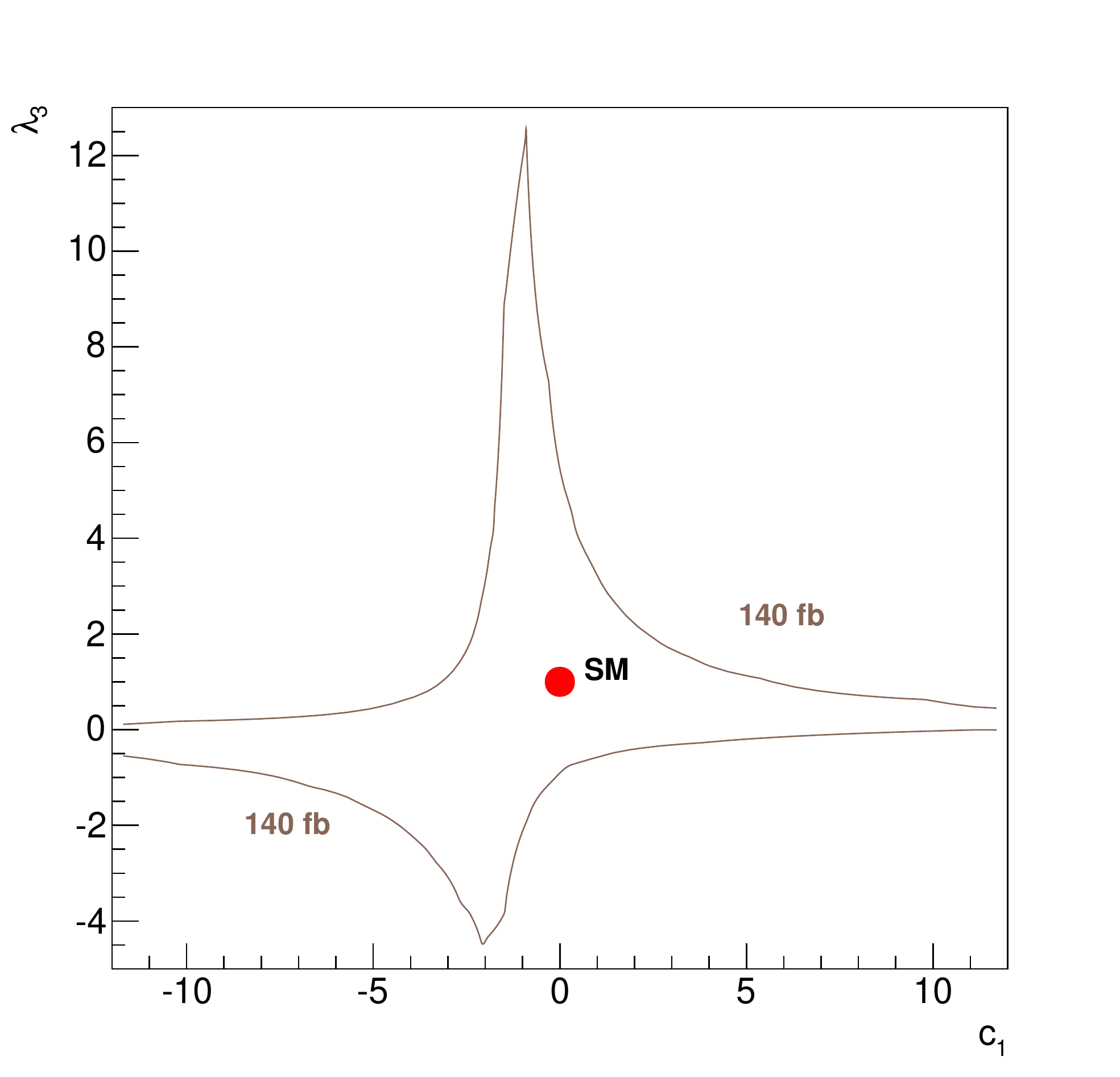}}
  \subfigure{
  \label{2hc2l314}\thesubfigure
  \includegraphics[width=0.4\textwidth]{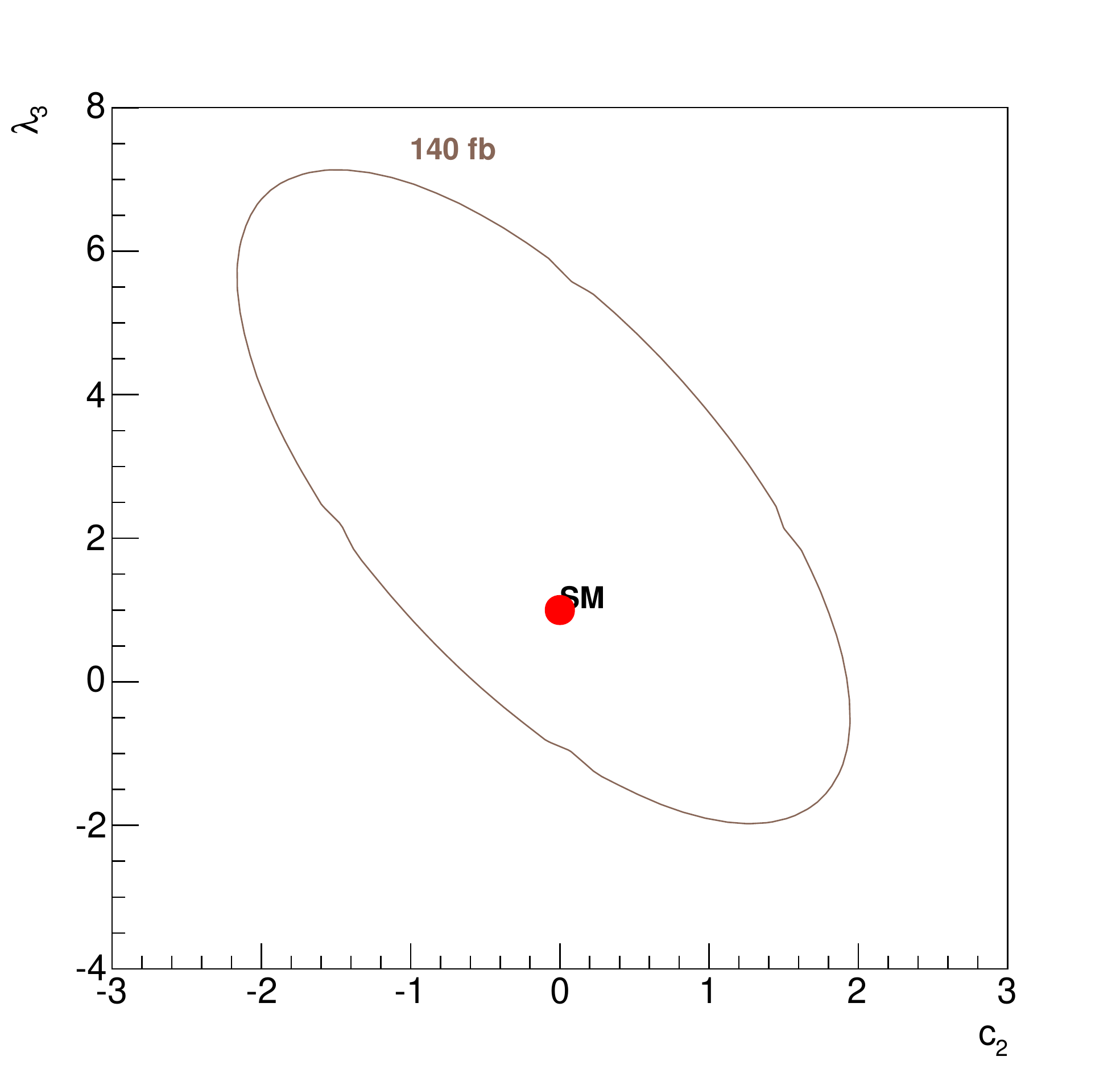}}
  \subfigure{
  \label{2hk5l314}\thesubfigure
  \includegraphics[width=0.4\textwidth]{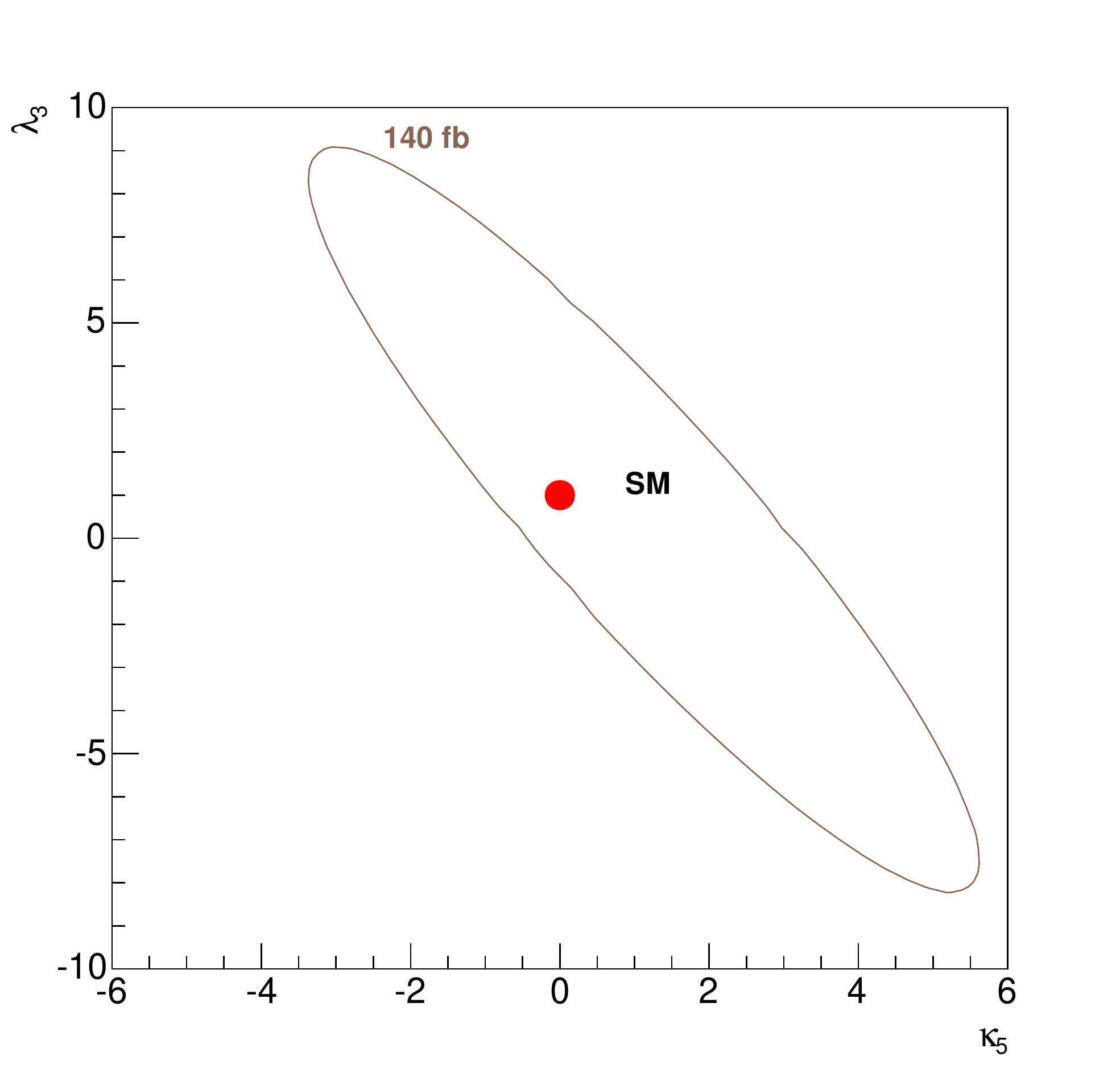}}
  \caption{Projected exclusion bounds in two-parameter planes between
    ($a_2$, $c_1$, $c_2$, $\kappa_5$) and $\lambda_3$, extracted from
    the process $g g \to h h$ at the LHC 14 TeV.  If the coefficient values are
    equal to the SM prediction, parameter values inside the
    contours are still allowed by the measurement.  The exclusion
    bounds correpond to a limit of 
    140 fb for the cross section. }\label{gg2h14bds}
\end{figure}

We now extend the study to the other parameters that enter $gg \to
hh$. In Fig.~\ref{gg2h14bds} we display the expected LHC bounds for
those in four planes, namely $a_2$-$\lambda_3$, $c_1$-$\lambda_3$,
$c_2$-$\lambda_3$, and $\kappa_5$-$\lambda_3$.  All bounds are derived
by requiring the cross section of $gg \to hh$ to be smaller than 140
fb, so the points inside the exclusion bounds will remain allowed by
the LHC 14 TeV data, if no deviation from the SM is detected.  The SM
reference points are also shown.  When we examine the correlations
between two parameters, here and in all later discussions we set the
remaining parameters equal to their SM values.

\begin{figure}[htbp]
  \centering
  \subfigure{
  \label{2ha2l3100}\thesubfigure
  \includegraphics[width=0.4\textwidth]{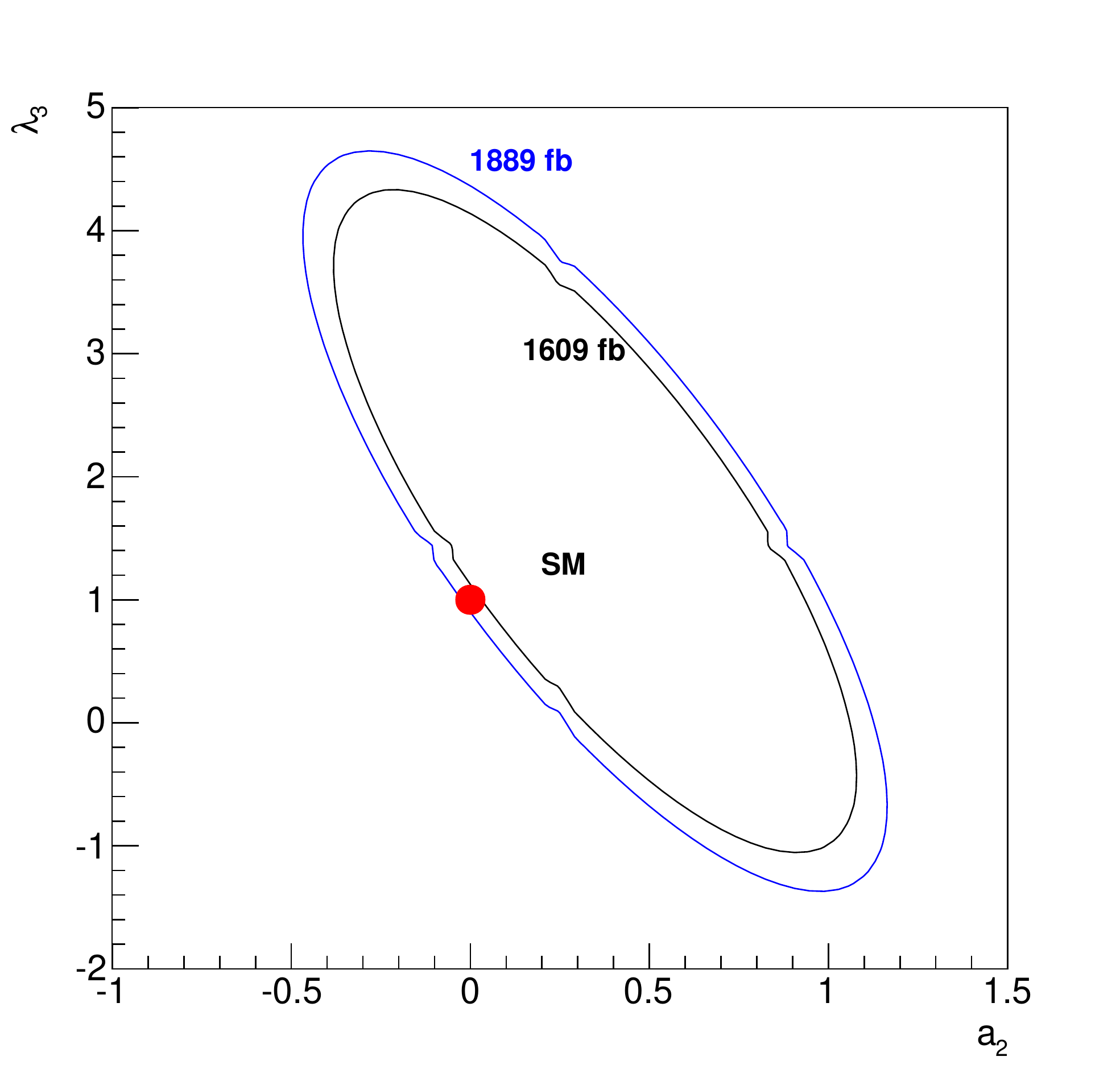}}
  \subfigure{
  \label{2hc1l3100}\thesubfigure
  \includegraphics[width=0.4\textwidth]{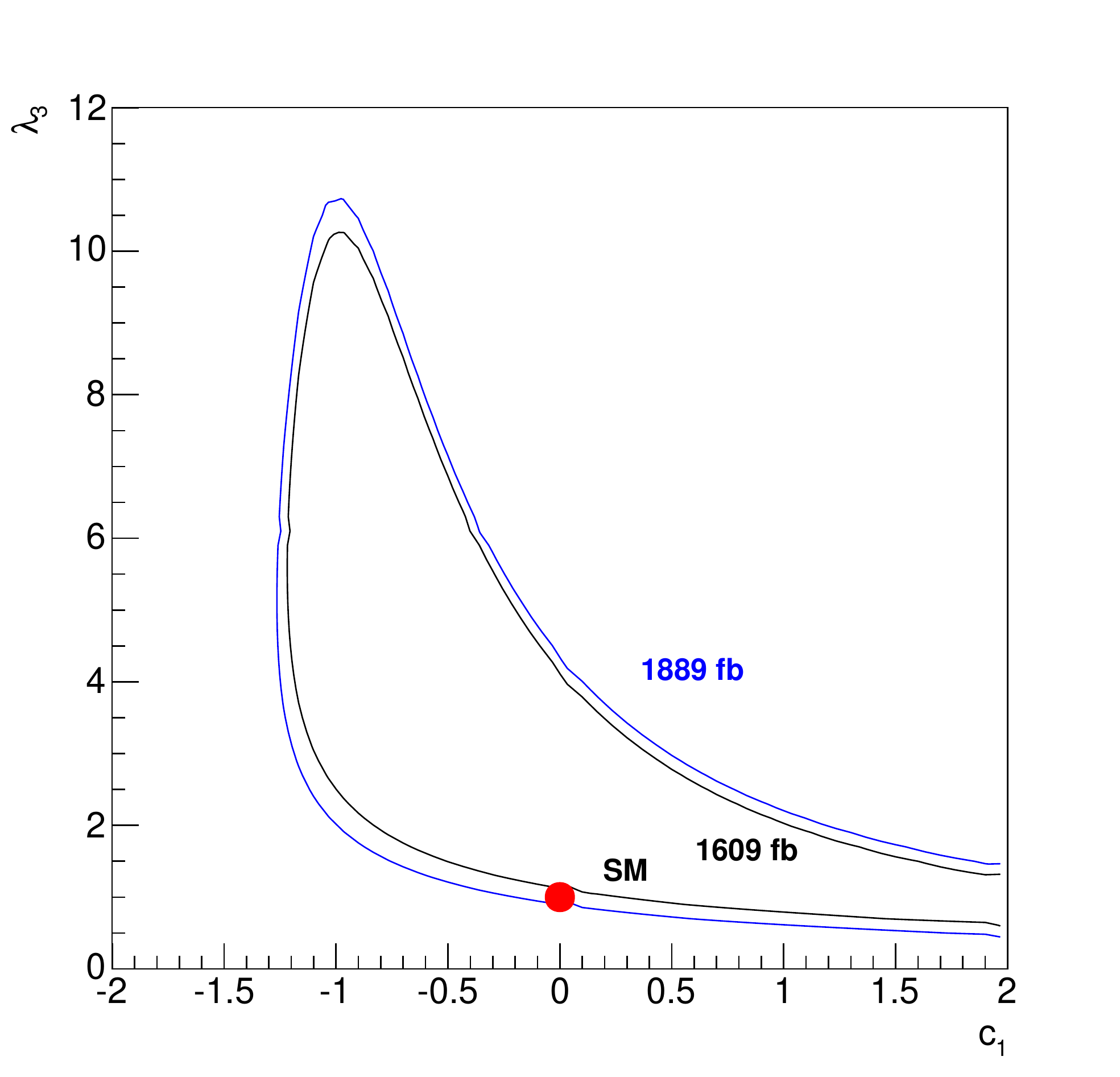}}
  \subfigure{
  \label{2hc2l3100}\thesubfigure
  \includegraphics[width=0.4\textwidth]{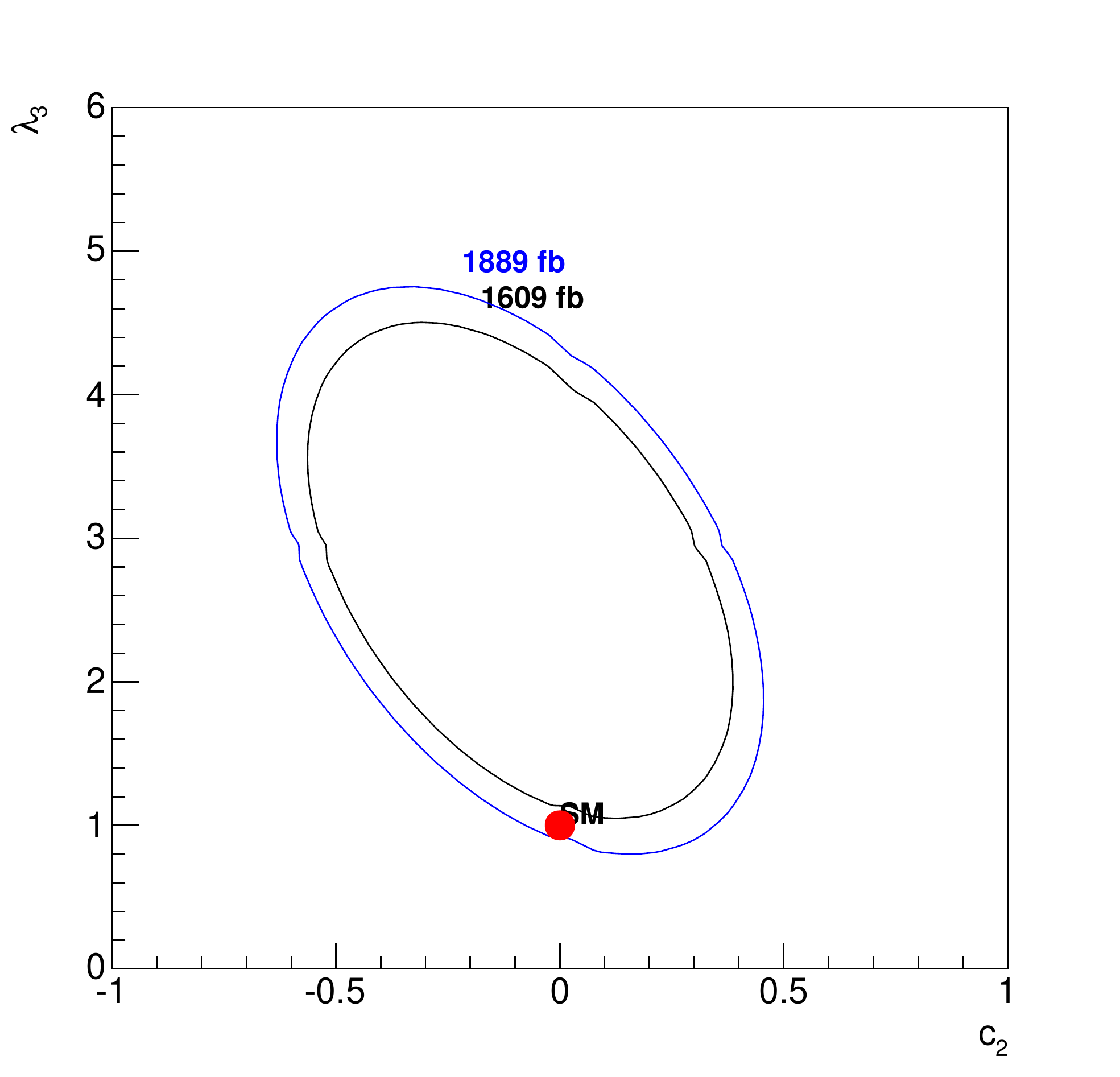}}
  \subfigure{
  \label{2hk5l3100}\thesubfigure
  \includegraphics[width=0.4\textwidth]{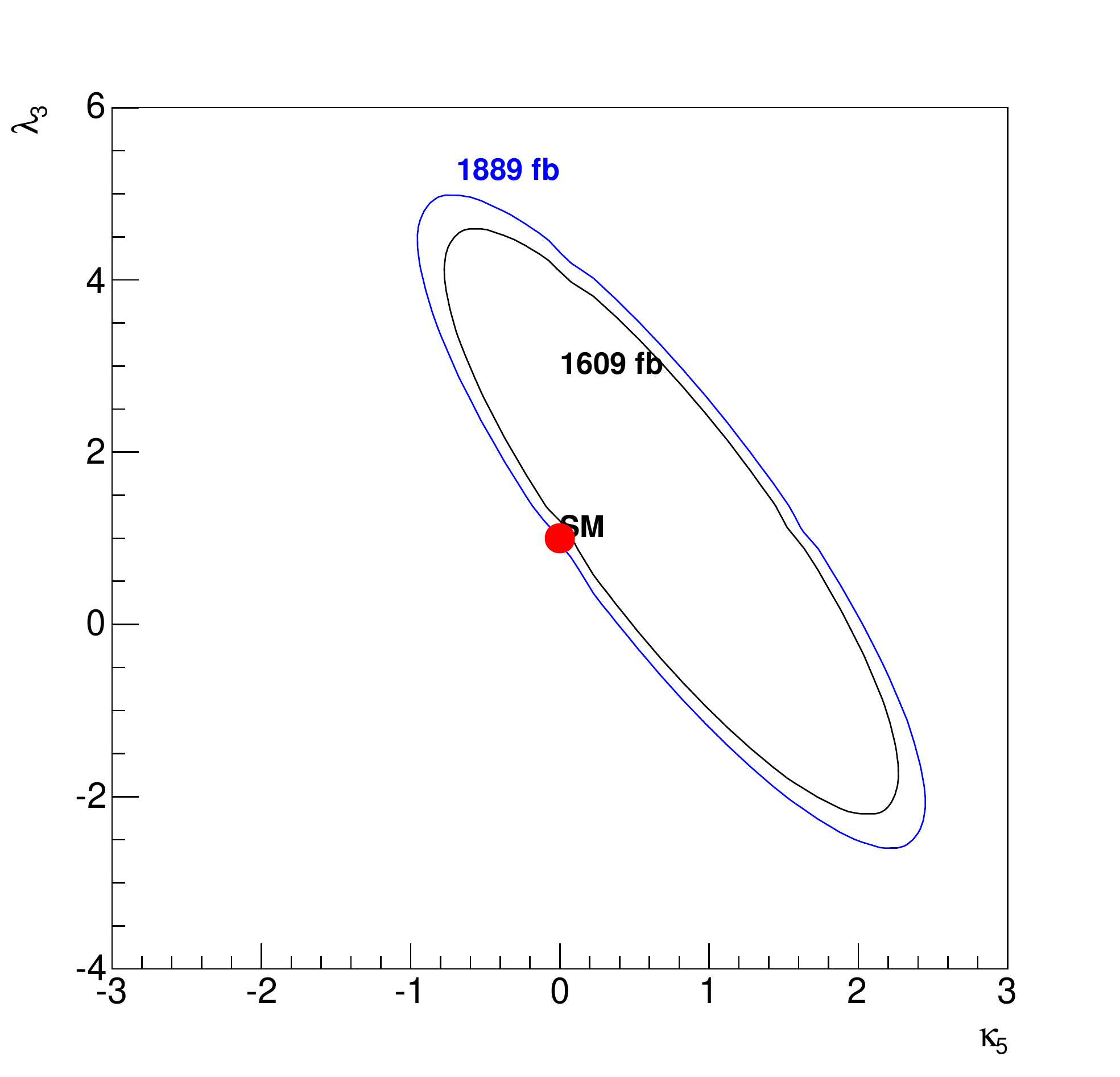}}
  \caption{Projected exclusion bounds in two-parameter planes between
    ($a_2$, $c_1$, $c_2$, $\kappa_5$) and $\lambda_3$, extracted from
    the process $g g \to h h$ a 100 TeV $pp$ collider. If the
    coefficient values are 
    equal to the SM prediction,  parameter values between the two
    contours are still allowed by the measurement.  The exclusion
    bounds correspond to a total error of $8\%$ on $\sigma(gg \to
    hh)$, theoretical and 
    experimental uncertainties combined. }\label{gg2h100bds} 
\end{figure}

In Fig.~\ref{gg2h100bds}, we draw the analogous bounds that we can
expect from analyzing $gg \to hh$ in $100$ TeV $pp$ data.  We assume
that the cross section of $gg \to hh$ can be measured to a precision
of~$8\%$. This $8\%$ uncertainty results from combining theoretical
and experimental uncertainties.  The allowed parameter regions shrink
considerably and become pinched between two contours, in each plot.
Comparing Fig.~\ref{gg2h14bds} and Fig.~\ref{gg2h100bds}, we conclude
that a 100 TeV collider can significantly improve the precision on
$a_2$, $c_2$, $\kappa_5$, and $\lambda_3$.

\begin{figure}[htbp]
  \centering
  \subfigure{
  \label{2ha2100}\thesubfigure
  \includegraphics[width=0.4\textwidth]{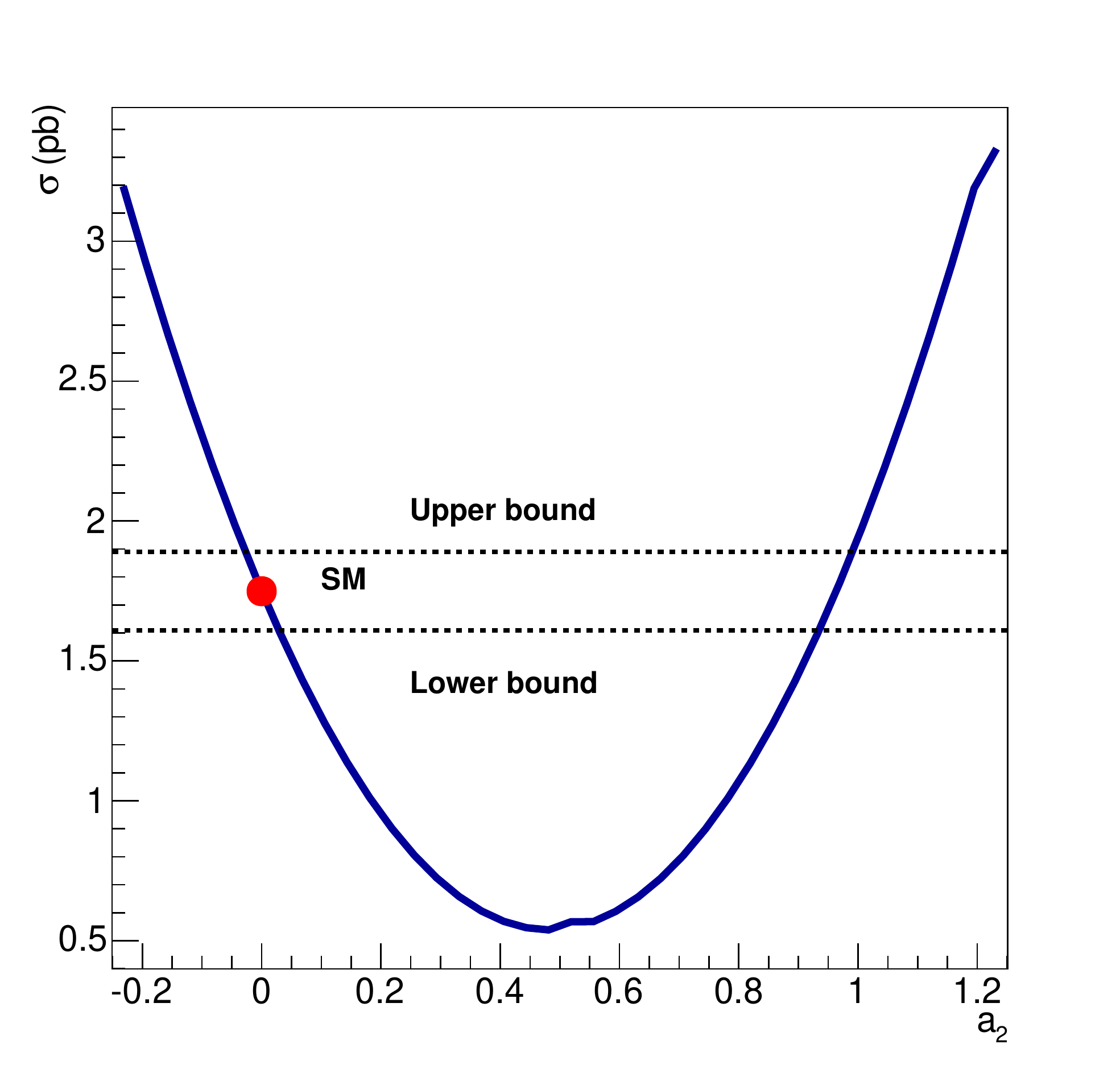}}
  \subfigure{
  \label{2hc2100}\thesubfigure
  \includegraphics[width=0.4\textwidth]{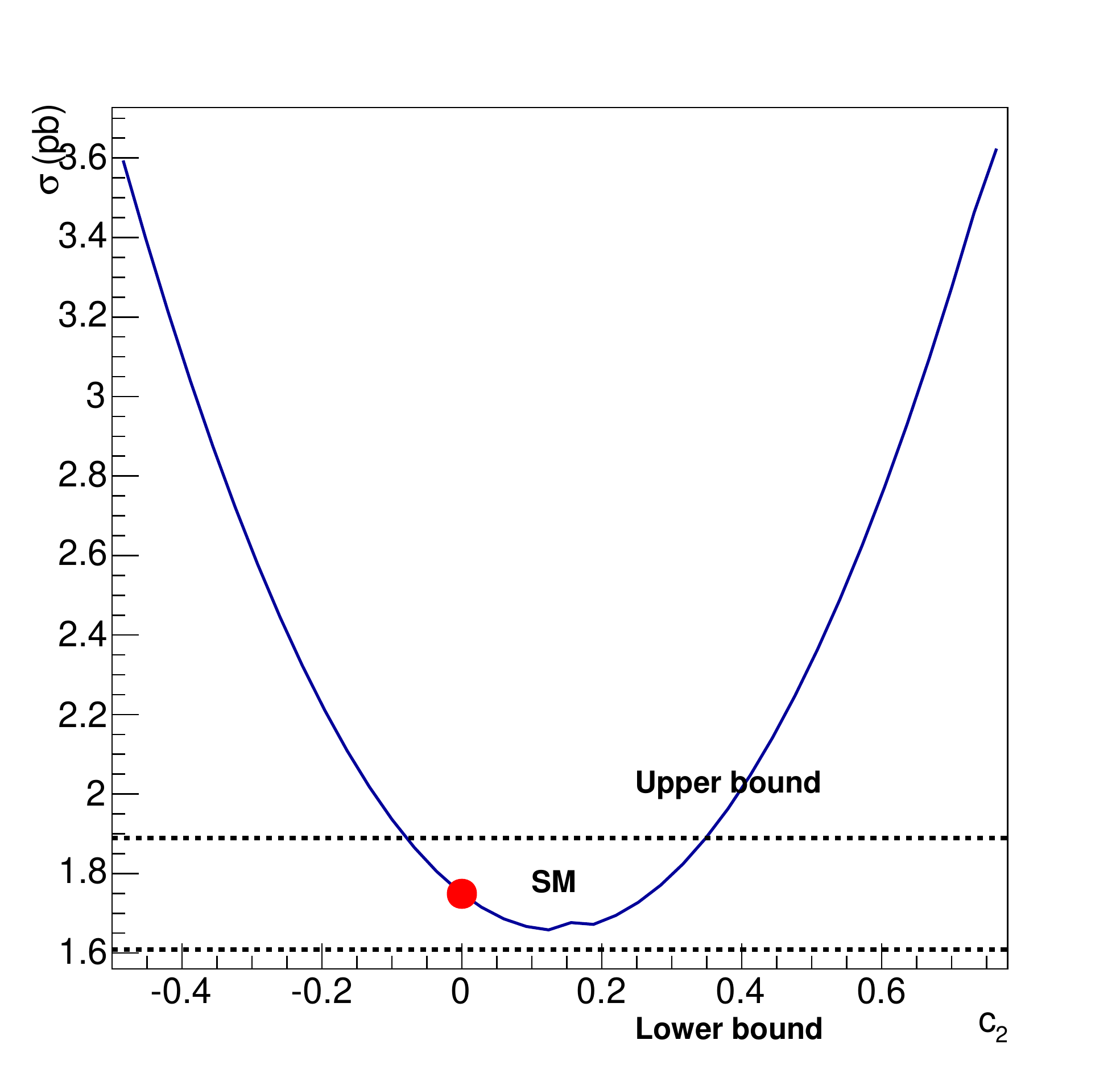}}
  \subfigure{
  \label{2hk5100}\thesubfigure
  \includegraphics[width=0.4\textwidth]{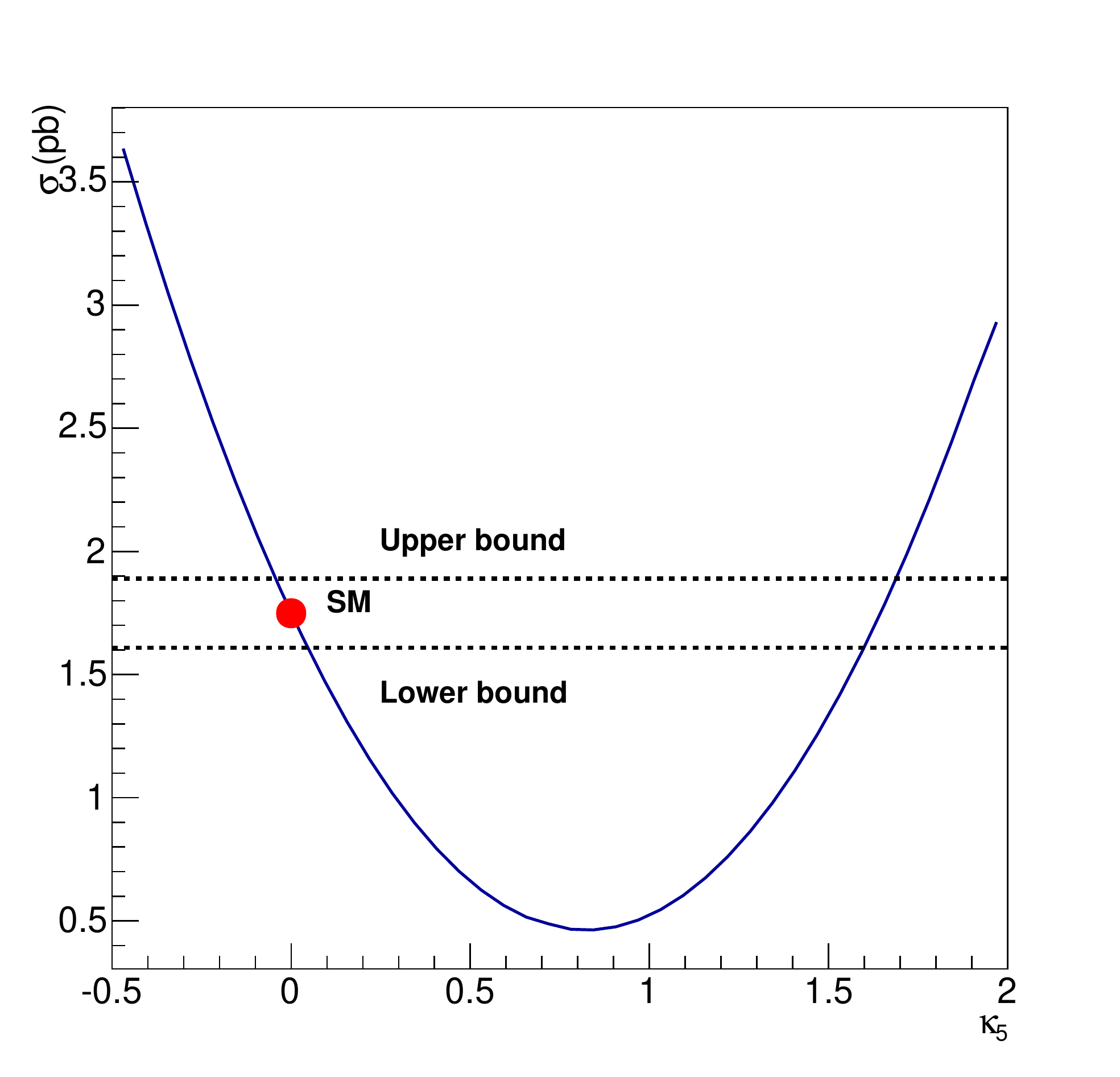}}
  \subfigure{
  \label{2hl3100}\thesubfigure
  \includegraphics[width=0.4\textwidth]{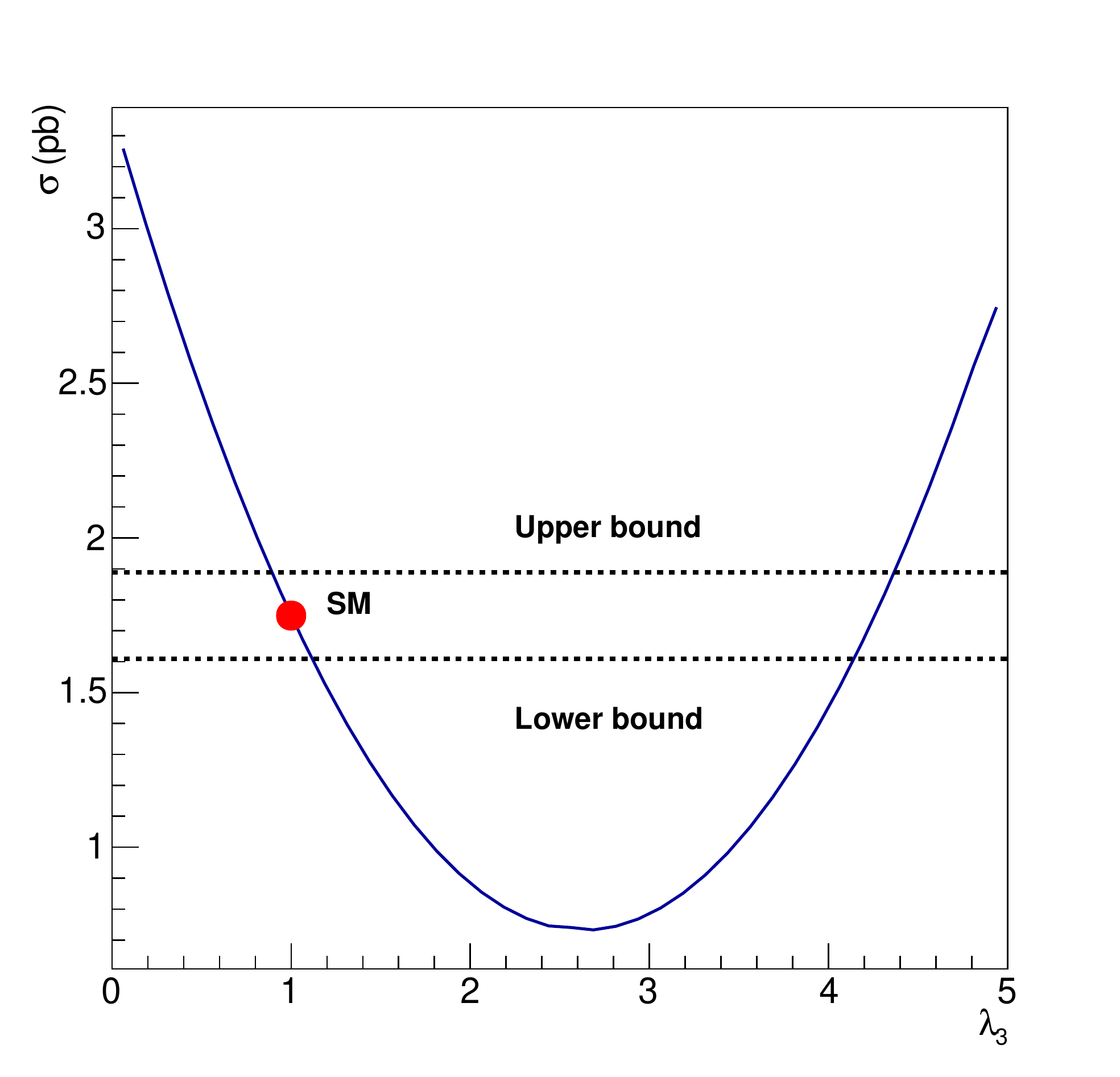}}
  \caption{Projected one-parameter exclusion bounds for $a_2$, $c_2$,
    $\kappa_5$ and $\lambda_3$, extracted from the process $g g \to h
    h$.  If the coefficient values are equal to the SM prediction,
    parameter values between the upper and lower bounds are allowed by the
    measurement.  The exclusion bounds correspond to a total error of
    $8\%$ on $\sigma(gg \to hh)$, theoretical and experimental
    uncertainties combined.}\label{gg2h1001dbds}
\end{figure}

We can also individually project out single-parameter bounds for each
of these four parameters, shown in Fig.~\ref{gg2h1001dbds}.  In this
approach, the parameters $a_2$, $\kappa_5$, and $\lambda_3$ can be
determined with a precision close to $10\%$. The two-fold ambiguities
in the solutions could possibly be removed by using the kinematics of
final states, as demonstrated in Ref.~\cite{Li:2015yia}.  The parameter
$c_2$ can be determined within the range $[-0.1,0.4]$.

\begin{figure}[htbp]
  \centering
  \subfigure{
  \label{3ha3l4100}\thesubfigure
  \includegraphics[width=0.4\textwidth]{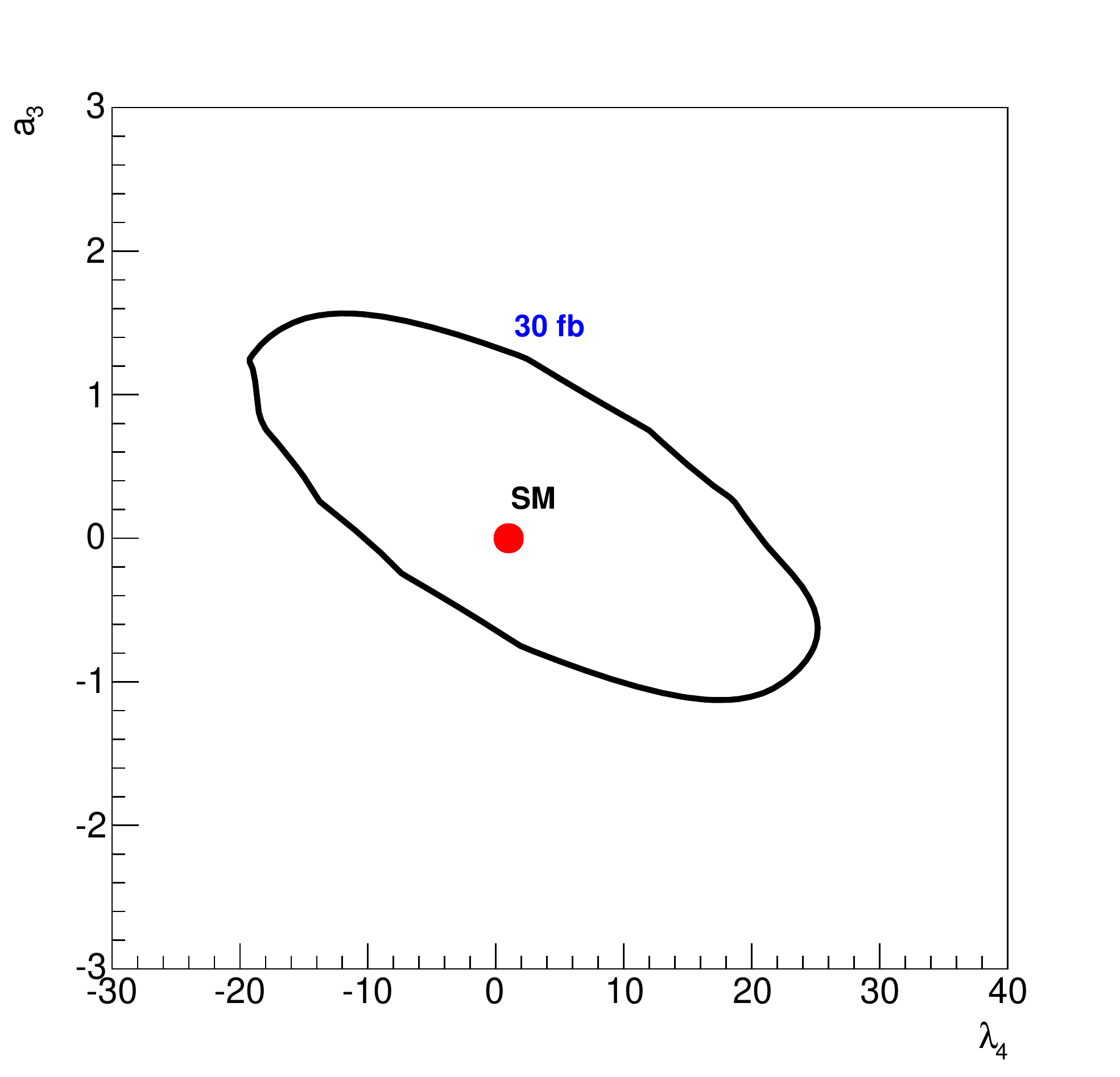}}
  \subfigure{
  \label{3h1da3100}\thesubfigure
  \includegraphics[width=0.4\textwidth]{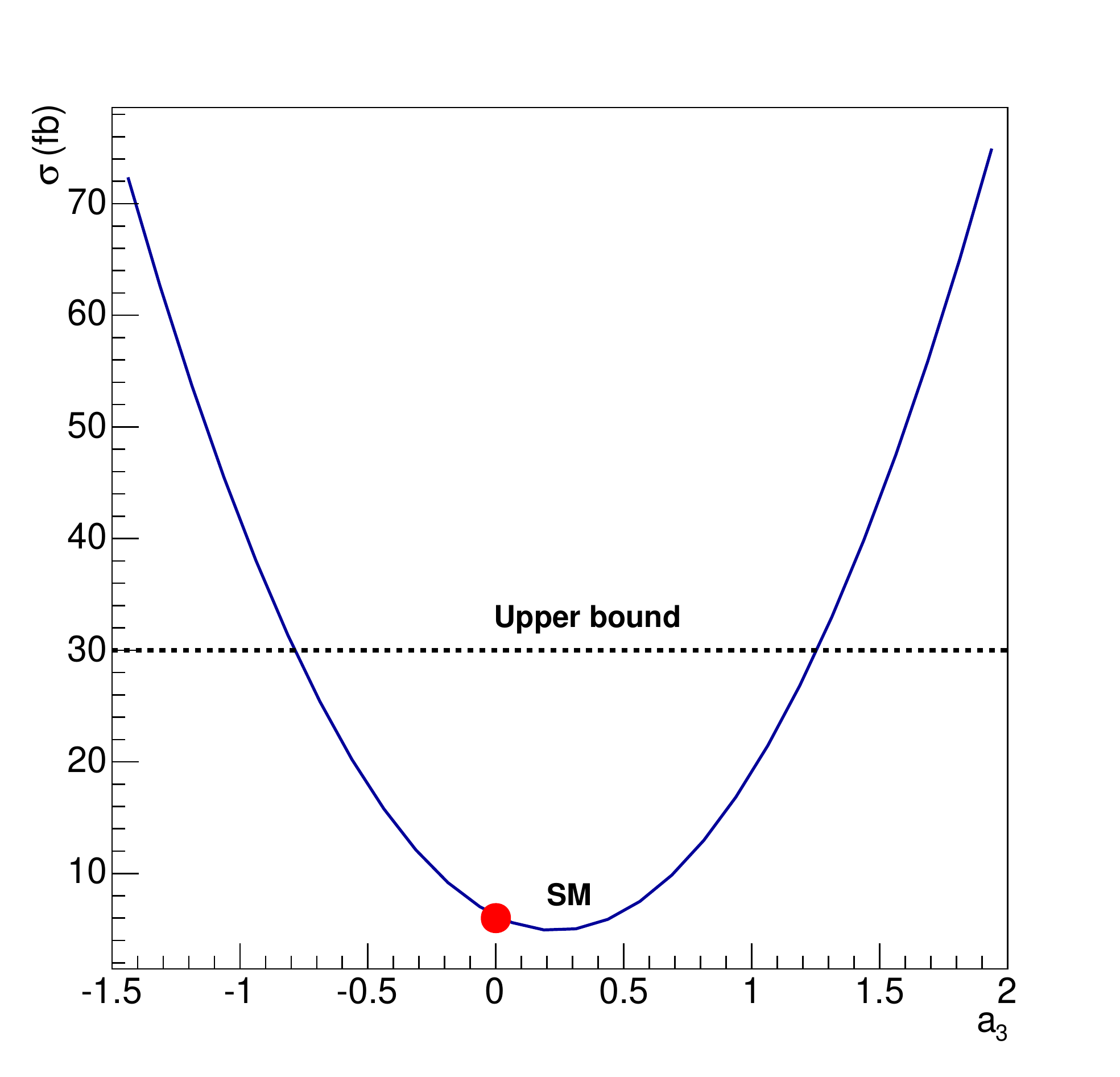}}
  \subfigure{
  \label{3hk6l4100}\thesubfigure
  \includegraphics[width=0.4\textwidth]{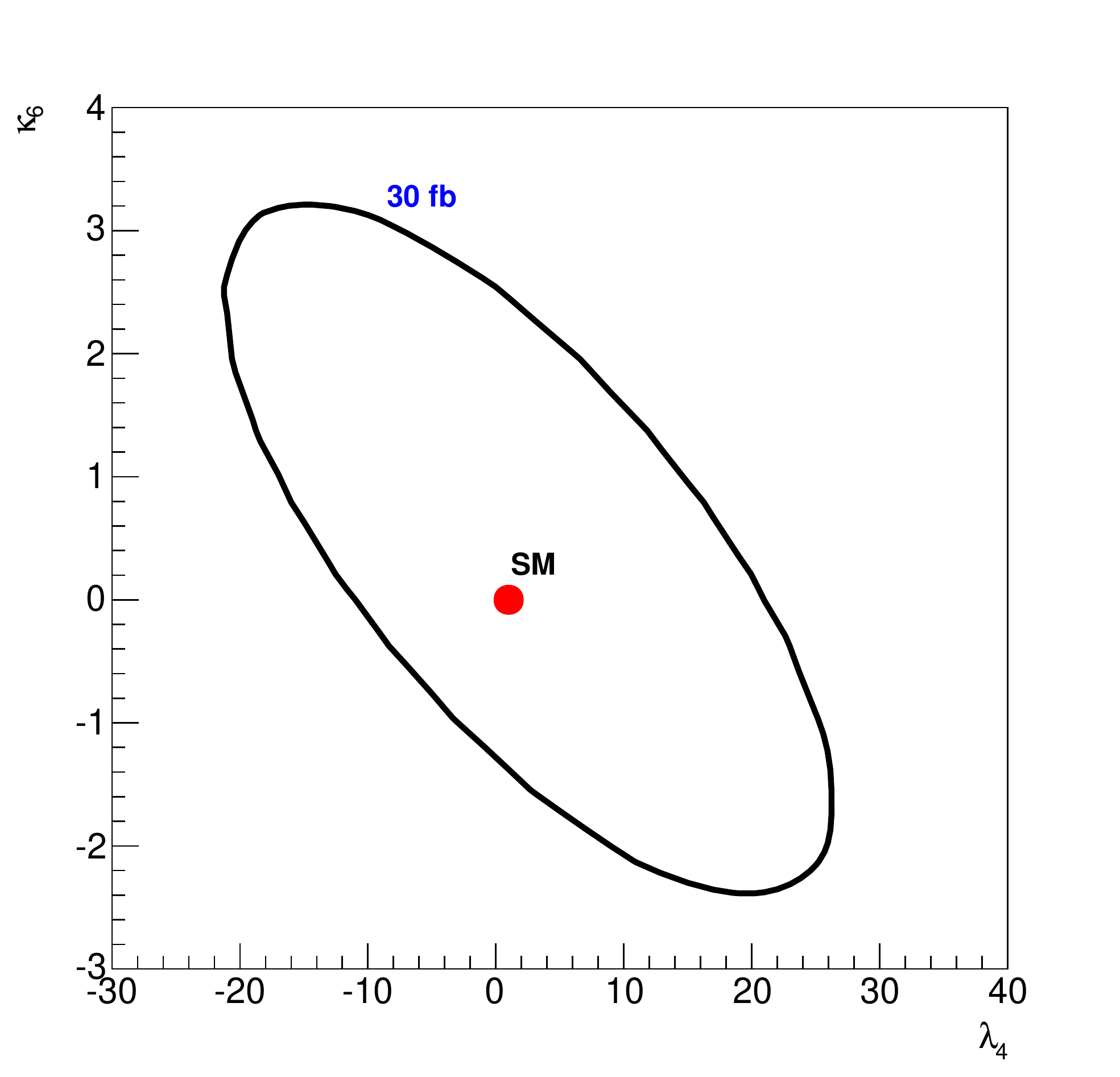}}
  \subfigure{
  \label{3h1dk6100}\thesubfigure
  \includegraphics[width=0.4\textwidth]{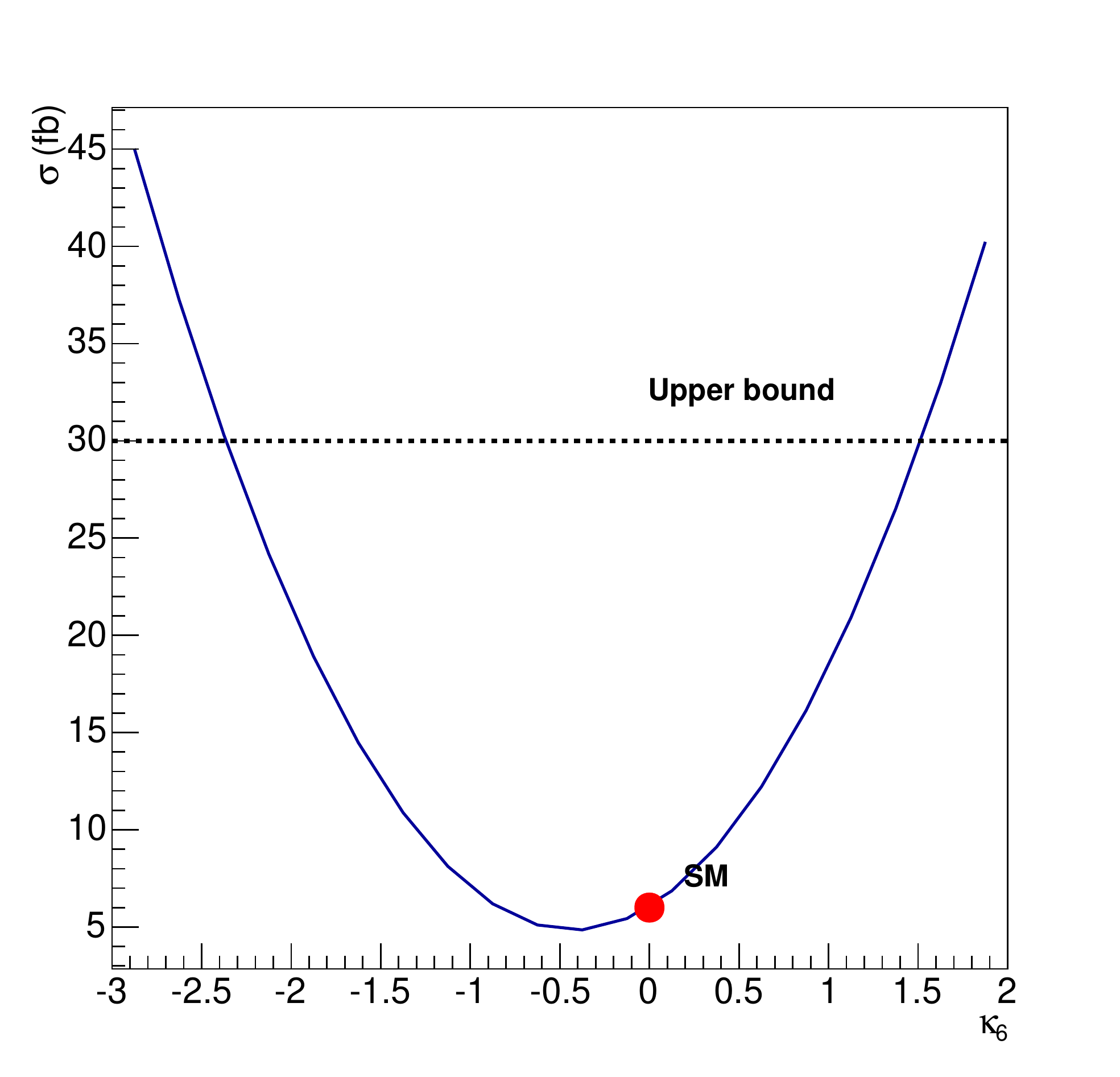}}
  \subfigure{
  \label{3ha3k6100}\thesubfigure
  \includegraphics[width=0.4\textwidth]{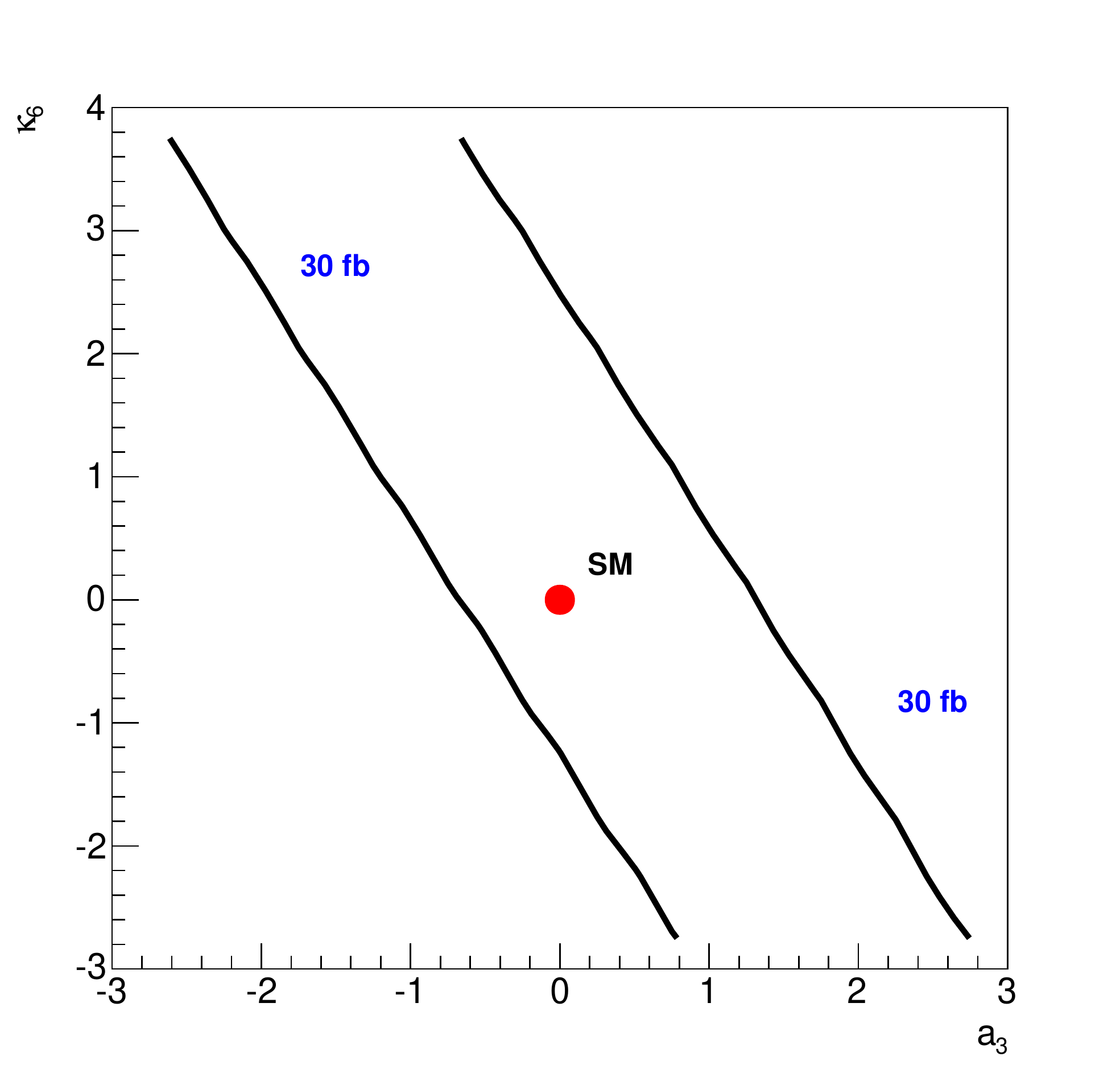}}
  \subfigure{
  \label{3h1dl4100}\thesubfigure
  \includegraphics[width=0.4\textwidth]{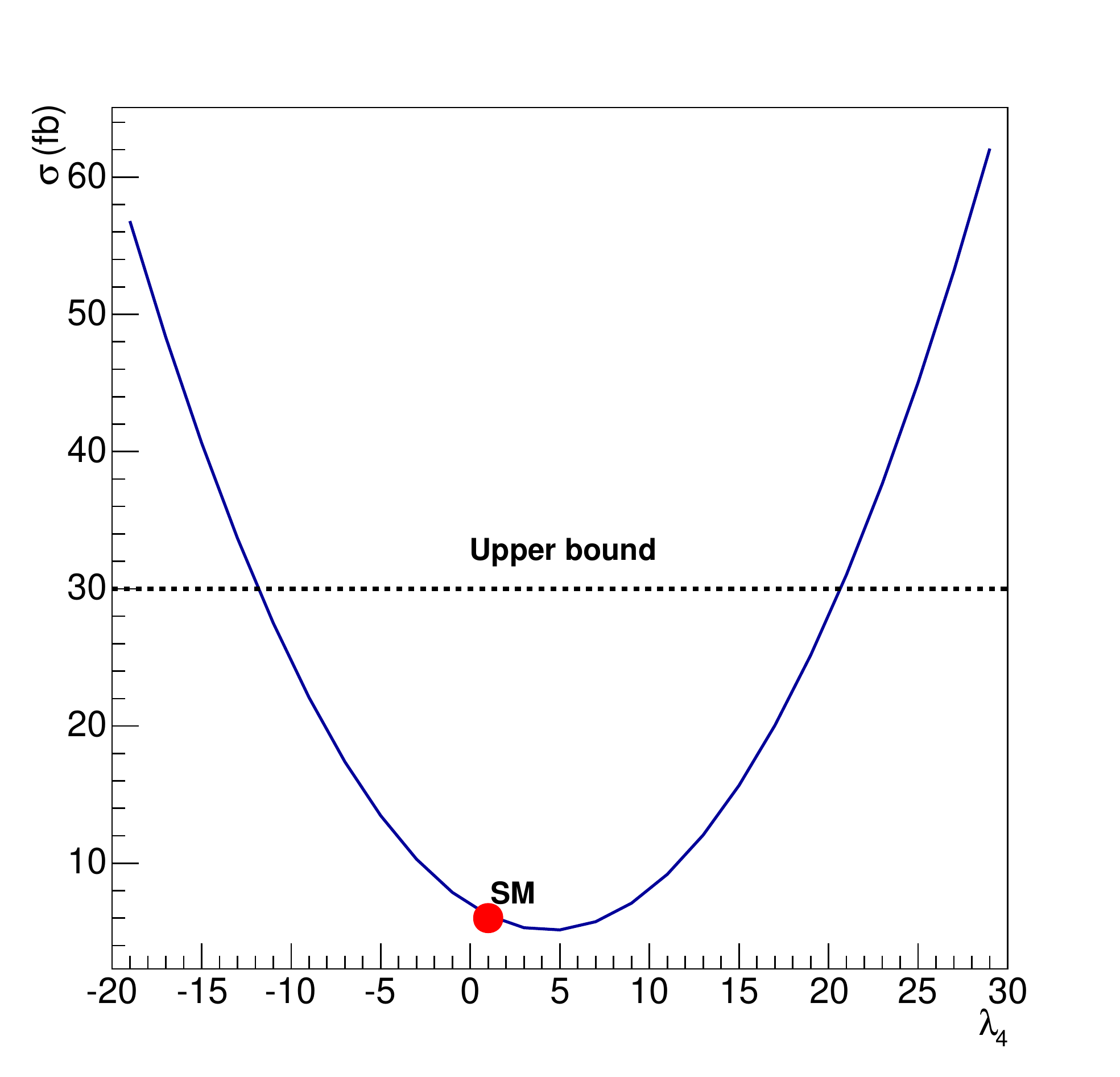}}
  \caption{Projected two-parameter (left) and one-parameter (right) exclusion
    bounds, extracted from the process $gg\to hhh$.  The left column
    shows the two-parameter planes $a_3$-$\lambda_4$,
    $\kappa_6$-$\lambda_4$, and $a_3$-$\kappa_6$, while the right
    column displays $a_3$, $\kappa_6$, and $\lambda_4$.}\label{gg3h1d4pbds}
\end{figure}

Finally, we add the exclusion bounds that we can expect from the
process $gg \to hhh$.  There are three parameters, namely $a_3$,
$\kappa_6$ and $\lambda_4$, which only contribute to the process $gg
\to hhh$.  We present two-dimensional bounds for all pairs of these
three independent four-point couplings in
Figs.~\ref{3ha3l4100}--\ref{3hk6l4100}.  The corresponding
one-dimensional bounds are given in
Figs.~\ref{3h1da3100}--\ref{3h1dk6100}. The parameter $a_3$ can be
constrained to the range [-0.8,1.2]; the parameter $\kappa_6$ can be
constrained to the range [-2.3, 1.5].  As shown
already in Ref.~\cite{Chen:2015gva}, the parameter $\lambda_4$ can only be
determined within a quite wide range [-13,20].  Clearly,
$\lambda_4$ is the most difficult parameter to measure in this
framework.

\begin{figure}[htbp]
  \centering
  \subfigure{
  \label{3ha2100}\thesubfigure
  \includegraphics[width=0.4\textwidth]{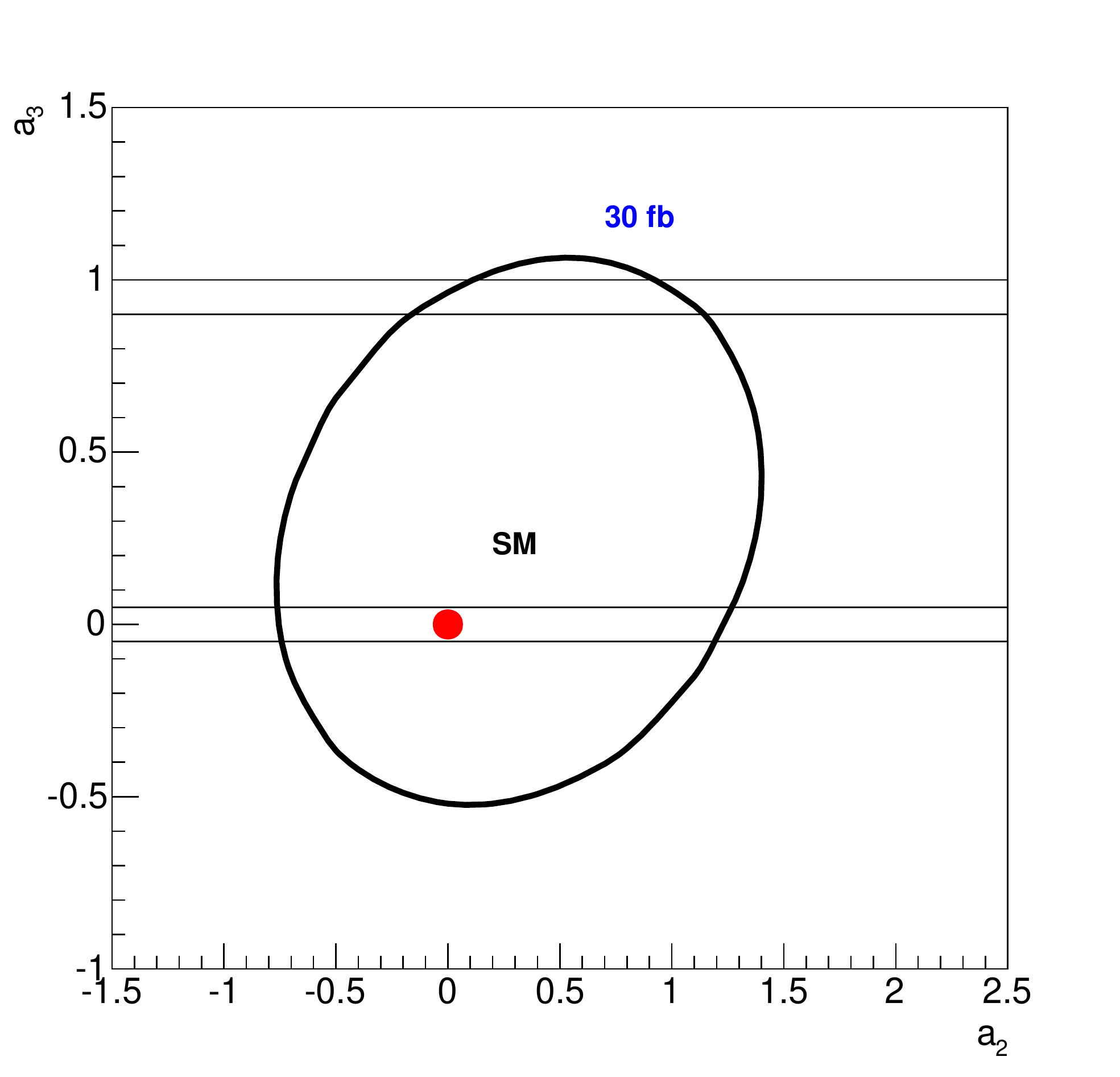}}
  \subfigure{
  \label{3hc2100}\thesubfigure
  \includegraphics[width=0.4\textwidth]{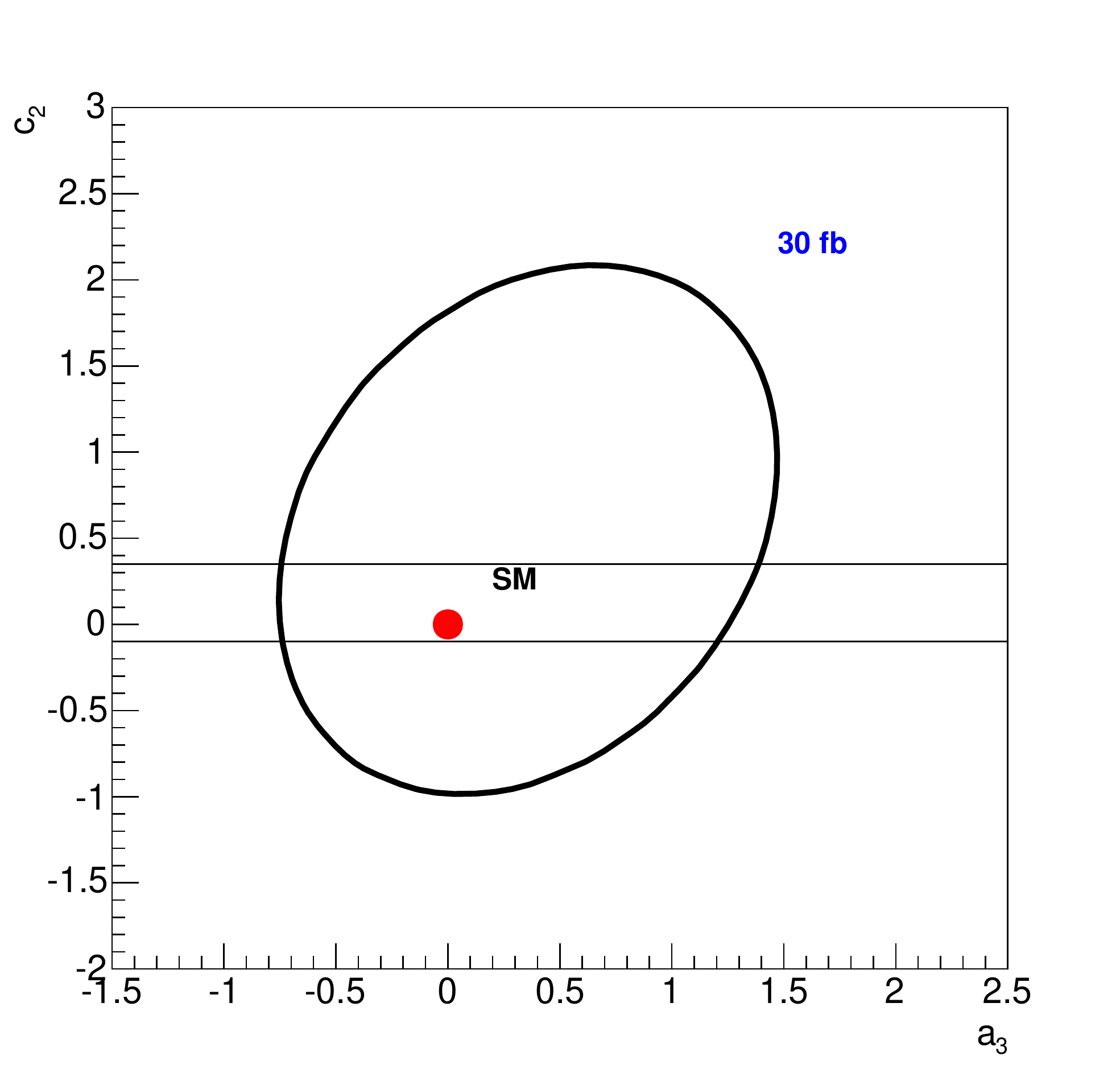}}
  \subfigure{
  \label{3hk5100}\thesubfigure
  \includegraphics[width=0.4\textwidth]{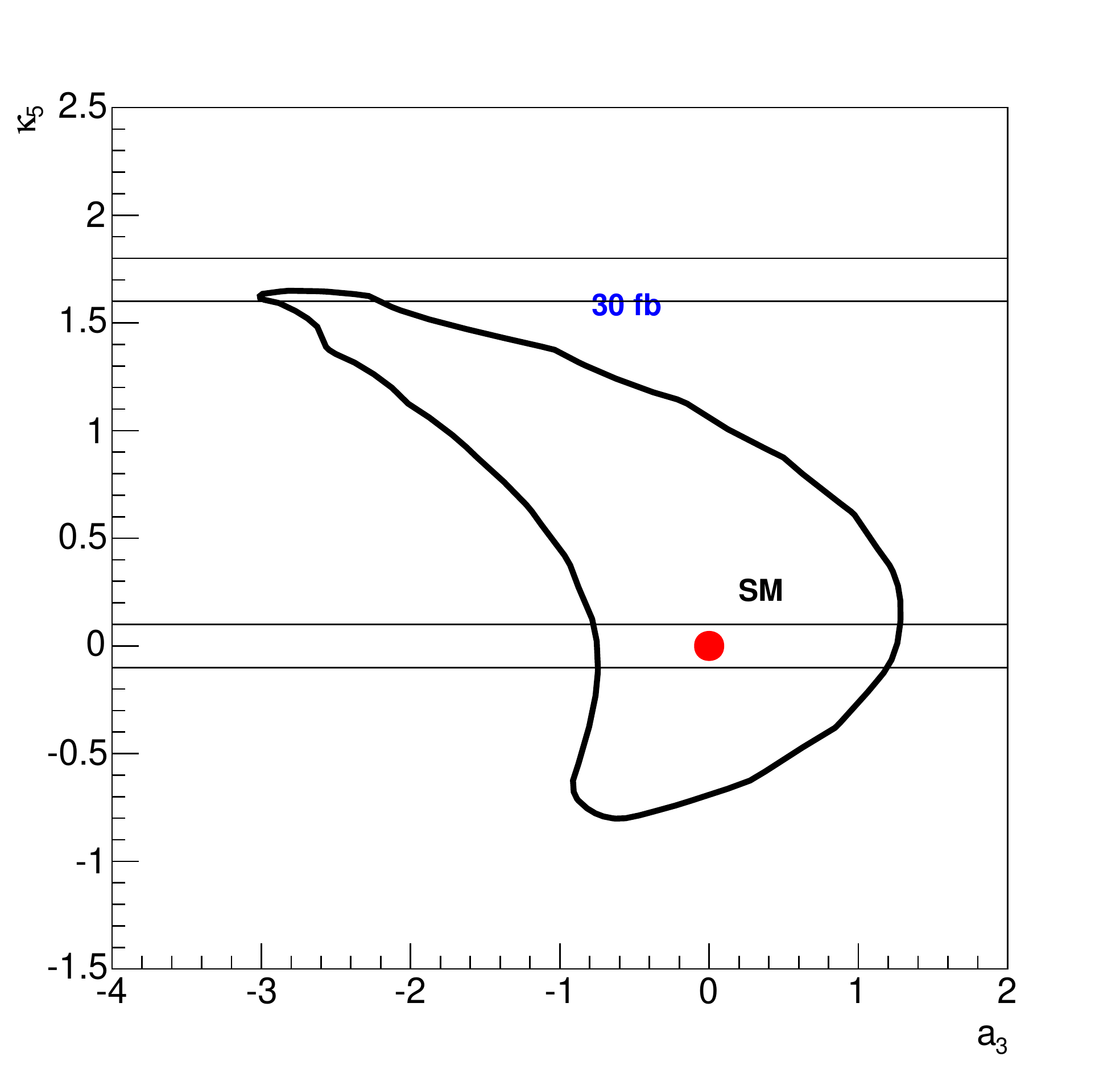}}
  \subfigure{
  \label{3hl3100}\thesubfigure
  \includegraphics[width=0.4\textwidth]{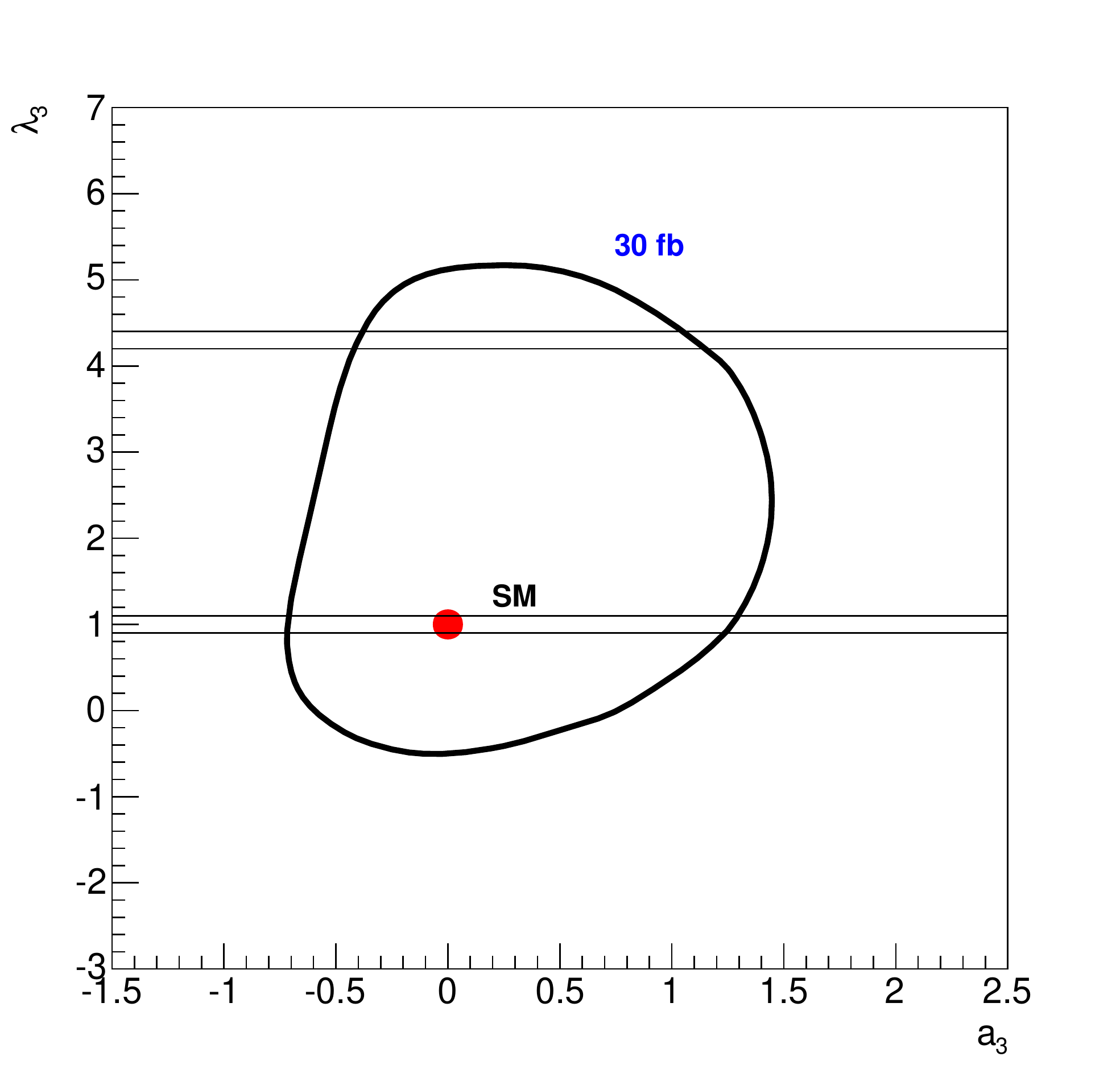}}
  \caption{Projected two-parameter exclusion
    bounds, extracted from the process $gg\to hhh$.  The straight-line
    pairs indicate the exclusion bounds extracted from $gg\to hh$.
    The plots show two-parameter correlations between $a_3$ and $a_2$,
    $c_2$, $\kappa_5$, and $\lambda_3$.}\label{gg3h100a3bds} 
\end{figure}

These results have to be combined with the parameter exclusion regions
derived from~$gg\to h h$.  In Fig.~\ref{gg3h100a3bds}, we show the
correlations between $a_3$ and $a_2$, $c_2$, $\kappa_5$, and
$\lambda_3$.  In these plots, we overlay the limits that follow from
Higgs-pair production, presented in Fig.~\ref{gg2h1001dbds}, to
the exclusion contours that follow from triple-Higgs production.  The
SM prediction is displayed for reference.  Clearly, the triple-Higgs
results yield weaker constraints, but they are nevertheless sensitive
to a different combination of parameters and thus cut off part of the
two-parameter exclusion regions.  As an example, we note that adding
in $gg \to hhh$ can help in resolving a two-fold ambiguity in the
$\kappa_5$-$a_3$ plane, cf.\ Fig.~\ref{3hk5100}.

\begin{figure}[htbp]
  \centering
  \subfigure{
  \label{3ha2k6100}\thesubfigure
  \includegraphics[width=0.4\textwidth]{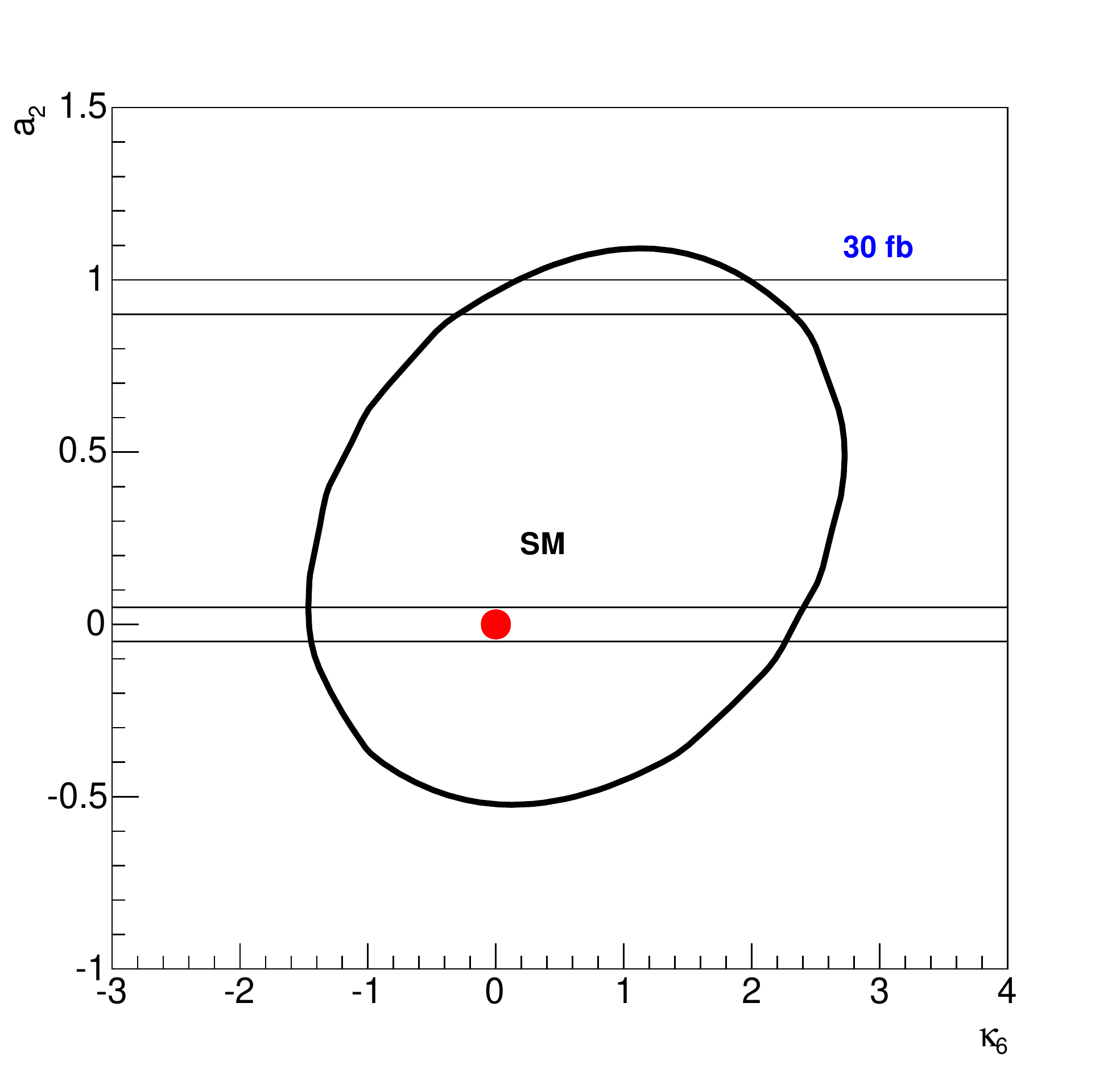}}
  \subfigure{
  \label{3hc2k6100}\thesubfigure
  \includegraphics[width=0.4\textwidth]{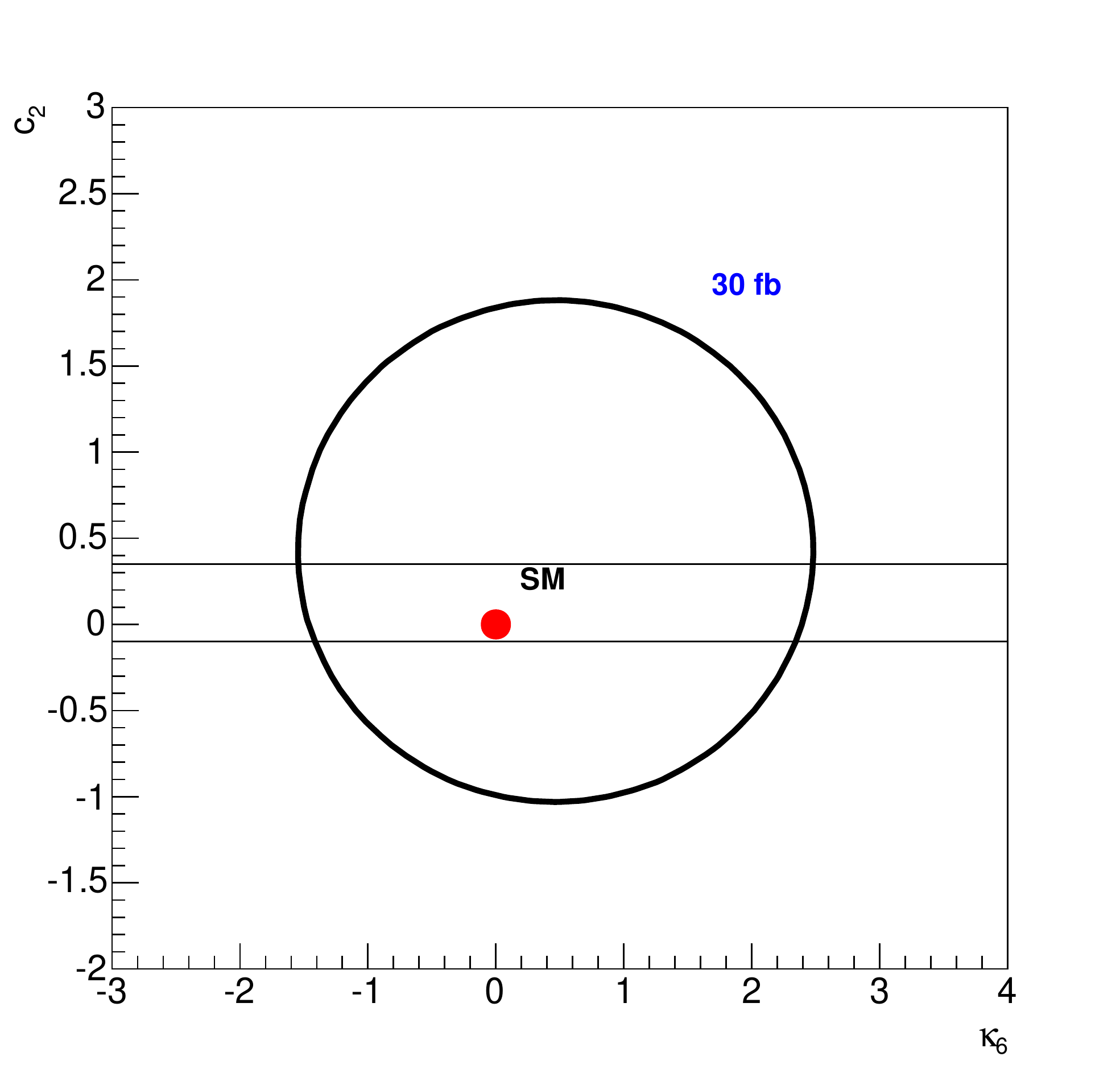}}
  \subfigure{
  \label{3hk5k6100}\thesubfigure
  \includegraphics[width=0.4\textwidth]{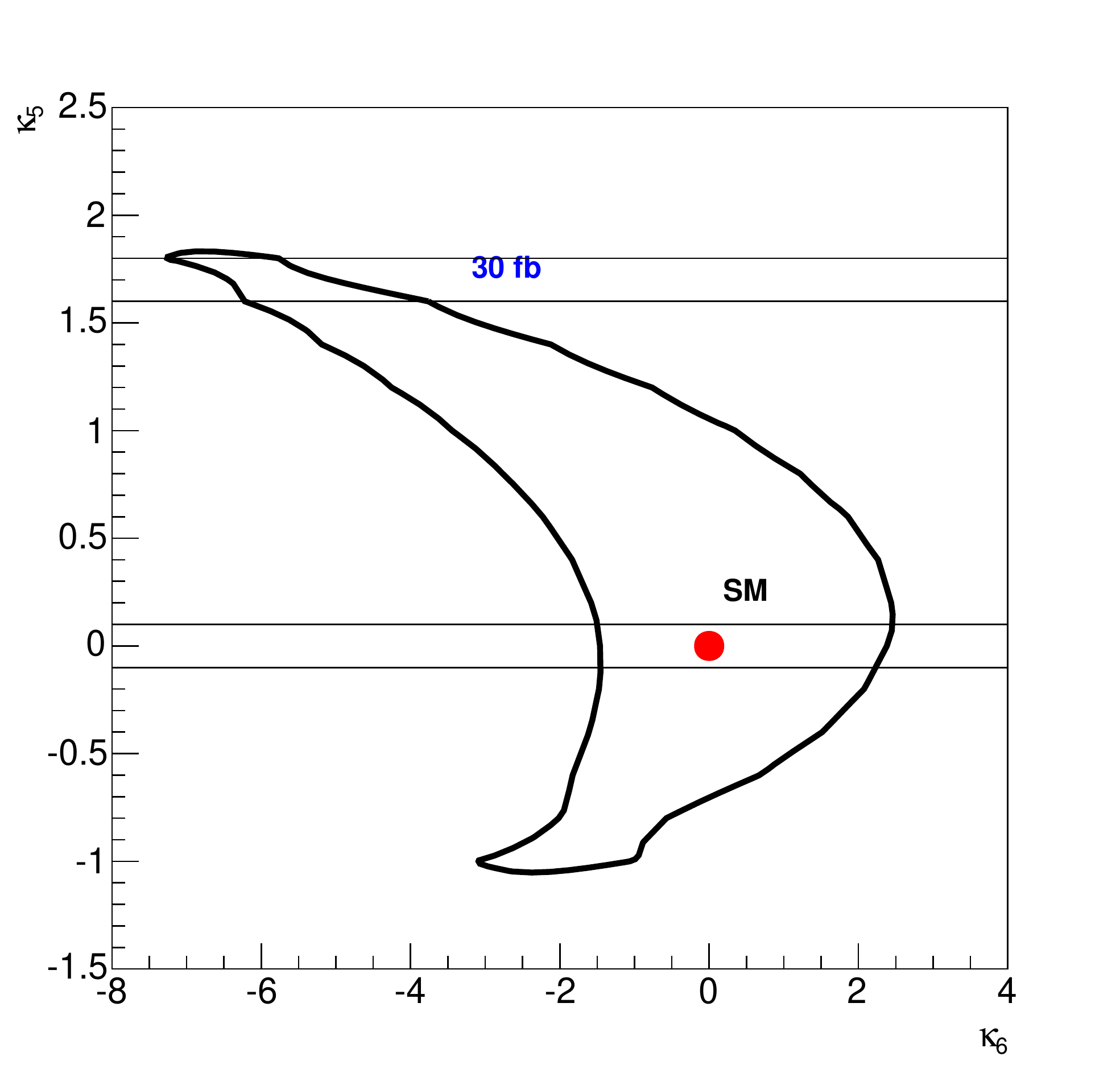}}
  \subfigure{
  \label{3hl3k6100}\thesubfigure
  \includegraphics[width=0.4\textwidth]{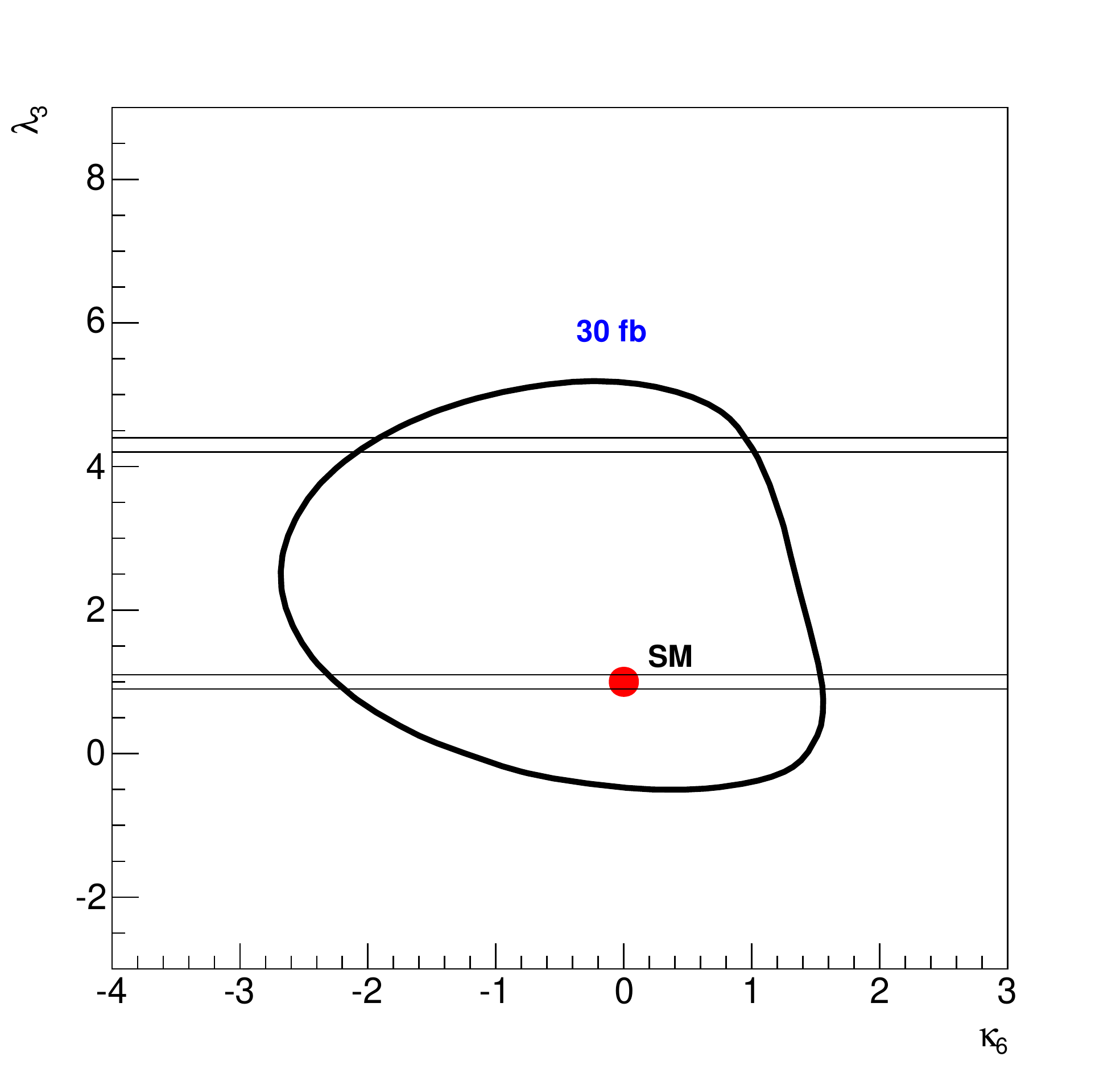}}
  \caption{Projected two-parameter exclusion bounds, extracted from
    the process $gg\to hhh$.  The straight-line pairs indicate the
    exclusion bounds extracted from $gg\to hh$.  The plots show
    two-parameter correlations between $\kappa_6$ and $a_2$, $c_2$,
    $\kappa_5$, and $\lambda_3$.}\label{gg3h100k6bds}
\end{figure}

\begin{figure}[htbp]
  \centering
  \subfigure{
  \label{3ha2l4100}\thesubfigure
  \includegraphics[width=0.4\textwidth]{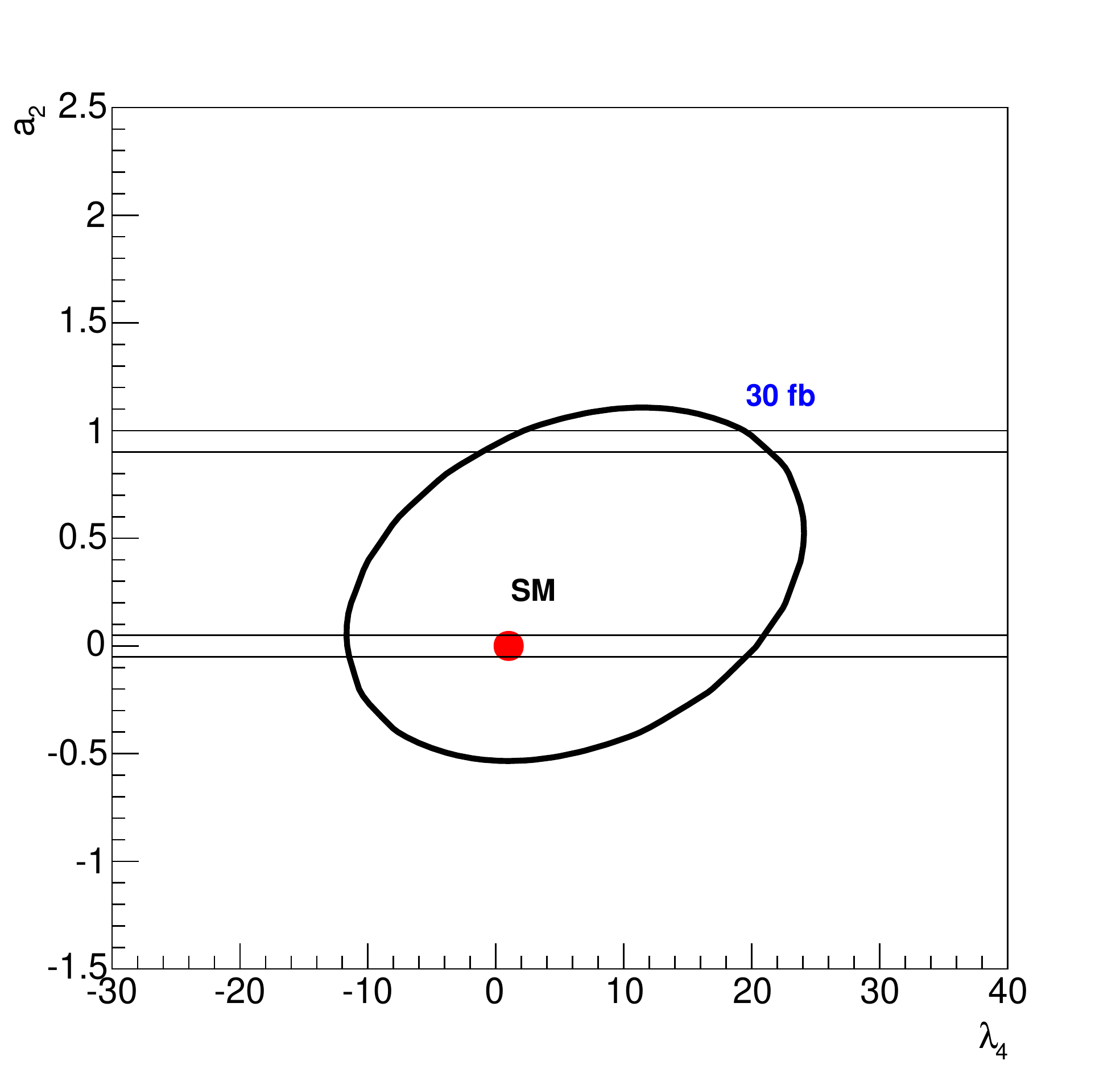}}
  \subfigure{
  \label{3hc2l4100}\thesubfigure
  \includegraphics[width=0.4\textwidth]{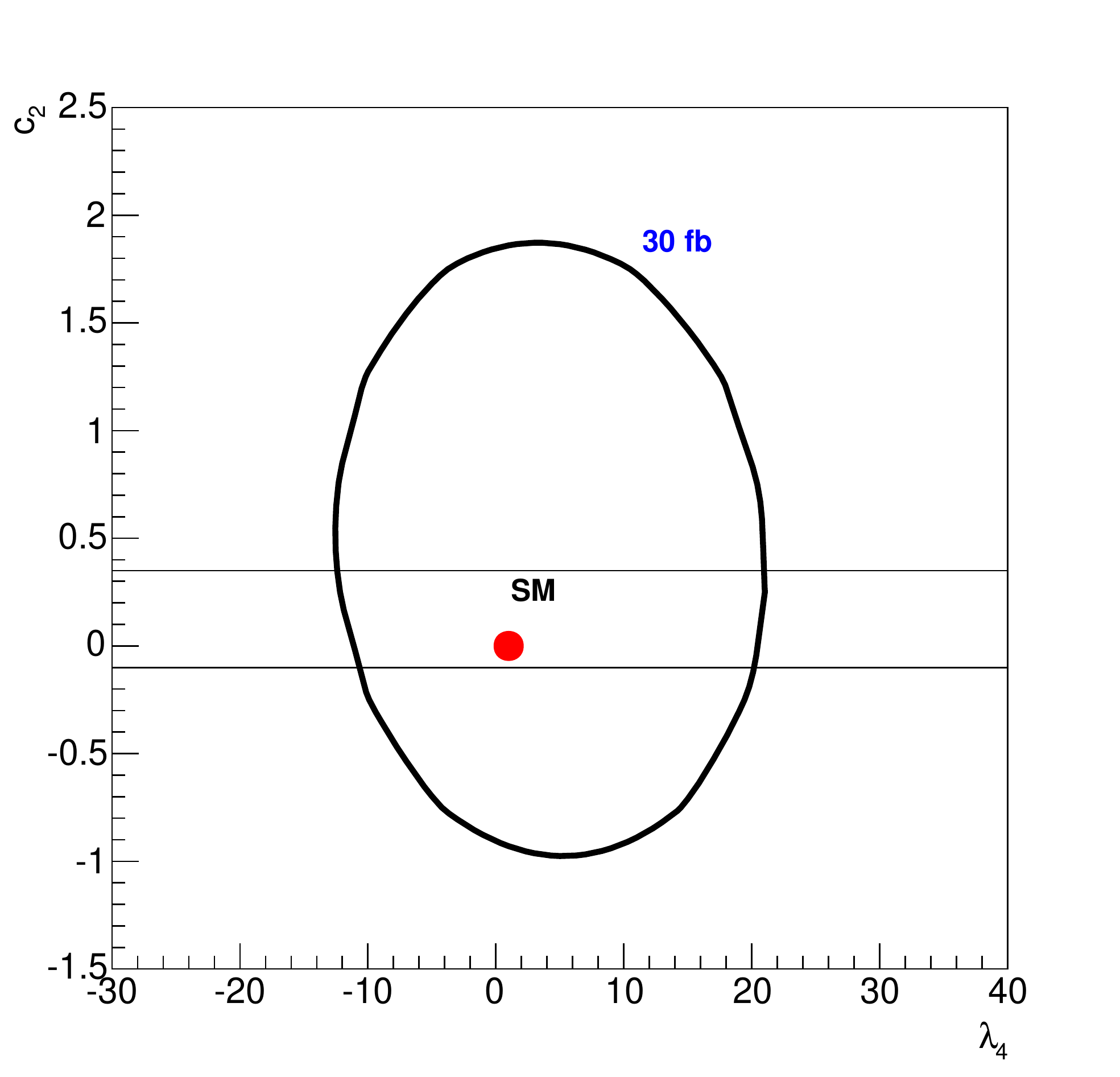}}
  \subfigure{
  \label{3hk5l4100}\thesubfigure
  \includegraphics[width=0.4\textwidth]{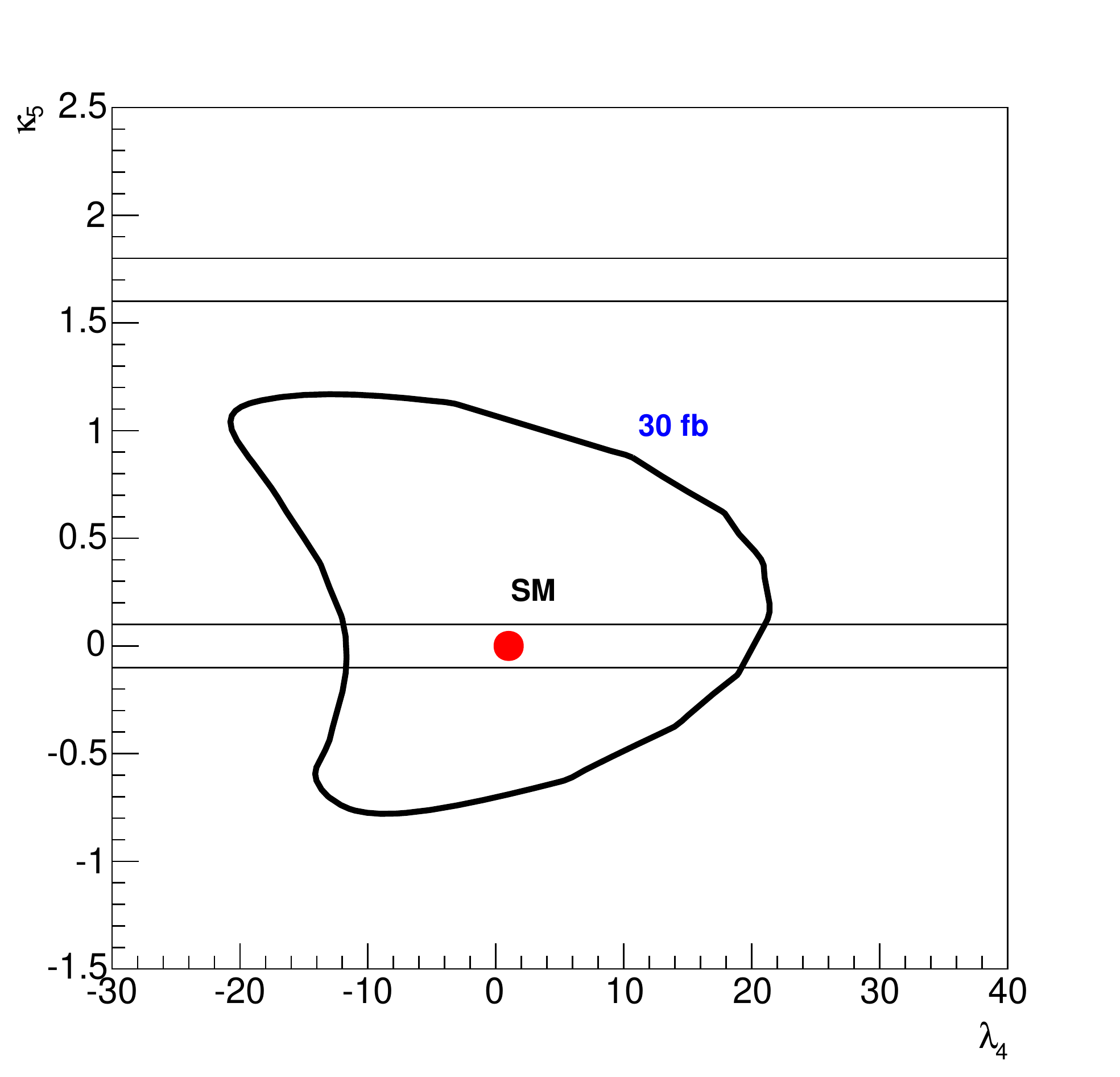}}
  \subfigure{
  \label{3hl3l4100}\thesubfigure
  \includegraphics[width=0.4\textwidth]{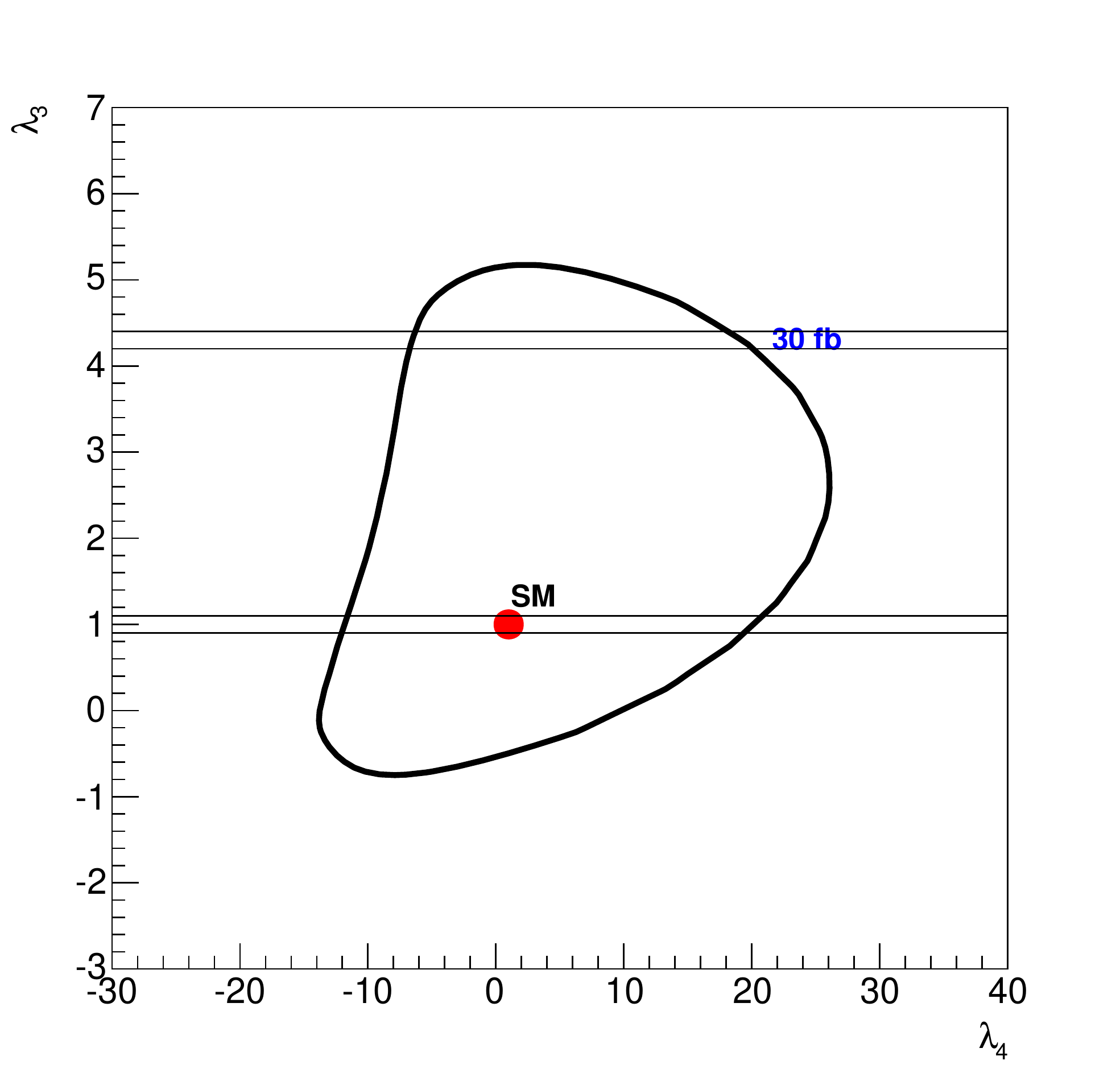}}
  \caption{Projected two-parameter exclusion bounds, extracted from
    the process $gg\to hhh$.  The straight-line pairs indicate the
    exclusion bounds extracted from $gg\to hh$.  The plots show
    two-parameter correlations between $\lambda_4$ and $a_2$, $c_2$,
    $\kappa_5$, and $\lambda_3$.}\label{gg3h100l4bds}
\end{figure}

Similarly, in Fig.~\ref{gg3h100k6bds}, we show the correlations
between $\kappa_6$ and $a_2$, $c_2$, $\kappa_5$, and $\lambda_3$. In
Fig.~\ref{gg3h100l4bds}, we show the correlations between
$\lambda_4$ and $a_2$, $c_2$, $\kappa_5$, and $\lambda_3$.  In
Fig.~\ref{3hk5l4100}, the $\kappa_5$-$\lambda_4$ plane, including
triple-Higgs production it becomes possible to separate the
$\kappa_5=0$ and $\kappa_5\neq 0$ regions.

\subsection{Analysis for models}

Finally, we can adapt the above results to more specific scenarios, as
we have introduced above in Sec.~\ref{Sec:EFT}.  In any given model,
the parameters of the unitarity-gauge Lagrangian~(\ref{eft}) can be
related to the original model parameters.  In particular, for a model
with a small set of independent parameters, we can recast the analysis
to a concrete prediction for the expected sensitivity to this model,
in a straightforward way.

\subsubsection{Strongly-interacting Higgs models}

\begin{figure}[htbp]
  \centering
  \subfigure{
  \label{mchm4}\thesubfigure
  \includegraphics[width=0.4\textwidth]{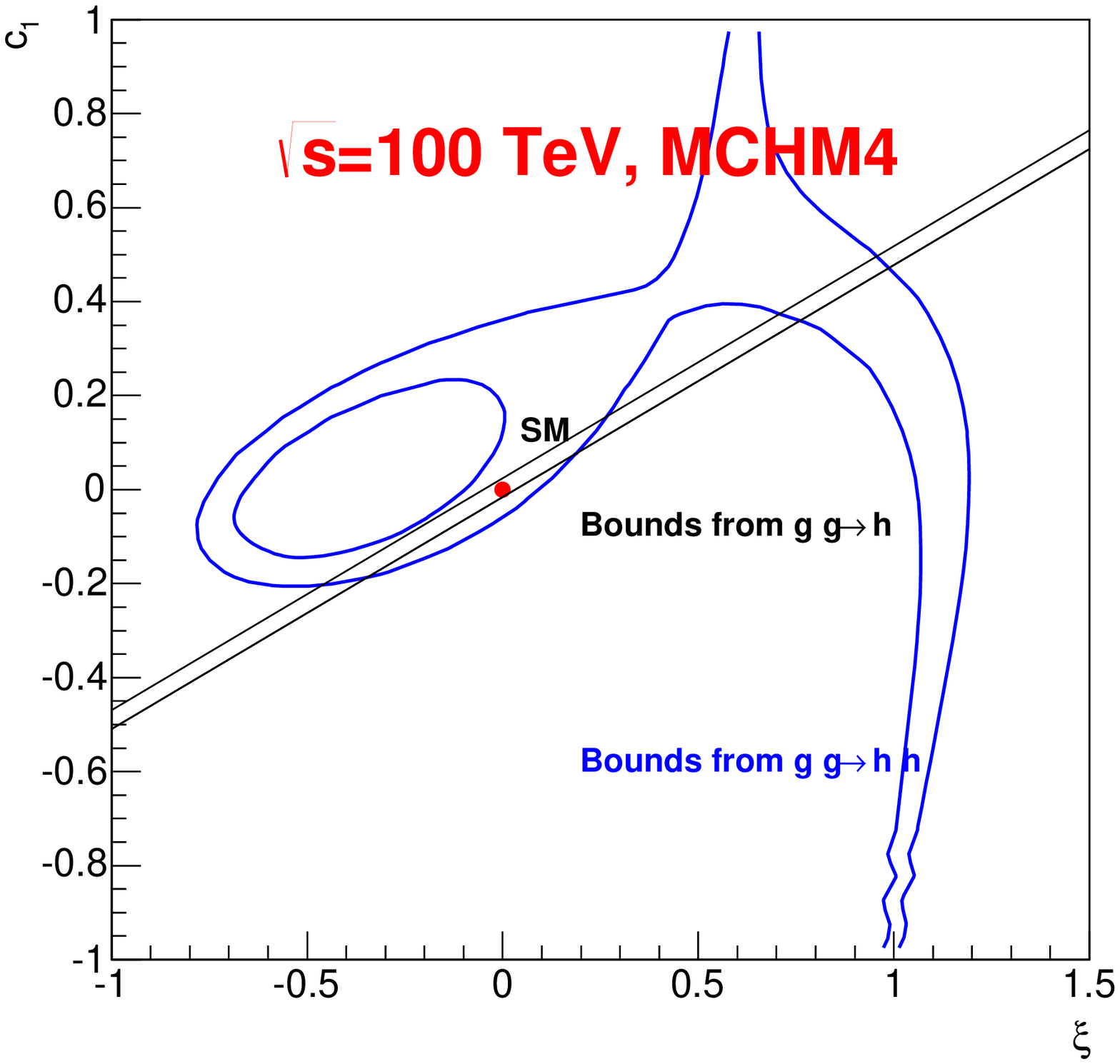}}
  \subfigure{
  \label{mchm5}\thesubfigure
  \includegraphics[width=0.4\textwidth]{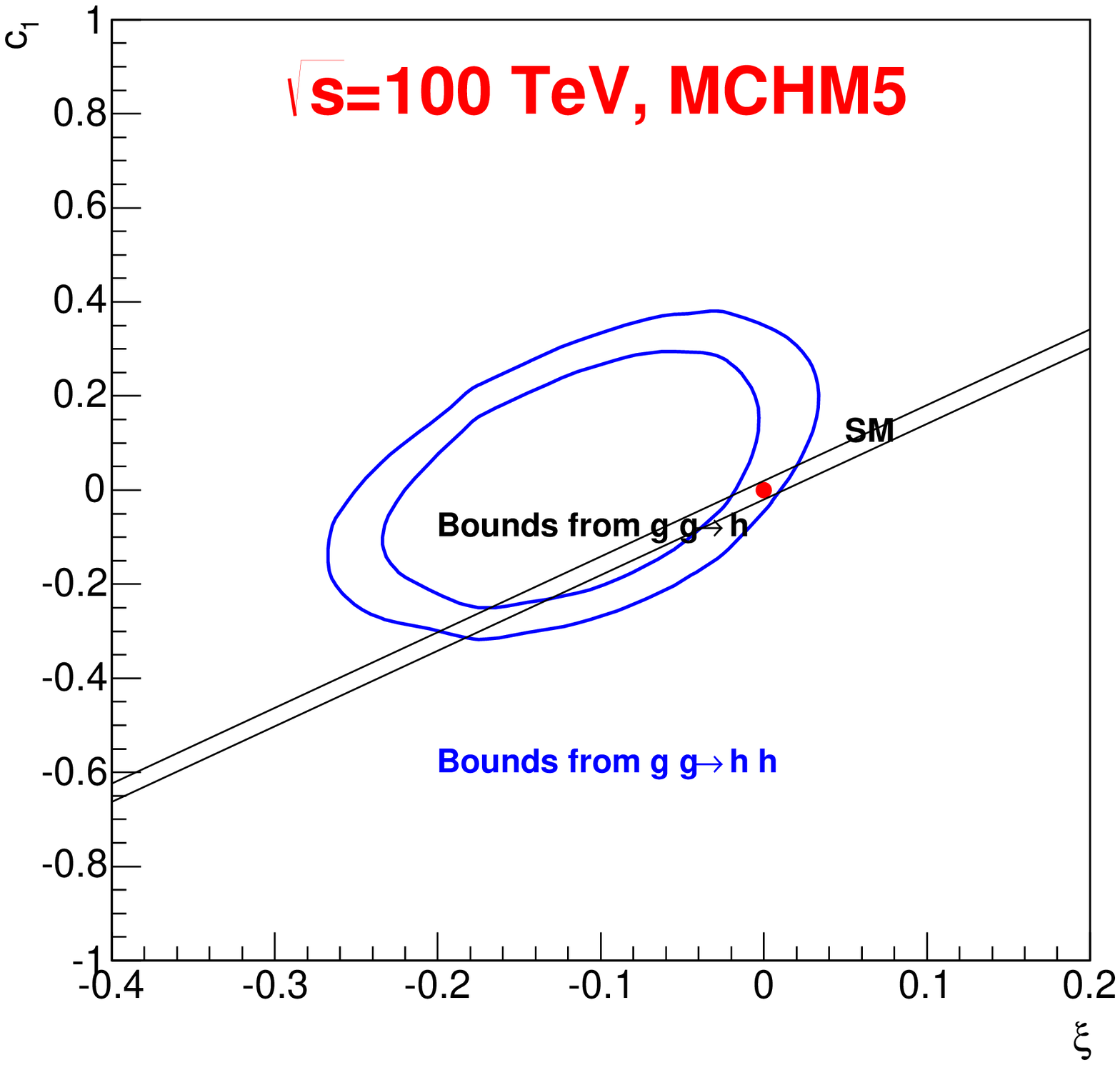}}
  \caption{Projected two-parameter exclusion bounds in the $\xi-c_1$ plane,
    extracted from the processes $gg \to h$ and $gg \to hh$ at a 100
    TeV collider for the models MCHM4 and MCHM5, respectively. }\label{mchms}
\end{figure}

In the generic dimension-six SILH Lagrangian, there are four free
parameters relevant to this stody, denoted by $C_y=c_y \xi$, $C_H = c_H \xi$,
$C_6 = c_6 \xi$, and $c_1$.  For both MCHM4 and MCHM5 as models which
reduce to SILH at low energy, there are only two independent
parameters, $\xi$ and $c_1$.  We apply the above analysis to those and
conclude that at a 100 TeV collider, data from $gg\to h$ and $gg \to
hh$ significantly constrain the allowed parameter space.  This is
demonstrated by Figs.~\ref{mchm4}--\ref{mchm5}.  Data from $gg \to h$
will result in an exclusion region bounded by two lines in the $\xi-c_1$
plane, while data from $gg \to hh$ will further reduce the allowed
parameter space to two small spots in the plane.

To illustrate the added value from triple-Higgs production, we consider two
benchmark points for MCHM4 and MCHM5 in Table~\ref{tab-mchms}. For
both benchmark points we obtain a large cross section for the $gg \to
hhh$ process, actually 80 and 55 times larger than that of the SM,
respectively.  These are examples of benchmark models that can not
just be detected, but also be distinguished from each other at a 100
TeV collider, given the result that the threshold cross section value
of $gg \to hhh$ for being sensitive to new physics is 30 fb or
smaller~\cite{Chen:2015gva}.  (We add the caveat that the benchmark
point of 
MCHM4 could be independently excluded by incorporating electroweak
precision data due to a large $\xi$ value.)

\begin{center}
\begin{table}
    \begin{center}
    \begin{tabular}{|c||c|c||c|c|c|}
        \hline
 No.       & $\xi$ & $c_1$ & $\sigma(gg\to h) [pb] $ &  $\sigma(gg \to h h)$ [fb] & $\sigma(gg \to h h h)$ [fb] \\
        \hline
        MCHM4 & 0.97 & 0.48& 764 & 1618 & 321 \\
        \hline
        MCHM5 & -0.20& -0.30 & 817 & 1854 & 122 \\
        \hline \hline
        GHM & $\hat{x}=0.02$& $\hat{r}=3.2$& 816 & 1786 & 37.78 \\
        \hline        
    \end{tabular}
    \end{center}
    \caption{Three representative points for the models MCHM4, MCHM5,
      and GHM, respectively, and the corresponding cross sections for
      Higgs production.\label{tab-mchms}} 
\end{table}
\end{center}

\subsubsection{The Gravity-Higgs Model}
The Gravity-Higgs model has only two free parameters, $\hat{x}$ and
$\hat{r}$.  The analysis is straightforward, since in this model,
single-Higgs production $gg \to h$ depends only on a single BSM
parameter $\hat{x}$, while double-Higgs production constrains the
second parameter $\hat{r}$. The cross section of $gg \to hhh$ is
completely determined once $\hat{x}$ and $\hat{r}$ are constrained.

\begin{figure}[htbp]
  \centering
  \subfigure{
  \label{mh14}\thesubfigure
  \includegraphics[width=0.4\textwidth]{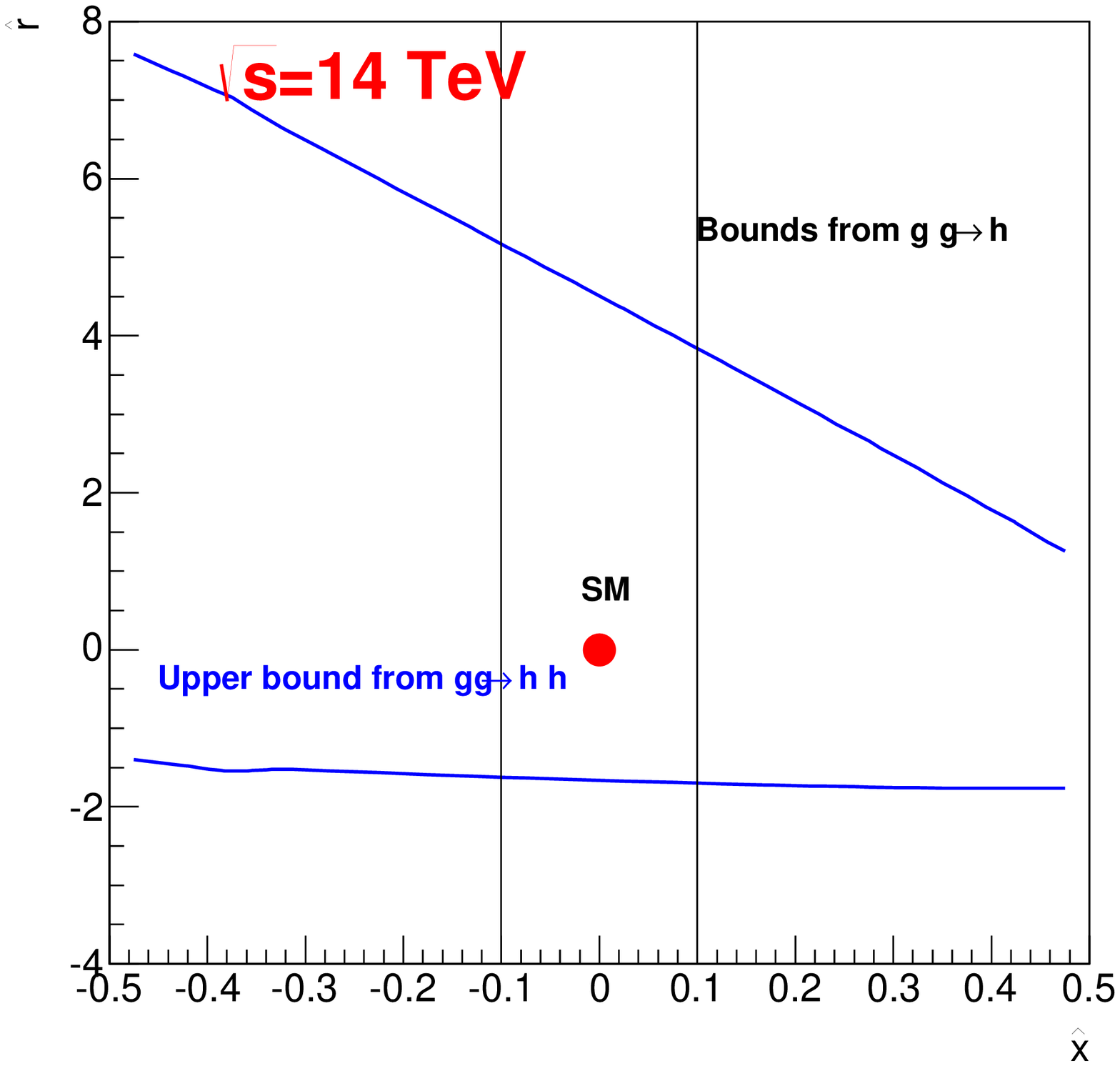}}
  \subfigure{
  \label{mh100}\thesubfigure
  \includegraphics[width=0.4\textwidth]{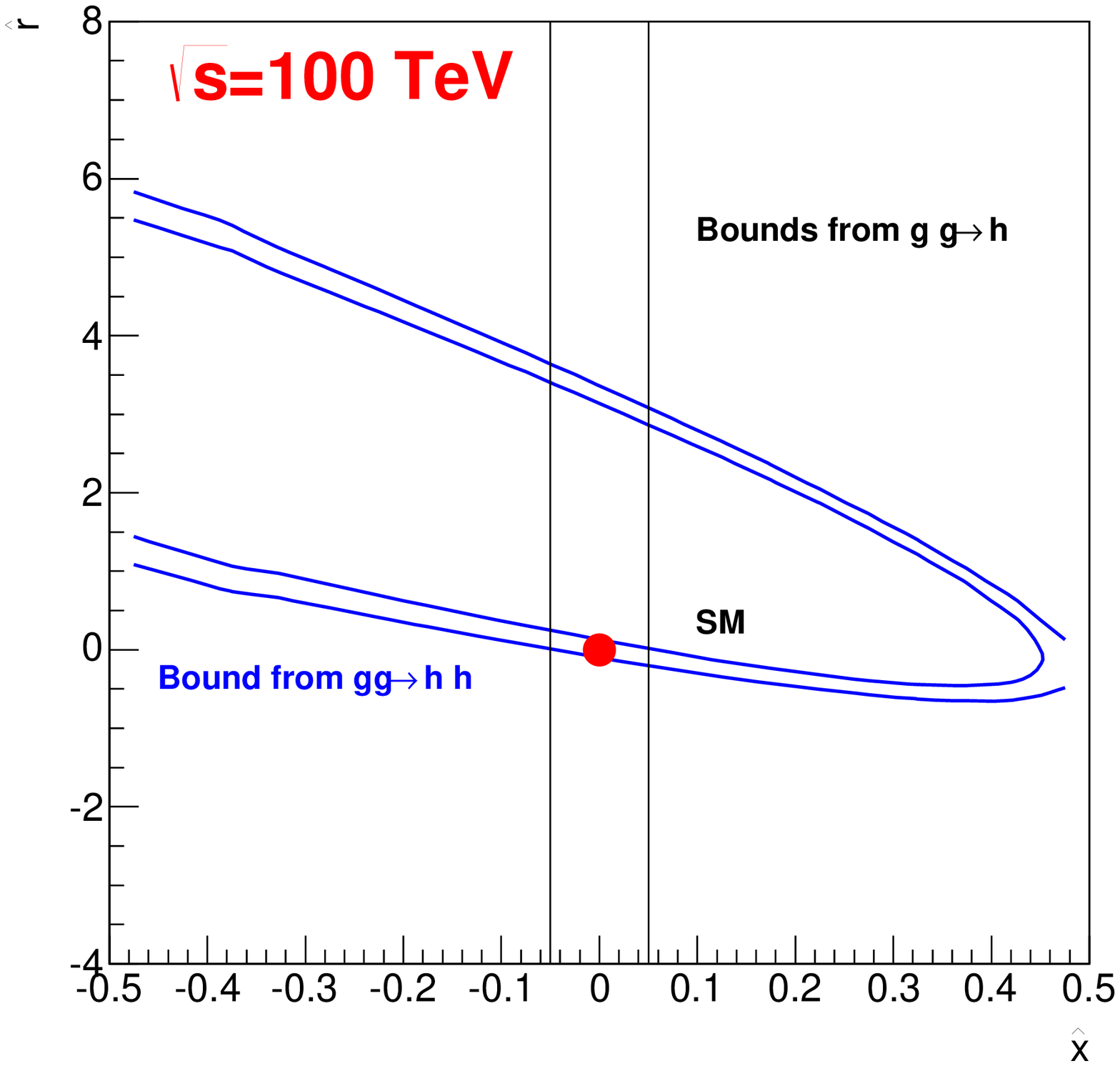}}
  \caption{Comparison of the projected exclusion bounds on
    $\hat{x}-\hat{r}$ from the LHC 14 TeV (left) and a 100 TeV
    collider (right).}\label{ghms}
\end{figure}

The expected LHC exclusion contours in the $\hat{x}-\hat{r}$ plane are
depicted in Fig.~\ref{mh14}.  From the process $gg \to h$ we obtain a
narrow band which is cut off by adding in the result from measuring
$gg\to hh$.  The latter constrains
the parameter $\hat{r}$ down to the range from -1.8 to 5.0 if we
assume that parameter space with a cross section of $\sigma(gg \to
hh)$ larger than 120 fb can be excluded.

Fig.~\ref{mh100} shows the analogous results for a $100$~TeV collider.
The measurements of $gg \to h$ and $gg \to hh$ let the allowed
parameter space shrink substantially compared to the LHC results.
From these measurements alone, the value of $\hat r$ is extracted with
a two-fold ambiguity.  However, for the solution with the larger value
of $\hat r$ ($\hat r\approx 3$) the cross section of $gg \to hhh$ is 5
times larger than near $\hat{r}=0$ due to the $\lambda_3^4$ dependence
in the cross section, cf.\ the benchmark points in
Table~\ref{tab-mchms}.  A measurement of triple-Higgs production could
therefore eliminate one of the solutions.

\section{Conclusion and discussion}
\label{Sec:conc}

In this paper, we have explored the discovery potential for
triple-Higgs production via the $2b2l^\pm4j+\missE$ decay channel at a
$100$ TeV hadron collider.  Despite the extremely small cross section
of the signal process in the SM, the parton-level study demonstrates
that the signal can be detected in principle.  We find that the $mT2$
variable is useful to find the correct combinations of the visible
objects that originate from Higgs boson decay, and to suppress
background efficiently.  However, once hadronic events and detector
effects are properly accounted for, extraction of the SM signal
becomes a real challenge.  The two main problems are that (1) the
transverse momentum of the softest jet from Higgs boson decay assumes
values of about
$10$ GeV, which makes it difficult to reconstruct; (2) since there are six
jets in the final state, the lepton isolation criteria reject most of
the signal events.

New-physics effects may enhance the cross section significantly and
distort kinematical distributions, so if the SM is not the true
theory, observation of the triple-Higgs production process becomes
more likely.  A measurement would then amount to a determination of
BSM parameters.  To discuss this possibility in a suitably generic
framework, we have employed an EFT approach and added a set
of dimension-6 operators that is taylored to represent new physics in
the Higgs sector.  In particular, we find that a sizable coefficient
for a derivative operator can modify the kinematical distributions of
the visible objects such that a reconstruction of the triple-Higgs
signal becomes feasible.  Using this information, we have investigated
the potential of such a measurement to improve on knowledge which can
already be gathered from single and double-Higgs production data.  It turns
out that while those processes are generally more powerful in constraining
BSM parameters, the triple-Higgs signal nevertheless reduces the
allowed parameter space and in some cases can eliminate
ambiguities in the parameter determination.

We would like to point out that our EFT approach
does not incorporate any new BSM particles which may be discovered in
the future. Our study also treats the SM particles,
especially the Higgs, as elementary degrees of freedom at the
energy scale relevant for 100 TeV collider.  We thus assume that, if
the SM Higgs is a composite particle, the compositeness scale is higher
than a
few TeV, at least.  If this assumption turns out to be invalid, i.e.,
qualitatively new phenomena become observable at lower energy, our
conservative approach would have to be revised to include explicit
model-dependent BSM effects in the calculation.

\appendix
\section{Derivation of parameters relations}
\label{appenda}

Expanding the SILH Lagragian in unitarity gauge and introducing the
physical Higgs scalar $h$, the derivative operator induces the
following term
\begin{eqnarray}
  \frac{c_H}{2f^2}\partial^\mu \left( H^\dagger H \right) \partial_\mu \left( H^\dagger H \right) &\to& \frac{c_H}{2f^2}(v+h)^2\partial^\mu{h}\partial_\mu{h}.
  \label{eq:a1}
\end{eqnarray}
In effect, the kinetic term of the Higgs field is modified to
\begin{eqnarray}
  \mathcal{L}_{kin} &=& \frac{1}{2}\left(1+c_H\xi\right)\partial^\mu{h}\partial_\mu{h},
  \label{eq:a2}
\end{eqnarray}
where $\xi\equiv v^2/f^2$. This means that the Higgs field should be
rescaled by $h\to\zeta h$, where
$\zeta=\left(1+c_H\xi\right)^{-1/2}$. Eq.~(\ref{eq:a1}) induces two further
derivative operators
\begin{eqnarray}
  \frac{c_H\xi}{v}\zeta^3 h\partial^\mu{h}\partial_\mu{h},   \\
  \frac{c_H\xi}{2v^2}\zeta^4 h^2\partial^\mu{h}\partial_\mu{h},
  \label{eq:a3}
\end{eqnarray}
which translate into the relations for $\kappa_5$ and $\kappa_6$ in
Table~\ref{tablecomposite}.

To find the relations for $\lambda_3$ and $\lambda_4$, we have to
consider the Higgs potential amended by the $c_6$ term:
\begin{eqnarray}
  \mathcal{V}\left( H^\dagger H \right) &=& -\mu^2\left( H^\dagger H \right)+\lambda\left( H^\dagger H \right)^2+\frac{c_6\lambda}{f^2}\left( H^\dagger H \right)^3
  \label{eq:a4}
\end{eqnarray}
In this case, the VEV is given by
\begin{eqnarray}
  -\mu^2+2\lambda v^2+\frac{3}{4}c_6\xi\lambda v^2 &=& 0,
  \label{eq:a5}
\end{eqnarray}
and the corresponding Higgs mass is defined by
\begin{eqnarray}
  \frac{1}{2}m^2_h &=& -\frac{1}{2}\mu^2+\frac{3}{2}\lambda v^2 + \frac{15}{8}c_6\xi\lambda v^2.
  \label{eq:a6}
\end{eqnarray}
After combining Eq.~(\ref{eq:a5}) and Eq.~(\ref{eq:a6}) and rescaling the
Higgs field, we obtain a modified Higgs mass
\begin{eqnarray}
  m^2_h &=& 2\lambda v^2\left(1+\frac{3}{2}c_6\xi\right)\zeta^2.
  \label{eq:a7}
\end{eqnarray}
With these definitions of Higgs field and Higgs mass, we can write
down the $h^3$ and $h^4$ terms:
\begin{eqnarray}
  \frac{m^2_h}{2v}\zeta\frac{1+5c_6\xi/2}{1+3c_6\xi/2} h^3,   \\
  \frac{m^2_h}{8v^2}\zeta^2\frac{1+15c_6\xi/2}{1+3c_6\xi/2} h^4,
  \label{eq:a8}
\end{eqnarray}
which yield the relations for $\lambda_3$ and $\lambda_4$ in
Table~\ref{tablecomposite}.

Finally, we consider the operator $\frac{c_yy_f}{f^2}H^\dagger H {\bar
  f}_L Hf_R$, which generates a term
\begin{eqnarray}
  \frac{c_yy_f}{f^2}H^\dagger H  {\bar f}_L Hf_R &\to& \frac{c_yy_f}{2\sqrt{2}f^2}(v+h)^3\bar{f}f.
  \label{eq:a9}
\end{eqnarray}
This term modifies the fermion mass by
\begin{eqnarray}
  m_f &=& \frac{y_fv}{\sqrt{2}}\left(1-\frac{1}{2}c_y\xi\right).
  \label{eq:a10}
\end{eqnarray}
After this redefinition, we obtain the following Higgs-fermion
interaction operators
\begin{eqnarray}
  && -\frac{m_f}{v}\zeta\frac{1-3c_y\xi/2}{1-c_y\xi/2}h\bar{f}f, \\
  && \frac{m_f}{v^2}\zeta^2\frac{3c_y\xi/2}{1-c_y\xi/2}hh\bar{f}f, \\
  && \frac{m_f}{v^3}\zeta^3\frac{c_y\xi/2}{1-c_y\xi/2}hhh\bar{f}f, 
  \label{eq:a11}
\end{eqnarray}
which yield the relations for $a_1$, $a_2$ and $a_3$ in
Table~\ref{tablecomposite}.

An alternative way of deriving such relations is performing a
non-linear transformation $h\to
h-\frac{c_H\xi}{2}(h+\frac{h^2}{v}+\frac{h^3}{3v^2})$~\cite{Giudice:2007fh}.
We compare the results of both approaches, up 
to $O(c_H\xi)$ terms, in Table~\ref{cmpredef} (the coefficients of
all other effective operators are set to zero).

\begin{center}
\begin{table}
    \begin{center}
    \begin{tabular}{|c|c|c|}
        \hline
        & Rescaling & Non-linear \\
        \hline
        $a_1$ & $1-\frac{1}{2}c_H\xi$ & $1-\frac{1}{2}c_H\xi$ \\
        $a_2$ & 0 & $-c_H\xi$ \\
        $a_3$ & 0 & $-c_H\xi$ \\
        $c_1$ & 0 & 0\\
        $c_2$ & 0 & 0\\
        $\lambda_3$ & $-\frac{1}{2}c_H\xi$ & $-\frac{3}{2}c_H\xi$ \\
        $\lambda_4$ & $-c_H\xi$ & $-\frac{25}{3}c_H\xi$ \\
        $\kappa_5$ & $-2c_H\xi$ & 0\\
        $\kappa_6$ & $-2c_H\xi$ & 0\\
        \hline
    \end{tabular}
    \end{center}
    \caption{Comparison of parameters relations between field rescaling and non-linear transformation.\label{cmpredef}}
\end{table}
\end{center}
Despite these differences, both transformations necessarily yield the
same cross section up to $O(c_H\xi)$.  However, in the non-linear
transformation approach, higher-order terms such as $O(\xi^2)$ are
much more complex than in the rescaling approach, and we have to deal
with vertices such as $tthh$ and $tthhh$ even if $c_y$ is
zero. Therefore, we adopt rescaling rather than the non-linear
transformation method for defining our phenomenological parameters.

\section{Numerical cross sections of $gg \to h$, $gg \to hh$, and $gg\to hhh$} 
\label{appendix}
The cross section for $gg \to h  $ can be put as
\bea
\sigma(gg \to h ) = K^{h} \times (\sum_{i=1}^{3}  F_i^{h} C^{i,h})\,, 
\label{xsech}
\eea
where the integrated form factors $F_i^h$ and the coefficients $C^{i,h}$ are given in Table \ref{numxsecht1}, and $K^h$ denotes the K-factor. The unit of $F_i^h$ is pb.
\begin{center}
\begin{table}{}
\begin{center}
\begin{tabular}{|c||c|c|c||c|c||c|c||}
\hline
 &  $K^h$  & $C^{1,h} = a_1^2$  & $C^{2,h} = a_1 c_1$& $ C^{3,h}=c_1^2$ \\
 \hline
  14 TeV & 2.85 & $F^{h}_{1}=19.15$ & $ F^{h}_{2} = 36.05$& $F^{h}_{3} = 17.14$ \\
 100 TeV &  2.24 & $F^{h}_{1}=357.53$ & $F^{h}_{2}=687.04$& $F^{h}_{3}=332.79$ \\
 \hline
\end{tabular}
\end{center}
\caption{\label{numxsecht1}The numerical value of $F^h_{1}-F^h_{3}$ at hadron colliders in Eq. (\ref{xsech}).}
\end{table}
\end{center}

It is found that values of $F^h_i$ given in Table (\ref{numxsecht1}) do produce a positive definite cross section of $gg \to h$.

The cross section for $gg \to h h $ at 14 TeV LHC and a 100 TeV collider can be put as
\bea
\sigma(gg \to h h ) = K^{2h} \times (\sum_{i=1}^{27}  F_i^{2h} C^{i,2h})\,, \label{xsechh}
\eea
where the integrated form factors $F_i^{2h}$ and the coefficients $C^{i,2h}$ are given in Table \ref{numxsechht1} and Table \ref{numxsechht2}, and $K^{2h}$ denotes the K-factor, which is equal to $2.20$ for the LHC 14 TeV and $2.17$ for the 100 TeV collision, respectively. The unit of $F_i^{2h}$ in these two tables is fb. 
\begin{center}
\begin{table}{}
\begin{center}
\begin{tabular}{|c|c||c|c||c|c||}
\hline
& $C^{2h} $& & $C^{2h} $& & $C^{2h}$ \\
\hline
$F^{2h}_{1}=36.17$ & $\power{a_1}{4}$ & 
$F^{2h}_{2}=-74.13$ & $\power{a_1}{2} a_2$ & 
$F^{2h}_{3}=44.58$ & $\power{a_2}{2}$ \\ 
$F^{2h}_{4}=-32.32$ & $\power{a_1}{2} c_2$ & 
$F^{2h}_{5}=43.31$ & $a_2 c_2$ & 
$F^{2h}_{6}=23.34$ & $\power{c_2}{2}$ \\ 
$F^{2h}_{7}=-48.98$ & $\power{a_1}{3} \kappa_5$ & 
$F^{2h}_{8}=56.9$ & $a_1 a_2 \kappa_5$ & 
$F^{2h}_{9}=-22.96$ & $\power{a_1}{2} c_1 \kappa_5$ \\ 
$F^{2h}_{10}=29.01$ & $a_2 c_1 \kappa_5$ & 
$F^{2h}_{11}=28.9$ & $a_1 c_2 \kappa_5$ & 
$F^{2h}_{12}=29.06$ & $c_1 c_2 \kappa_5$ \\ 
$F^{2h}_{13}=18.54$ & $\power{a_1}{2} \power{\kappa_5}{2}$ & 
$F^{2h}_{14}=19.78$ & $a_1 c_1 \power{\kappa_5}{2}$ & 
$F^{2h}_{15}=9.32$ & $\power{c_1}{2} \power{\kappa_5}{2}$ \\ 
$F^{2h}_{16}=-23.87$ & $\power{a_1}{3} \lambda_3$ & 
$F^{2h}_{17}=24.71$ & $a_1 a_2 \lambda_3$ & 
$F^{2h}_{18}=-13.7$ & $\power{a_1}{2} c_1 \lambda_3$ \\ 
$F^{2h}_{19}=14.78$ & $a_2 c_1 \lambda_3$ & 
$F^{2h}_{20}=14.53$ & $a_1 c_2 \lambda_3$ & 
$F^{2h}_{21}=11.36$ & $c_1 c_2 \lambda_3$ \\ 
$F^{2h}_{22}=17.28$ & $\power{a_1}{2} \kappa_5 \lambda_3$ & 
$F^{2h}_{23}=21.3$ & $a_1 c_1 \kappa_5 \lambda_3$ & 
$F^{2h}_{24}=8.19$ & $\power{c_1}{2} \kappa_5 \lambda_3$ \\ 
$F^{2h}_{25}=4.94$ & $\power{a_1}{2} \power{\lambda_3}{2}$ & 
$F^{2h}_{26}=6.68$ & $a_1 c_1 \power{\lambda_3}{2}$ & 
$F^{2h}_{27}=2.53$ & $\power{c_1}{2} \power{\lambda_3}{2}$\\
 \hline
\end{tabular}
\end{center}
\caption{\label{numxsechht1}The numerical value of $F^{2h}_{1}-F^{2h}_{27}$ at the LHC 14 TeV in Eq. (\ref{xsechh}).}
\end{table}
\end{center}

\begin{center}
\begin{table}{}
\begin{center}
\begin{tabular}{|c|c||c|c||c|c||}
\hline
& $C^{2h} $& & $C^{2h}$& & $C^{2h} $ \\
\hline
$F^{2h}_{1}=1565.1$ & $\power{a_1}{4}$ & 
$F^{2h}_{2}=-3346.56$ & $\power{a_1}{2} a_2$ & 
$F^{2h}_{3}=2274.94$ & $\power{a_2}{2}$ \\ 
$F^{2h}_{4}=-1232.43$ & $\power{a_1}{2} c_2$ & 
$F^{2h}_{5}=1790.73$ & $a_2 c_2$ & 
$F^{2h}_{6}=2407.17$ & $\power{c_2}{2}$ \\ 
$F^{2h}_{7}=-2133.02$ & $\power{a_1}{3} \kappa_5$ & 
$F^{2h}_{8}=2781.8$ & $a_1 a_2 \kappa_5$ & 
$F^{2h}_{9}=-857.36$ & $\power{a_1}{2} c_1 \kappa_5$ \\ 
$F^{2h}_{10}=1174.17$ & $a_2 c_1 \kappa_5$ & 
$F^{2h}_{11}=1202.36$ & $a_1 c_2 \kappa_5$ & 
$F^{2h}_{12}=2651.46$ & $c_1 c_2 \kappa_5$ \\ 
$F^{2h}_{13}=866.44$ & $\power{a_1}{2} \power{\kappa_5}{2}$ & 
$F^{2h}_{14}=797.2$ & $a_1 c_1 \power{\kappa_5}{2}$ & 
$F^{2h}_{15}=745.46$ & $\power{c_1}{2} \power{\kappa_5}{2}$ \\ 
$F^{2h}_{16}=-924.06$ & $\power{a_1}{3} \lambda_3$ & 
$F^{2h}_{17}=1014.84$ & $a_1 a_2 \lambda_3$ & 
$F^{2h}_{18}=-494.$ & $\power{a_1}{2} c_1 \lambda_3$ \\ 
$F^{2h}_{19}=567.$ & $a_2 c_1 \lambda_3$ & 
$F^{2h}_{20}=604.62$ & $a_1 c_2 \lambda_3$ & 
$F^{2h}_{21}=510.85$ & $c_1 c_2 \lambda_3$ \\ 
$F^{2h}_{22}=679.86$ & $\power{a_1}{2} \kappa_5 \lambda_3$ & 
$F^{2h}_{23}=817.06$ & $a_1 c_1 \kappa_5 \lambda_3$ & 
$F^{2h}_{24}=342.33$ & $\power{c_1}{2} \kappa_5 \lambda_3$ \\ 
$F^{2h}_{25}=172.4$ & $\power{a_1}{2} \power{\lambda_3}{2}$ & 
$F^{2h}_{26}=232.94$ & $a_1 c_1 \power{\lambda_3}{2}$ & 
$F^{2h}_{27}=88.15$ & $\power{c_1}{2} \power{\lambda_3}{2}$ \\
\hline
\end{tabular}
\end{center}
\caption{\label{numxsechht2}The numerical value of $F^{2h}_{1}-F^{2h}_{27}$ at a 100 TeV collider in Eq. (\ref{xsechh}).}
\end{table}
\end{center}

The largest absolute value goes to the coefficient $F_2^{2h}$, which is $74.13$ and $3346.56$ for either 14 TeV or 100 TeV cases. The minimal absolute value goes to the coefficient $F^{2h}_{27}$, which is $2.53$ for 14 TeV and $88.15$ for 100 TeV case.

Compared with those of the 14 TeV case, most of coefficients can be enhanced by a factor around 40 or so for the 100 TeV case. Among them, the coefficients $F_6^{2h}$, $F^{2h}_{12}$ and $F^{2h}_{15}$ have the largest enhancements from 14 TeV to 100 TeV collisions, and they are $103.1$, $91.2$, and $79.9$, respectively. 

In order to guarantee the positive and definite results of the cross section of all points in the parameter space, the contribution of b quark should be removed from the diagrams. Otherwise, a more general parameterisation of the cross section should be introduced. Furthermore, we have used more than 5,000 points in the parameter space of $a_2$, $c_2$, $\kappa_5$, and $\lambda_3$ to determine these $F^{2h}_i$ after taking into account the constraints on parameters $a_1$ and $c_1$ from the projected precision in the measurement  of $\sigma(gg \to h)$. The positivity and definiteness of the cross sections are examined to be hold in a random scan in the parameter space of $a_2$, $c_2$, $\kappa_5$, and $\lambda_3$ with a total number of points $10^7$. If $a_1$ and $c_1$ can significantly deviate from the values of the SM, these results might not be valid anymore.

The cross section for $gg \to h h h $ at a 100 TeV collider can be put as
\bea
\sigma(gg \to h h h) = K^{3h} \times (\sum_{i=1}^{154}  F_i^{3h} C^{i,3h})\,, \label{xsechhh}
\eea
where the integrated form factors $F^{3h}_i$ and the coefficients $C^{i,3h}$ are given in Table \ref{numxsechhht1} and Table \ref{numxsechhht2}, and K denotes the K-factor which is taken as $2.1$. The unit of $F^{3h}_i$ is fb. We have used more than 12,000 points to determine these $F^{3h}_i$.

The largest absolute coefficient is $F^{3h}_{30}$. In contrast, the smallest absolute coefficients are $F^{3h}_{81}$ and $F^{3h}_{83}$.

After taking into account the constraints on parameters $a_1$ and $c_1$ from the projected precision data  of $\sigma(gg \to h)$ and the constraints on parameters $a_2$, $c_2$, $\lambda_3$ and $\kappa_5$ from the projected precision data of $\sigma(gg \to hh)$, the positivity and definiteness of the cross sections are examined to be hold in a random scan in the parameter space of $a_3$, $\lambda_4$ and $\kappa_6$ with a total number of points $10^7$.
\begin{table}{}
\begin{tabular}{|c|c||c|c||c|c||c|c|}
\hline
& $C^{3h} $& & $C^{3h} $& & $C^{3h}$\\
\hline
$F^{3h}_{1}=7.47$ & $\power{a_1}{6}$ & 
$F^{3h}_{2}=-19.58$ & $\power{a_1}{4} a_2$ & 
$F^{3h}_{3}=31.31$ & $\power{a_1}{2} \power{a_2}{2}$ \\ 
$F^{3h}_{4}=-1.57$ & $\power{a_1}{3} a_3$ & 
$F^{3h}_{5}=-18.57$ & $a_1 a_2 a_3$ & 
$F^{3h}_{6}=11.62$ & $\power{a_3}{2}$ \\ 
$F^{3h}_{7}=-13.69$ & $\power{a_1}{5} \kappa_5$ & 
$F^{3h}_{8}=38.21$ & $\power{a_1}{3} a_2 \kappa_5$ & 
$F^{3h}_{9}=-35.02$ & $a_1 \power{a_2}{2} \kappa_5$ \\ 
$F^{3h}_{10}=-11.93$ & $\power{a_1}{2} a_3 \kappa_5$ & 
$F^{3h}_{11}=40.49$ & $a_2 a_3 \kappa_5$ & 
$F^{3h}_{12}=-12.54$ & $\power{a_1}{3} c_2 \kappa_5$ \\ 
$F^{3h}_{13}=68.7$ & $a_1 a_2 c_2 \kappa_5$ & 
$F^{3h}_{14}=-43.97$ & $a_3 c_2 \kappa_5$ & 
$F^{3h}_{15}=12.32$ & $\power{a_1}{4} \power{\kappa_5}{2}$ \\ 
$F^{3h}_{16}=-48.4$ & $\power{a_1}{2} a_2 \power{\kappa_5}{2}$ & 
$F^{3h}_{17}=35.74$ & $\power{a_2}{2} \power{\kappa_5}{2}$ & 
$F^{3h}_{18}=28.13$ & $a_1 a_3 \power{\kappa_5}{2}$ \\ 
$F^{3h}_{19}=-9.72$ & $\power{a_1}{3} c_1 \power{\kappa_5}{2}$ & 
$F^{3h}_{20}=46.85$ & $a_1 a_2 c_1 \power{\kappa_5}{2}$ & 
$F^{3h}_{21}=-28.8$ & $a_3 c_1 \power{\kappa_5}{2}$ \\ 
$F^{3h}_{22}=44.85$ & $\power{a_1}{2} c_2 \power{\kappa_5}{2}$ & 
$F^{3h}_{23}=-61.34$ & $a_2 c_2 \power{\kappa_5}{2}$ & 
$F^{3h}_{24}=935.73$ & $\power{c_2}{2} \power{\kappa_5}{2}$ \\ 
$F^{3h}_{25}=-16.66$ & $\power{a_1}{3} \power{\kappa_5}{3}$ & 
$F^{3h}_{26}=49.63$ & $a_1 a_2 \power{\kappa_5}{3}$ & 
$F^{3h}_{27}=30.55$ & $\power{a_1}{2} c_1 \power{\kappa_5}{3}$ \\ 
$F^{3h}_{28}=-40.08$ & $a_2 c_1 \power{\kappa_5}{3}$ & 
$F^{3h}_{29}=-33.7$ & $a_1 c_2 \power{\kappa_5}{3}$ & 
$F^{3h}_{30}=1244.83$ & $c_1 c_2 \power{\kappa_5}{3}$ \\ 
$F^{3h}_{31}=17.29$ & $\power{a_1}{2} \power{\kappa_5}{4}$ & 
$F^{3h}_{32}=-21.77$ & $a_1 c_1 \power{\kappa_5}{4}$ & 
$F^{3h}_{33}=414.36$ & $\power{c_1}{2} \power{\kappa_5}{4}$ \\ 
$F^{3h}_{34}=-0.57$ & $\power{a_1}{4} \kappa_6$ & 
$F^{3h}_{35}=-10.98$ & $\power{a_1}{2} a_2 \kappa_6$ & 
$F^{3h}_{36}=12.06$ & $a_1 a_3 \kappa_6$ \\ 
$F^{3h}_{37}=-2.73$ & $\power{a_1}{3} c_1 \kappa_6$ & 
$F^{3h}_{38}=20.91$ & $a_1 a_2 c_1 \kappa_6$ & 
$F^{3h}_{39}=-14.48$ & $a_3 c_1 \kappa_6$ \\ 
$F^{3h}_{40}=-6.95$ & $\power{a_1}{3} \kappa_5 \kappa_6$ & 
$F^{3h}_{41}=21.22$ & $a_1 a_2 \kappa_5 \kappa_6$ & 
$F^{3h}_{42}=13.81$ & $\power{a_1}{2} c_1 \kappa_5 \kappa_6$ \\ 
$F^{3h}_{43}=-20.28$ & $a_2 c_1 \kappa_5 \kappa_6$ & 
$F^{3h}_{44}=-22.57$ & $a_1 c_2 \kappa_5 \kappa_6$ & 
$F^{3h}_{45}=609.84$ & $c_1 c_2 \kappa_5 \kappa_6$ \\ 
$F^{3h}_{46}=14.81$ & $\power{a_1}{2} \power{\kappa_5}{2} \kappa_6$ & 
$F^{3h}_{47}=-25.91$ & $a_1 c_1 \power{\kappa_5}{2} \kappa_6$ & 
$F^{3h}_{48}=406.05$ & $\power{c_1}{2} \power{\kappa_5}{2} \kappa_6$ \\ 
$F^{3h}_{49}=3.17$ & $\power{a_1}{2} \power{\kappa_6}{2}$ & 
$F^{3h}_{50}=-7.38$ & $a_1 c_1 \power{\kappa_6}{2}$ & 
$F^{3h}_{51}=99.6$ & $\power{c_1}{2} \power{\kappa_6}{2}$ \\ 
$F^{3h}_{52}=-7.66$ & $\power{a_1}{5} \lambda_3$ & 
$F^{3h}_{53}=19.44$ & $\power{a_1}{3} a_2 \lambda_3$ & 
$F^{3h}_{54}=-15.69$ & $a_1 \power{a_2}{2} \lambda_3$ \\ 
$F^{3h}_{55}=-5.8$ & $\power{a_1}{2} a_3 \lambda_3$ & 
$F^{3h}_{56}=11.98$ & $a_2 a_3 \lambda_3$ & 
$F^{3h}_{57}=-6.43$ & $\power{a_1}{3} c_2 \lambda_3$ \\ 
$F^{3h}_{58}=13.84$ & $a_1 a_2 c_2 \lambda_3$ & 
$F^{3h}_{59}=-0.21$ & $a_3 c_2 \lambda_3$ & 
$F^{3h}_{60}=14.43$ & $\power{a_1}{4} \kappa_5 \lambda_3$ \\ 
$F^{3h}_{61}=-37.05$ & $\power{a_1}{2} a_2 \kappa_5 \lambda_3$ & 
$F^{3h}_{62}=22.98$ & $\power{a_2}{2} \kappa_5 \lambda_3$ & 
$F^{3h}_{63}=9.86$ & $a_1 a_3 \kappa_5 \lambda_3$ \\ 
$F^{3h}_{64}=-5.78$ & $\power{a_1}{3} c_1 \kappa_5 \lambda_3$ & 
$F^{3h}_{65}=10.78$ & $a_1 a_2 c_1 \kappa_5 \lambda_3$ & 
$F^{3h}_{66}=0.4$ & $a_3 c_1 \kappa_5 \lambda_3$ \\ 
$F^{3h}_{67}=9.62$ & $\power{a_1}{2} c_2 \kappa_5 \lambda_3$ & 
$F^{3h}_{68}=2.49$ & $a_2 c_2 \kappa_5 \lambda_3$ & 
$F^{3h}_{69}=73.58$ & $\power{c_2}{2} \kappa_5 \lambda_3$ \\ 
$F^{3h}_{70}=-18.8$ & $\power{a_1}{3} \power{\kappa_5}{2} \lambda_3$ & 
$F^{3h}_{71}=35.59$ & $a_1 a_2 \power{\kappa_5}{2} \lambda_3$ & 
$F^{3h}_{72}=7.37$ & $\power{a_1}{2} c_1 \power{\kappa_5}{2} \lambda_3$ \\ 
$F^{3h}_{73}=3.23$ & $a_2 c_1 \power{\kappa_5}{2} \lambda_3$ & 
$F^{3h}_{74}=4.32$ & $a_1 c_2 \power{\kappa_5}{2} \lambda_3$ & 
$F^{3h}_{75}=97.65$ & $c_1 c_2 \power{\kappa_5}{2} \lambda_3$ \\ 
\hline
\end{tabular}
\caption{\label{numxsechhht1}The numerical value of $F_{1}-F_{75}$ at 100TeV hadron collider in Eq. (\ref{xsechhh}).}
\end{table}

\begin{table}{}
\begin{tabular}{|c|c||c|c||c|c||c|c|}
\hline
& $C^{3h}$ & & $C^{3h} $& & $C^{3h}$ \\
\hline
$F^{3h}_{76}=13.75$ & $\power{a_1}{2} \power{\kappa_5}{3} \lambda_3$ & 
$F^{3h}_{77}=4.16$ & $a_1 c_1 \power{\kappa_5}{3} \lambda_3$ & 
$F^{3h}_{78}=32.71$ & $\power{c_1}{2} \power{\kappa_5}{3} \lambda_3$ \\ 
$F^{3h}_{79}=-3.37$ & $\power{a_1}{3} \kappa_6 \lambda_3$ & 
$F^{3h}_{80}=6.86$ & $a_1 a_2 \kappa_6 \lambda_3$ & 
$F^{3h}_{81}=0.01$ & $\power{a_1}{2} c_1 \kappa_6 \lambda_3$ \\ 
$F^{3h}_{82}=0.57$ & $a_2 c_1 \kappa_6 \lambda_3$ & 
$F^{3h}_{83}=-0.01$ & $a_1 c_2 \kappa_6 \lambda_3$ & 
$F^{3h}_{84}=22.56$ & $c_1 c_2 \kappa_6 \lambda_3$ \\ 
$F^{3h}_{85}=5.58$ & $\power{a_1}{2} \kappa_5 \kappa_6 \lambda_3$ & 
$F^{3h}_{86}=1.29$ & $a_1 c_1 \kappa_5 \kappa_6 \lambda_3$ & 
$F^{3h}_{87}=14.88$ & $\power{c_1}{2} \kappa_5 \kappa_6 \lambda_3$ \\ 
$F^{3h}_{88}=4.32$ & $\power{a_1}{4} \power{\lambda_3}{2}$ & 
$F^{3h}_{89}=-8.46$ & $\power{a_1}{2} a_2 \power{\lambda_3}{2}$ & 
$F^{3h}_{90}=5.24$ & $\power{a_2}{2} \power{\lambda_3}{2}$ \\ 
$F^{3h}_{91}=0.99$ & $a_1 a_3 \power{\lambda_3}{2}$ & 
$F^{3h}_{92}=-0.53$ & $\power{a_1}{3} c_1 \power{\lambda_3}{2}$ & 
$F^{3h}_{93}=0.37$ & $a_1 a_2 c_1 \power{\lambda_3}{2}$ \\ 
$F^{3h}_{94}=0.29$ & $a_3 c_1 \power{\lambda_3}{2}$ & 
$F^{3h}_{95}=1.19$ & $\power{a_1}{2} c_2 \power{\lambda_3}{2}$ & 
$F^{3h}_{96}=2.32$ & $a_2 c_2 \power{\lambda_3}{2}$ \\ 
$F^{3h}_{97}=7.71$ & $\power{c_2}{2} \power{\lambda_3}{2}$ & 
$F^{3h}_{98}=-7.67$ & $\power{a_1}{3} \kappa_5 \power{\lambda_3}{2}$ & 
$F^{3h}_{99}=11.39$ & $a_1 a_2 \kappa_5 \power{\lambda_3}{2}$ \\ 
$F^{3h}_{100}=0.94$ & $\power{a_1}{2} c_1 \kappa_5 \power{\lambda_3}{2}$ & 
$F^{3h}_{101}=3.02$ & $a_2 c_1 \kappa_5 \power{\lambda_3}{2}$ & 
$F^{3h}_{102}=3.06$ & $a_1 c_2 \kappa_5 \power{\lambda_3}{2}$ \\ 
$F^{3h}_{103}=12.29$ & $c_1 c_2 \kappa_5 \power{\lambda_3}{2}$ & 
$F^{3h}_{104}=5.69$ & $\power{a_1}{2} \power{\kappa_5}{2} \power{\lambda_3}{2}$ & 
$F^{3h}_{105}=3.39$ & $a_1 c_1 \power{\kappa_5}{2} \power{\lambda_3}{2}$ \\ 
$F^{3h}_{106}=5.$ & $\power{c_1}{2} \power{\kappa_5}{2} \power{\lambda_3}{2}$ & 
$F^{3h}_{107}=0.58$ & $\power{a_1}{2} \kappa_6 \power{\lambda_3}{2}$ & 
$F^{3h}_{108}=0.41$ & $a_1 c_1 \kappa_6 \power{\lambda_3}{2}$ \\ 
$F^{3h}_{109}=0.36$ & $\power{c_1}{2} \kappa_6 \power{\lambda_3}{2}$ & 
$F^{3h}_{110}=-0.96$ & $\power{a_1}{3} \power{\lambda_3}{3}$ & 
$F^{3h}_{111}=1.18$ & $a_1 a_2 \power{\lambda_3}{3}$ \\ 
$F^{3h}_{112}=-0.06$ & $\power{a_1}{2} c_1 \power{\lambda_3}{3}$ & 
$F^{3h}_{113}=0.44$ & $a_2 c_1 \power{\lambda_3}{3}$ & 
$F^{3h}_{114}=0.41$ & $a_1 c_2 \power{\lambda_3}{3}$ \\ 
$F^{3h}_{115}=0.69$ & $c_1 c_2 \power{\lambda_3}{3}$ & 
$F^{3h}_{116}=1.14$ & $\power{a_1}{2} \kappa_5 \power{\lambda_3}{3}$ & 
$F^{3h}_{117}=0.85$ & $a_1 c_1 \kappa_5 \power{\lambda_3}{3}$ \\ 
$F^{3h}_{118}=0.6$ & $\power{c_1}{2} \kappa_5 \power{\lambda_3}{3}$ & 
$F^{3h}_{119}=0.09$ & $\power{a_1}{2} \power{\lambda_3}{4}$ & 
$F^{3h}_{120}=0.07$ & $a_1 c_1 \power{\lambda_3}{4}$ \\ 
$F^{3h}_{121}=0.04$ & $\power{c_1}{2} \power{\lambda_3}{4}$ & 
$F^{3h}_{122}=0.16$ & $\power{a_1}{4} \lambda_4$ & 
$F^{3h}_{123}=-1.54$ & $\power{a_1}{2} a_2 \lambda_4$ \\ 
$F^{3h}_{124}=0.96$ & $a_1 a_3 \lambda_4$ & 
$F^{3h}_{125}=-0.59$ & $\power{a_1}{3} c_1 \lambda_4$ & 
$F^{3h}_{126}=0.9$ & $a_1 a_2 c_1 \lambda_4$ \\ 
$F^{3h}_{127}=0.11$ & $a_3 c_1 \lambda_4$ & 
$F^{3h}_{128}=-1.08$ & $\power{a_1}{3} \kappa_5 \lambda_4$ & 
$F^{3h}_{129}=1.86$ & $a_1 a_2 \kappa_5 \lambda_4$ \\ 
$F^{3h}_{130}=0.56$ & $\power{a_1}{2} c_1 \kappa_5 \lambda_4$ & 
$F^{3h}_{131}=0.28$ & $a_2 c_1 \kappa_5 \lambda_4$ & 
$F^{3h}_{132}=0.24$ & $a_1 c_2 \kappa_5 \lambda_4$ \\ 
$F^{3h}_{133}=3.71$ & $c_1 c_2 \kappa_5 \lambda_4$ & 
$F^{3h}_{134}=1.35$ & $\power{a_1}{2} \power{\kappa_5}{2} \lambda_4$ & 
$F^{3h}_{135}=0.47$ & $a_1 c_1 \power{\kappa_5}{2} \lambda_4$ \\ 
$F^{3h}_{136}=2.54$ & $\power{c_1}{2} \power{\kappa_5}{2} \lambda_4$ & 
$F^{3h}_{137}=0.54$ & $\power{a_1}{2} \kappa_6 \lambda_4$ & 
$F^{3h}_{138}=0.17$ & $a_1 c_1 \kappa_6 \lambda_4$ \\ 
$F^{3h}_{139}=1.15$ & $\power{c_1}{2} \kappa_6 \lambda_4$ & 
$F^{3h}_{140}=-0.65$ & $\power{a_1}{3} \lambda_3 \lambda_4$ & 
$F^{3h}_{141}=0.91$ & $a_1 a_2 \lambda_3 \lambda_4$ \\ 
$F^{3h}_{142}=0.08$ & $\power{a_1}{2} c_1 \lambda_3 \lambda_4$ & 
$F^{3h}_{143}=0.28$ & $a_2 c_1 \lambda_3 \lambda_4$ & 
$F^{3h}_{144}=0.22$ & $a_1 c_2 \lambda_3 \lambda_4$ \\ 
$F^{3h}_{145}=0.85$ & $c_1 c_2 \lambda_3 \lambda_4$ & 
$F^{3h}_{146}=0.83$ & $\power{a_1}{2} \kappa_5 \lambda_3 \lambda_4$ & 
$F^{3h}_{147}=0.48$ & $a_1 c_1 \kappa_5 \lambda_3 \lambda_4$ \\ 
$F^{3h}_{148}=0.68$ & $\power{c_1}{2} \kappa_5 \lambda_3 \lambda_4$ & 
$F^{3h}_{149}=0.11$ & $\power{a_1}{2} \power{\lambda_3}{2} \lambda_4$ & 
$F^{3h}_{150}=0.08$ & $a_1 c_1 \power{\lambda_3}{2} \lambda_4$ \\ 
$F^{3h}_{151}=0.06$ & $\power{c_1}{2} \power{\lambda_3}{2} \lambda_4$ & 
$F^{3h}_{152}=0.04$ & $\power{a_1}{2} \power{\lambda_4}{2}$ & 
$F^{3h}_{153}=0.03$ & $a_1 c_1 \power{\lambda_4}{2}$ \\ 
$F^{3h}_{154}=0.03$ & $\power{c_1}{2} \power{\lambda_4}{2}$
&&&&\\
\hline
\end{tabular}
\caption{\label{numxsechhht2} The numerical value of $F_{76}-F_{154}$ at 100TeV hadron collider in Eq. (\ref{xsechhh}).}
\end{table}

\begin{acknowledgments}
We would like to thank Ning Chen for contributing to this project at an early
stage, and Jiunn-Wei Chen for discussions.  Z.J. Zhao is supported by a
Nikolai Uraltsev Fellowship of the Center for Particle Physics,
University of Siegen.  S.C. Sun is supported by the MOST of Taiwan
under grant number of 105-2811-M-002-130 and the CRF Grants of the
Government of the Hong Kong SAR under HUKST4/CRF/13G.  Q.S. Yan and
X.R. Zhao are supported by the Natural Science Foundation of China
under the grant NO. 11175251 and NO. 11575005. X.R. Zhao is also
partially supported by the European Union as part of the FP7 Marie
Curie Initial Training Network MCnetITN (PITN-GA-2012-315877).
\end{acknowledgments}



\end{document}